\documentclass[reprint,prd,onecolumn,notitlepage,11pt]{revtex4-1}
\usepackage[english]{babel}
\usepackage{amsmath,amssymb,graphicx,bm,lipsum}
\usepackage[colorlinks=true,citecolor=blue,linkcolor=blue,urlcolor=blue]{hyperref} 
\usepackage[font={small},flushleft,indent]{caption}
\usepackage{setspace}

\begin{document}

\title{Flattened bispectrum of the scalar-induced gravitational waves}

\author{Qing-Hua Zhu}
\email{zhuqh@cqu.edu.cn}
\affiliation{School of Physics, Chongqing University, Chongqing 401331, China} 
  
\begin{abstract} 
  Recent pulsar timing array collaborations have reported evidence of the stochastic gravitational wave background. The gravitational waves induced by primordial curvature perturbations, referred to as scalar-induced gravitational waves (SIGWs), could potentially be the physical origins of the gravitational wave background. Due to nonlinearity of Einstein's gravity, there is non-Gaussianity of SIGWs even  when the sourced primordial curvature perturbation is Gaussian.   This paper investigates the intrinsic non-Gaussianity of SIGWs influenced by formation of primordial black holes. Specifically, we examine whether spectral width of Gaussian primordial curvature perturbations can affect non-Gaussianity of SIGWs. In order to ensure us to correctly quantify the degree of non-Gaussianity, we introduce an oscillation average scheme that can conserve the exact results of skewness of SIGWs. In this framework, the oscillation of SIGWs not only suppresses the bispectrum amplitude but also leads to a flattened-type bispectrum. Based on our results of skewness, it is found that the primordial curvature power spectrum with a narrower width can enhance the intrinsic non-Gaussianity.
\end{abstract} 
 
\maketitle

\section{Introduction}     

Recent pulsar timing array (PTA) collaborations have reported evidence of the stochastic gravitational wave background (SGWB) by observing the Hellings-Downs curves \cite{Hellings:1983fr, NANOGrav:2023gor, EPTA:2023fyk, Reardon:2023gzh, Xu:2023wog}. Revealing the origins of the observed SGWB can be considerably interesting because it is promising as a probe for studying cosmology at the early time \cite{NANOGrav:2023hvm, EPTA:2023xxk} or phenomena in astrophysics \cite{NANOGrav:2023pdq, NANOGrav:2023hfp, EPTA:2023gyr}. Notably, recent statistical analyses from PTA collaborations have suggested that the cosmological origin of SGWB, such as the scalar-induced gravitational waves (SIGWs) \cite{Ananda:2006af, Baumann:2007zm, Espinosa:2018eve, Kohri:2018awv}, seem to be more favored than the standard interpretation in astrophysics, i.e., the superposition of the gravitational waves produced by inspiring supermassive black hole binaries \cite{EPTA:2023xxk, NANOGrav:2023hvm}. Although further evidence would be required, it was expected by pioneers recently that we may be at the beginning of the era of observational gravitational wave cosmology \cite{Harigaya:2023pmw}. 

Because of the stochastic nature of SGWB, the statistical properties can provide lots of information on GW sources. The central limit theorem indicates that SGWB from astrophysical origin should be Gaussian because the individual source, such as supermassive black hole binaries, is reckoned as an independent system. On the cosmological side, the inflation theory suggests that the cosmological perturbations originate from quantum fluctuations \cite{Guth:1982ec, Starobinsky:1982ee}, and the statistical properties of SGWBs here are derived from the primordial fluctuations \cite{Grishchuk:1974ny, Starobinsky:1979ty, Thomas:2023poh, Cai:2022lec, Peng:2022ttg, Cai:2022nqv, Fumagalli:2021mpc, Cai:2020qpu, Ito:2020neq}. Therefore, it is motivated to explore the non-Gaussianity of the inflationary GWs \cite{Maldacena:2002vr, Bartolo:2004if, Bartolo:2020gsh, Aoki:2020ila, Adshead:2009bz}, as well as SIGWs generated by non-Gaussian primordial curvature perturbations \cite{Cai:2018dig, Inomata:2020xad, Atal:2021jyo, Yuan:2020iwf, Adshead:2021hnm, Ragavendra:2021qdu, Rezazadeh:2021clf, Zhang:2021rqs, Lin:2021vwc, Meng:2022ixx, Chen:2022dqr, Abe:2022xur, Chang:2023aba, Garcia-Saenz:2022tzu}. This was also extensively considered in the studies on the anisotropies of SIGWs \cite{Bartolo:2019oiq, Bartolo:2019zvb, Li:2023qua, Li:2023xtl}. In addition to the non-Gaussianity related to the primordial universe, SIGW itself is a non-Gaussian stochastic variable due to nonlinearity of Einstein's gravity \cite{Espinosa:2018eve, Bartolo:2018evs, Bartolo:2018rku}. In other words, there is intrinsic non-Gaussianity of SIGWs even when the sourced primordial curvature perturbation is Gaussian. 
This properties of SIGWs could provide a distinct way to extract physical information about the origins of SGWBs.

The feasibility of detecting non-Gaussian SGWB was explored in the GW detectors, such as LISA satellites \cite{Bartolo:2018qqn, Bartolo:2018evs, Bartolo:2018rku}, and PTAs \cite{Tsuneto:2018tif, Powell:2019kid, Tasinato:2022xyq, Zhu:2022bwf}. The presence of non-Gaussianity in SGWB might lead to a non-vanishing bispectrum for the detector's output \cite{Bartolo:2018qqn, Tsuneto:2018tif, Powell:2019kid} and can influence the Hellings-Downs curves \cite{Tasinato:2022xyq, Zhu:2022bwf}. The detectability serves as a motivation of our study on the non-Gaussianity of SIGWs. 
Theoretically, recent interest in SIGWs stems from the potential existence of a large primordial curvature perturbation on a small scale \cite{Ananda:2006af, Baumann:2007zm, Espinosa:2018eve, Kohri:2018awv, Sasaki:2018dmp, Domenech:2021ztg}. It significantly enhance the amplitude of SIGWs, thereby improving the detectability of GW strains. Moreover, the large curvature perturbation can also result in an enhancement of the anisotropies of SIGWs \cite{Bartolo:2019zvb, Li:2023qua, Li:2023xtl}, and might lead to primordial black hole overproduction \cite{NANOGrav:2023hvm, Franciolini:2023pbf, Dandoy:2023jot, Gorji:2023sil, Balaji:2023ehk, Harigaya:2023pmw}. It motivates us to examine whether the large curvature perturbations can enhance the intrinsic non-Gaussianity of SIGWs as well.

In this study, we investigate the intrinsic non-Gaussianity of SIGWs within a phenomenological manner. We employ the log-normal power spectrum for the curvature perturbations and subsequently compute the bispectrum and skewness. 
For SIGWs generated during the early time universe, such as the radiation-dominated era in $\Lambda$CDM cosmology, the spectrum and bispectrum are highly oscillating with frequency. To  reduce computational costs, the oscillation average is employed for calculating SIGW energy density spectrum \cite{Espinosa:2018eve, Bartolo:2018evs, Bartolo:2018rku,Kohri:2018awv,Maggiore:2018sht}. The oscillation average is expected to be a computational treatment, and should not affect statistical properties of SIGWs. However, extending it to bispectrum of SIGWs \cite{Espinosa:2018eve, Bartolo:2018evs, Bartolo:2018rku}, our calculation finds that pioneers' oscillation average scheme does not conserve the skewness of SIGWs. This poses a challenge for us to examine whether a large curvature perturbation induced by primordial black hole can affect the intrinsic non-Gaussianity.
In this context, we introduce alternative oscillation average scheme that can conserve the skewness.  In this framework, the oscillation of SIGWs not only suppresses the bispectrum amplitude but also leads to a flattened-type bispectrum ($k_3\approx k_1+k_2$). It inevitably differs from the results in previous studies that the equilateral configuration dominates the bispectrum \cite{Espinosa:2018eve}.  Additionally, our results show that the bispectrum in the collinear configuration, as well as the $\times\times\times$, $++\times$, $+\times+$, and $\times++$ components of the bispectrum vanish. Based on our results of skewness of SIGWs, it shows that the curvature power spectrum with a  narrower width can indeed enhance the intrinsic non-Gaussianity.

The rest of the paper is organized as follows. In Sec.~\ref{II}, we enumerate relevant formulas about the dynamics of SIGWs from previous studies. These formulas will be utilized for computing the bispectrum.  In Sec.~\ref{IIIa}, we introduce an oscillation average scheme for SIGWs generated during radiation-dominated era and present its motivation. In Sec.~\ref{III}, we compute the bispectra of SIGWs, including collinear  and triangle configuration. For the triangular bispectrum, the results for the matter-dominated era and radiation-dominated era are both presented. In Sec.~\ref{IV}, we calculate the skewness using the derived bispectrum and study its relation with spectral width of the curvature perturbations. In Sec.~\ref{V}, conclusions and discussions are summarized.

\

\section{Scalar-induced gravitational waves \label{II}}

Since we have little knowledge about the universe at the moment after the exit of the inflationary era, there are possibilities of the existence of enhanced curvature perturbations on a small scale, which can consequently lead to a large energy density of SIGWs \cite{Sasaki:2018dmp, Domenech:2021ztg}. In this section, we will enumerate the essential formulas that will be used in the computation of the bispectrum. Most of them can be found in pioneers' studies \cite{Ananda:2006af, Baumann:2007zm, Espinosa:2018eve, Kohri:2018awv}.
\subsection{Evolution of scalar-induced gravitational waves}

To evaluate the Einstein field equation in cosmological perturbation theory, we pertubing the metric in the conformal Newtonian gauge,
  ${\rm d} s^2  =  a^2 (\eta) \Big( - (1 + 2 \phi) {\rm d} t^2 + \big(
  \delta_{i   j} (1 + 2 \psi) + \frac{1}{2} h_{i   j} \big)
  {\rm d} x^i {\rm d} x^j \Big)$, 
where \(a\) is the conformal scalar factor, \(h_{i j}\) is the secondary gravitational wave, \(\psi\) and \(\phi\) are the curvature and Newton potential perturbations, respectively. They are also referred to as the scalar perturbations. 
After the exit of the inflationary era, the matter filed in the universe could be effectively considered as perfect fluids. Thus, one can obtain the Einstein field equations in the second order for SIGWs as $h_{i   j}'' + 2\mathcal{H}   h_{i   j}' - \Delta h_{i
    j}  =  - 4 \Lambda^{a   b}_{i   j} \mathcal{S}_{a
    b} ,  \label{eq:h}$
where $\mathcal{H}$ is the conformal Hubble parameter, and the effective source term is
\begin{eqnarray}
  \mathcal{S}_{a   b} & = & \frac{2 (5 + 3 w)}{3 (1 + w)} \partial_{a
   } \psi \partial_b \psi + \frac{4}{3 (1 + w) \mathcal{H}} (\partial_a
  \psi \partial_b \psi' + \partial_a \psi' \partial_b \psi') + \frac{4}{3 (1 +
  w) \mathcal{H}^2} \partial_a \psi' \partial_b \psi' ~. \nonumber\\
\end{eqnarray}
Given a constant equation of state parameter $w$, the conformal Hubble parameter can be evaluated to be $\mathcal{H}= 2 ((1 + 3 w) \eta)^{- 1}$. 
The evolution equation of the curvature perturbations is
$\psi'' + 3 (1 + c_s^2) \mathcal{H} \psi' + 3 (c_s^2 - w) \mathcal{H}^2 \psi
  - c_s^2 \Delta \psi  =  0$, 
where $c_s$ is the speed of sound.
 
In the momentum space, as equation for $h_{ij}$ reduces to an ordinary differential equation with respect to conformal time $\eta$, the Fourier modes of $h_{ij}$ can take the form of
\begin{eqnarray}
  h_{i   j, \textbf{k}} & = & \int \frac{{\rm d}^3 p}{(2 \pi)^3} \left\{
  \Psi_{\textbf{k} - \textbf{p}} \Psi_{\textbf{p}} \Theta_{i   j}
  \left( \textbf{k}, \textbf{p} \right) I_h \left( \left| \textbf{k} -
  \textbf{p} \right|, \left| \textbf{p} \right|, k, \eta \right) \right\}
  ~, \label{hGW}
\end{eqnarray}
where $\Theta_{ij}(\mathbf{k}, \mathbf{p}) \equiv k^{-2} \Lambda^{ab}_{ij}(\hat{k}) p_a p_b$, $\Lambda^{ab}_{ij}(\hat{k})$ is the transverse-traceless operator, $\Psi_{\mathbf{k}}$ is the initial curvature perturbations. One might find that $\Theta_{ij}(\mathbf{k}, \mathbf{p}) = \Theta_{ij}(\pm \mathbf{k}, \pm \mathbf{p})$ and $\Theta_{ij}(\mathbf{k}, \mathbf{p}) = \Theta_{ij}(\mathbf{k}, \mathbf{p} + \alpha \mathbf{k})$ for an arbitrary number $\alpha$. Here, $\Psi$ is related to primordial curvature perturbations $\zeta$ through $\Psi = ({3(1 + w)}/{(5 + 3w)}) \zeta$. The kernel function $I_h(|\mathbf{k} - \mathbf{p}|, p, k, \eta)$ can be given by
\begin{eqnarray}
  I_h \left( \left| \textbf{k} - \textbf{p} \right|, p, k, \eta \right) & = &
  \int_0^{\eta} {\rm d} \bar{\eta} \Bigg\{ 4 k^2 G_{k  } (\eta,  \bar{\eta})
  \nonumber\\ 
  &&  \times \left( \frac{2 (5 + 3 w) T_{\textbf{k} - \textbf{p}} T_{\textbf{p}}}{3 (1
  + w)} + \frac{4 \left( T_{ \textbf{k}-\textbf{p}} T'_{\textbf{p}} + T'_{\textbf{k -
  p}} T_{\textbf{p}} \right)}{3 (1 + w) \mathcal{H}} + \frac{4 T'_{\textbf{k} -
  \textbf{p}} T'_{\textbf{p}}}{3 (1 + w) \mathcal{H}^2} \right) \Bigg\} ~, \nonumber \\
\end{eqnarray} 
where $T_{\textbf{k}} (\eta)$ is the transfer function of the curvature
perturebations, namely, $\psi_{\textbf{k}} \equiv \Psi_{\textbf{k}}
T_{\textbf{k}} (\eta)$. In this sense, one can find $I_h \left( \left|
\textbf{k} - \textbf{p} \right|, p, k, \eta \right) = I_h \left( p, \left|
\textbf{k} - \textbf{p} \right|, k, \eta \right)$.

In radiation-dominated era, the secondary gravitational wave exhibits behavior akin to relativistic radiation as the conformal time $\eta$ approaches infinity. In this case, the kernel function reduces to
\begin{eqnarray}
  I_h \left( \left| \textbf{k} - \textbf{p} \right|, p, k, \eta \right) &
  \simeq & \frac{1}{k \eta} \left( \cos (k \eta) I_\text{A} \left( \left| \textbf{k}
  - \textbf{p} \right|, \left| \textbf{p} \right|, k \right) + \sin (k \eta)
  I_\text{B} \left( \left| \textbf{k} - \textbf{p} \right|, \left| \textbf{p}
  \right|, k \right) \right) \nonumber \\
  &\equiv& I^{({\rm RD})}_h \left( \left| \textbf{k}
  - \textbf{p} \right|, \left| \textbf{p} \right|, k, \eta \right) ~, \label{defIRD}
\end{eqnarray} 
where
\begin{subequations}
  \begin{eqnarray}
    I_\text{A} \left( \left| \textbf{k} - \textbf{p} \right|, p, k \right) & \equiv &
    \frac{27 \pi \left( 3 k^2 - \left| \textbf{k} - \textbf{p} \right|^2 - p^2
    \right)^2}{8 \left| \textbf{k} - \textbf{p} \right|^3 p^3} \left(
    {\rm sign} \left( 3 k - \sqrt{3} \left| \textbf{k} - \textbf{p} \right| -
    \sqrt{3} p \right) - 1 \right)~, \nonumber \\ \\
    I_\text{B} \left( \left| \textbf{k} - \textbf{p} \right|, p, k \right) & \equiv &
    \frac{27 \left( 3 k^2 - \left| \textbf{k} - \textbf{p} \right|^2 - p^2
    \right)}{4 \left| \textbf{k} - \textbf{p} \right|^3 p^3} \Bigg( 4 \left|
    \textbf{k} - \textbf{p} \right| p \nonumber\\ && + \left( 3 k^2 - \left| \textbf{k} -
    \textbf{p} \right|^2 - p^2 \right) \ln \left| \frac{3 k^2 + \left( \left|
    \textbf{k} - \textbf{p} \right| + p \right)^2}{3 k^2 + \left( \left|
    \textbf{k} - \textbf{p} \right| - p \right)^2} \right| \Bigg)~.
  \end{eqnarray} \label{defIRD2}
\end{subequations}
Similarly, in the matter-dominated era, we have 
\begin{eqnarray}
  I_h \left( \left| \textbf{k} - \textbf{p} \right|, p, k, \eta \right) & = &
  \frac{40 ((k \eta)^3 + 3 k \eta \cos (k \eta) - 3 \sin (k \eta))}{3 (k
  \eta)^3} \equiv I^{({\rm MD})}_h (k, \eta)~. \label{defIMD}
\end{eqnarray}
In contrast to the situation for the radiation-dominated era, the $I_h^{({\rm MD})}(k, \eta)$ tends to be a constant at the late time. 

\

\subsection{Statistics of the scalar-induced gravitational waves \label{II.B}}

The power spectrum $P_h^{\lambda_1\lambda_2}$ for the gravitational waves $h_{ij, \textbf{k}}$ can be obtained through the two-point functions, namely,
\begin{eqnarray}
  \left\langle h_{\textbf{k}_1}^{\lambda_1} h_{\textbf{k}_2}^{\lambda_2}
  \right\rangle & = & (2 \pi)^3 \delta \left( \textbf{k}_1 + \textbf{k}_2
  \right) P_h^{\lambda_1 \lambda_2} \left( \textbf{k}_2 \right) ~, \label{2PGW}
\end{eqnarray}
where the polarization component of the gravitational wave is given by
$h^{\lambda}_{\textbf{k}} \equiv e^{\lambda}_{ij} (
{\textbf{k}} ) h_{ij, \textbf{k}}$, and
$e^{\lambda}_{ij} ( {\textbf{k}} )$ represents
the polarization tensor.
The three-point functions of the $h^\lambda_{\textbf{k}}$ can then introduce the bispectrum $B_h^{\lambda_0\lambda_1\lambda_2}$ as follows,
\begin{eqnarray}
  \left\langle h^{\lambda_0}_{\textbf{k}_0} h^{\lambda_1}_{\textbf{k}_1}
  h_{\textbf{k}_2}^{\lambda_2} \right\rangle & = & (2 \pi)^3 \delta \left(
  \textbf{k}_0 + \textbf{k}_1 + \textbf{k}_2 \right) B_h^{\lambda_0 \lambda_1
  \lambda_2} \left( \textbf{k}_1, \textbf{k}_2 \right) ~. \label{3PGW}
\end{eqnarray}
In the context of a Gaussian stochastic variable, the bispectrum is expected to vanish.

For the zero-mean variable $h^\lambda_{\textbf{k}}$, its amplitude can be characterized by the variance $\sigma^2$, namely,
\begin{eqnarray}
  \sigma^2_{\lambda_1 \lambda_2} & = & \int \frac{{\rm d}^3 k}{(2 \pi)^3}
  P_h^{\lambda_1 \lambda_2} \left( \textbf{k} \right) = \int {\rm d}  
  \ln   k \int \frac{{\rm d} \Omega}{4 \pi} \mathcal{P}^{\lambda_1
  \lambda_2}_h \left( \textbf{k} \right)~,\label{var}
\end{eqnarray}
where the dimensionless spectrum can be given by
$\mathcal{P}_h^{\lambda_1 \lambda_2} \left( \textbf{k} \right) = (k^3 / 2
\pi^2) P_h^{\lambda_1 \lambda_2} \left( \textbf{k} \right)$. Under the
assumption of an isotropic spectrum, one can obtain $\int \frac{{\rm d} \Omega}{4 \pi}
\mathcal{P}^{\lambda_1 \lambda_2}_h (\textbf{k}) =\mathcal{P}_h^{\lambda_1 \lambda_2}
(k)$. To quantify the non-Gaussianity of $h^\lambda_{\textbf{k}}$, one can consider the third moment $\mu^3$, which can be computed by making use of the bispectrum, namely,
\begin{equation}
  \mu^3_{\lambda_0 \lambda_1 \lambda_2} = \int \frac{{\rm d}^3 k_1}{(2 \pi)^3}
  \frac{{\rm d}^3 k_2}{(2 \pi)^3} B_h^{\lambda_0 \lambda_1 \lambda_2} \left(
  \textbf{k}_1, \textbf{k}_2 \right) = \int {\rm d}   \ln   k_1
  \int {\rm d}   \ln   k_2 \int \frac{{\rm d} \Omega_1}{4 \pi}
  \int \frac{{\rm d} \Omega_2}{4 \pi} \mathcal{B}_h^{\lambda_0 \lambda_1
  \lambda_2} \left( \textbf{k}_1, \textbf{k}_2 \right) ~. \label{trimon}
\end{equation}
From Eq.~(\ref{trimon}), the dimensionless bispectrum can be given by 
\begin{eqnarray}
  \mathcal{B}_h^{\lambda_1 \lambda_2 \lambda_3} \left( \textbf{k}_1,
  \textbf{k}_2 \right) & = & \frac{k_1^3}{2 \pi^2} \frac{k_2^3}{2 \pi^2}
  B_h^{\lambda_1 \lambda_2 \lambda_3} \left( \textbf{k}_1, \textbf{k}_2
  \right) ~.
\end{eqnarray}
If the bispectrum only depends on the norms $k_1$, $k_2$ and the angle
$\theta_k \equiv \arccos ( \hat{\textbf{k}}_1 \cdot
\hat{\textbf{k}}_2 )$, Eq.~(\ref{trimon}) can reduce to
\begin{eqnarray}
  \mu^3_{\lambda_0 \lambda_1 \lambda_2} & = & \int {\rm d}   \ln
    k_1 \int {\rm d}   \ln   k_2 \int_{- 1}^1 \frac{{\rm d}
  \cos \theta_k}{2} \mathcal{B}_h^{\lambda_0 \lambda_1 \lambda_2} (k_1, k_2,
  \theta_k) ~. \label{12}
\end{eqnarray}
With the third moment in Eq.~(\ref{trimon}) and variance in Eq.~(\ref{var}), one can give the normalized third moment, also referred to as skewness, as follows,
\begin{eqnarray}
  \Gamma & \equiv & \frac{\mu^3}{\sigma^3} ~. \label{defSkew}
\end{eqnarray}
In this study, we will utilize the skewness as a measure to quantify the degree of non-Gaussianity for SIGWs.

The power spectra $P_h^{\lambda_1 \lambda_2} \left( \textbf{k}_2 \right)$ and bispectra $B_h^{\lambda_0 \lambda_1 \lambda_2} \left( \textbf{k}_1, \textbf{k}_2 \right)$ are derived from the statistical properties of the $\psi$, since SIGWs are generated by the curvature perturbation $\psi$.  In this study, we consider a Gaussian primordial curvature perturbation. As a result, the curvature perturbation is inherently Gaussian due to $\Psi = ({3(1 + w)}/{(5 + 3w)}) \zeta|_{\eta=0}$. Thus we can only consider the two-point functions of the curvature perturbation as follows,
\begin{eqnarray}
  \left\langle \Psi_{\textbf{k}} \Psi_{\textbf{k}'} \right\rangle & = & (2
  \pi)^3 \delta \left( \textbf{k} + \textbf{k}' \right) P_{\Psi}(k) ~, \label{2Pcurvature}
\end{eqnarray}
where $P_{\Psi}(k)$ is the power spectrum of the initial curvature perturbations.
Because the Gaussian stochastic variable $\Psi$, the higher-order correlations of $\Psi$ are deemed unnecessary for deriving the bispectrum of SIGWs.

\ 

\subsection{Spectrum and Bispectrum of scalar-induced gravitational waves }

Due to the interest in the energy density spectrum of SIGWs, the power spectrum has been studied by pioneers \cite{Ananda:2006af, Baumann:2007zm, Espinosa:2018eve, Kohri:2018awv, Sasaki:2018dmp,Domenech:2021ztg}. By making use of Eqs.~(\ref{hGW}), (\ref{2PGW}) and (\ref{2Pcurvature}), the power spectrum can be given by
\begin{eqnarray}
  P_h^{\lambda \lambda'} \left( \textbf{k} \right) & = & e^{\lambda}_{i
    j} \left( \textbf{k} \right) e^{\lambda'}_{a   b} \left(
  \textbf{k} \right) \int \frac{{\rm d}^3 p}{(2 \pi)^3} \left\{ 2 P_{\Psi}
  \left( \left| \textbf{k} - \textbf{p} \right| \right) P_{\Psi}\left( \left|
  \textbf{p} \right| \right) \Theta_{i   j} \left( \textbf{k},
  \textbf{p} \right) \Theta_{a   b} \left( \textbf{k}, \textbf{p}
  \right) I_h \left( \left| \textbf{k} - \textbf{p} \right|, \left| \textbf{p}
  \right|, k, \eta \right)^2 \right\} \nonumber \\
  & = & \frac{1}{2} \delta^{\lambda \lambda'} \int \frac{{\rm d}^3 p}{(2
  \pi)^3} \left\{ \left( 1 - \left( \hat{\textbf{k}} \cdot
  \hat{\textbf{p}} \right)^2 \right)^2 \left( \frac{\left| \textbf{p}
  \right|}{k} \right)^4 P_{\Psi}\left( \left| \textbf{k} - \textbf{p} \right|
  \right) P_{\Psi}\left( \left| \textbf{p} \right| \right) I_h \left( \left|
  \textbf{k} - \textbf{p} \right|, \left| \textbf{p} \right|, k, \eta
  \right)^2 \right\}~.\nonumber \\ \label{spectrah} 
\end{eqnarray}
Becuase the SIGW is a non-Gaussian stochastic variable, there could be a non-vanishing three-point function of $h_{i   j, \textbf{k}}$, which can be evaluated  with Eqs.~(\ref{hGW}) and (\ref{2Pcurvature}), namely
\begin{eqnarray}
  \left\langle h_{i   j, \textbf{k}_0} h_{a   b, \textbf{k}_1}
  h_{c   d, \textbf{k}_2} \right\rangle & = & \int \frac{{\rm d}^3 p_0
  {\rm d}^3 p_1 {\rm d}^3 p_2}{(2 \pi)^9} \left\{ \left\langle
  \Psi_{\textbf{k}_0 - \textbf{p}_0} \Psi_{\textbf{p}_0} \Psi_{\textbf{k}_1 -
  \textbf{p}_1} \Psi_{\textbf{p}_1} \Psi_{\textbf{k}_2 - \textbf{p}_2}
  \Psi_{\textbf{p}_2} \right\rangle \right. \nonumber\\
  && \left. \times \Theta_{i   j} \left( \textbf{k}_0,
  \textbf{p}_0 \right) \Theta_{a   b} \left( \textbf{k}_1, \textbf{p}_1
  \right) \Theta_{c   d} \left( \textbf{k}_2, \textbf{p}_2 \right)
  \right. \nonumber\\ && \left. \times I_h \left( \left| \textbf{k}_0 - \textbf{p}_0 \right|, p_0, k_0, \eta
  \right) I_h \left( \left| \textbf{k}_1 - \textbf{p}_1 \right|, p_1, k_1,
  \eta \right) I_h \left( \left| \textbf{k}_2 - \textbf{p}_2 \right|, p_2,
  k_1, \eta \right) \right\} \nonumber\\
  & = & (2 \pi)^3 \delta \left( \textbf{k}_0 + \textbf{k}_1 + \textbf{k}_2
  \right) \Bigg( \left. 4 \int \frac{{\rm d}^3 p}{(2 \pi)^3} \left. \Big\{ P_{\Psi}\left(
  \left| \textbf{k}_2 - \textbf{p} \right| \right) P_{\Psi}\left( \left|
  \textbf{k}_1 + \textbf{k}_2 - \textbf{p} \right| \right) P_{\Psi}(p) \right.\right. \nonumber\\ &&  \times
  \Theta_{i   j} \left( \textbf{k}_1 + \textbf{k}_2, \textbf{p} \right)
  \Theta_{a   b} \left( \textbf{k}_1, - \textbf{k}_2 + \textbf{p}
  \right) \Theta_{c   d} \left( \textbf{k}_2, \textbf{p} \right) \nonumber\\ && \times I_h
  \left( \left| \textbf{k}_1 + \textbf{k}_2 - \textbf{p} \right|, \left|
  \textbf{p} \right|, \left| \textbf{k}_1 + \textbf{k}_2 \right|, \eta \right)
   \nonumber\\ && \times I_h \left( \left| \textbf{k}_1 + \textbf{k}_2 - \textbf{p} \right|, \left|
  \textbf{k}_2 - \textbf{p} \right|, \left| \textbf{k}_1 \right|, \eta \right)
  I_h \left( \left| \textbf{k}_2 - \textbf{p} \right|, \left| \textbf{p}
  \right|, \left| \textbf{k}_2 \right|, \eta \right)  \Big\} \nonumber\\ &&
  + \left(
  \textbf{k}_1 \leftrightarrow \textbf{k}_2, a \leftrightarrow b, c\leftrightarrow d \right)  \Bigg) ~. \label{3Phij}
\end{eqnarray}
The delta function $\delta (\textbf{k}_0 + \textbf{k}_1 + \textbf{k}_2)$ in Eq.~(\ref{3Phij}) indicates conservation of the momenta. The triangular configurations are formed by the momenta, referred to as the shape of the bispectrum. Associating Eq.~(\ref{3Phij}) with Eq.~(\ref{3PGW}), we obtain the bispectrum as
\begin{eqnarray}
  B_h^{\lambda_0 \lambda_1 \lambda_2} \left( \textbf{k}_1, \textbf{k}_2
  \right) & \equiv & B^{\lambda_0 \lambda_1 \lambda_2} \left(
  \textbf{k}_1, \textbf{k}_2 \right) + B^{\lambda_0 \lambda_2 \lambda_1}
  \left( \textbf{k}_2, \textbf{k}_1 \right) ~,\label{defBh}
\end{eqnarray}
where
\begin{eqnarray}
  B^{\lambda_0 \lambda_1 \lambda_2} \left( \textbf{k}_1, \textbf{k}_2 \right)
  & = & 4 \int \frac{{\rm d}^3 p}{(2 \pi)^3} \Big\{ P_{\Psi}\left( \left|
  \textbf{k}_2 - \textbf{p} \right| \right) P_{\Psi}\left( \left|
  \textbf{k}_1 + \textbf{k}_2 - \textbf{p} \right| \right) P_{\Psi}(p)
  \mathbb{P}^{\lambda_0 \lambda_1 \lambda_2} \nonumber \\ && \times
  I_h \left( \left| \textbf{k}_1 +
  \textbf{k}_2 - \textbf{p} \right|, p, \left| \textbf{k}_1 + \textbf{k}_2
  \right|, \eta \right) \nonumber \\ && \times  I_h \left( \left| \textbf{k}_1 + \textbf{k}_2 -
  \textbf{p} \right|, \left| \textbf{k}_2 - \textbf{p} \right|, k_1, \eta
  \right) I_h \left( \left| \textbf{k}_2 - \textbf{p} \right|, p, k_2, \eta
  \right) \Big\} ~.  \label{defBispectra}
\end{eqnarray}
The kernel functions $I_h$ for the radiation-dominated era and matter-dominated era have been shown in Eqs.~(\ref{defIMD}) and (\ref{defIRD}), and 
\begin{eqnarray}
  \mathbb{P}^{\lambda_0 \lambda_1 \lambda_2} & \equiv & e^{\lambda_0}_{i
    j} \left(- \textbf{ k}_1 - \textbf{k}_2 \right) e^{\lambda_1}_{a
    b} \left( \textbf{k}_1 \right) e^{\lambda_2}_{c   d} \left(
  \textbf{k}_2 \right) \Theta_{i   j} \left( \textbf{k}_1 +
  \textbf{k}_2, \textbf{p} \right) \Theta_{a   b} \left( \textbf{k}_1,
  - \textbf{k}_2 + \textbf{p} \right) \Theta_{c   d} \left(
  \textbf{k}_2, \textbf{p} \right) ~. \nonumber \\ \label{defPP}
\end{eqnarray}
The $\mathbb{P}^{\lambda_0 \lambda_1 \lambda_2}$ can be obtained by providing the representation of the polarization tensors. We will present explicit expression of $\mathbb{P}^{\lambda_0 \lambda_1 \lambda_2}$ latter in Sec.~\ref{III.B}.

\ 

\section{An oscillation average scheme conserving skewness \label{IIIa}}
From Eqs.~(\ref{defIRD}) and (\ref{defIMD}), the kernel functions for matter-dominated era tends to be a constant as $\eta\rightarrow\infty$, while those for radiation-dominated era are highly oscillating with frequency. The latter one can be shown through expanding the kernel function in Eq.~(\ref{defIRD}) with a small fraction of $k$, which leads to
\begin{eqnarray}
  I_h  &  \simeq & \frac{\cos ((k+\delta k) \eta) I_\text{A}  + \sin ((k+\delta k) \eta)   I_\text{B}}{k \eta} + \mathcal{O}(\delta k)~. \label{15}
\end{eqnarray}
In the case of $1/\eta \ll \delta k$, the kernel function with a large $\eta$ highly oscillate with respect to $\delta k$. In PTA observations, frequency resolution $\delta f(=\delta k/2\pi)$ is proportional to inverse of the observation time $1/T_\text{obs}\simeq 10^{-8}\text{s}^{-1}$, and the inverse of current time of the universe is $1/\eta_0\simeq 10^{-18}\text{s}^{-1}$. It indicates that kernel function in Eq.~(\ref{15}) varies very fast with respect $\delta k$. 
In order to obtain a result reflecting the output on GW detctors like PTAs, the oscillation average is well-motivated \cite{Maggiore:2018sht}. 


In this context, the pioneers replace the oscillating kernel functions with its envelope, namely, $I_h\rightarrow \sqrt{I^2_\text{A}+I^2_\text{B}}/(k\eta)$, in order to obtain  smoothed spectrum and bispectrum of SIGWs \cite{Espinosa:2018eve, Bartolo:2018evs, Bartolo:2018rku}. It is expected that utilization of the oscillation averages does not change statistical properties of the $h_{ij,\textbf{k}}$. However, it is found that above envelope scheme for oscillation averages can result in a different value of skewness of SIGWs. It motivated us to introduce an alternative oscillation average scheme. In the subsequent parts, we will elaborate on these points.  

\ 

\subsection{Skewness of SIGWs as the conformal time tends to be infinity }

As given in Eqs.~(\ref{var}), (\ref{trimon}) and (\ref{defSkew}),  the skewness can be obtained via integrals of spectrum and bispectrum over the momentum space. For highly oscillatory kernel functions in Eq.~(\ref{defIRD}), the skewness can be calculated in approximation under $\eta\rightarrow\infty$, in accordance with the Riemann-Lebesgue lemma, which states that for an integrable function $g(\textbf{k})$, we have
\begin{eqnarray}
  \lim_{|\xi|\rightarrow\infty}\int \textrm{d}^3k \big\{\cos{(\textbf{k}\cdot \xi})g(\textbf{k})\big\} = 0~, & & \lim_{|\xi|\rightarrow\infty}\int \textrm{d}^3k \big\{\sin{(\textbf{k}\cdot \xi})g(\textbf{k})\big\} = 0~. \label{21}
\end{eqnarray}
The highly oscillatory terms from the trigonometric functions can result in cancellation of the integrals. This lemma has been utilized to address highly oscillatory terms in Helling-Downs curves in the studies of detecting SIGWs in PTAs \cite{Chamberlin:2011ev, Mingarelli:2018kgp}.
Here, it implies that highly oscillatory terms in kernel functions lead to negligible contributions to the variance and skewness.  

In the following, we will calculate the skewness of SIGWs based on Riemann-Lebesgue lemma in Eq.~(\ref{21}). 
By making use of Eqs.~(\ref{defIRD}) and (\ref{spectrah}), the spectrum of SIGWs can be evaluated to be
\begin{eqnarray}
  P_h^{\lambda \lambda'} \left( \textbf{k} \right) & = & \frac{1}{2}
  \delta^{\lambda \lambda'} \int \frac{\mathrm{d}^3 p}{(2 \pi)^3} \Bigg\{ \left( 1
  - \left( \hat{\textbf{k}} \cdot \hat{\textbf{p}} \right)^2 \right)^2
  \left( \frac{\left| \textbf{p} \right|}{k} \right)^4 P_{\Psi} \left( \left|
  \textbf{k} - \textbf{p} \right| \right) P_{\Psi} (p) \nonumber \\ && \times \left(\frac{I_\text{A}
  \cos (k \eta) + I_\text{B} \sin (k \eta)}{k \eta} \right)^2 \Bigg\} \nonumber \\
  & = & \frac{1}{2} \delta^{\lambda \lambda'} \int \frac{\mathrm{d}^3 p}{(2
  \pi)^3} \Bigg\{ \left( 1 - \left( \hat{\textbf{k}} \cdot
  \hat{\textbf{p}} \right)^2 \right)^2 \left( \frac{\left| \textbf{p}
  \right|}{k} \right)^4 P_{\Psi} \left( \left| \textbf{k} - \textbf{p} \right|
  \right) P_{\Psi} (p) \nonumber \\ && \times  \frac{1}{(k \eta)^2} \left( \frac{1}{2} (I_\text{A}^{(2)} + I_\text{B}^{(2)}) +
  \frac{1}{2} \cos (2 k \eta) (I_\text{A}^{(2)} - I_\text{B}^{(2)}) + I_\text{A} I_\text{B} \sin (2 k \eta)
  \right) \Bigg\}~. \label{22}\nonumber \\ &&
\end{eqnarray}
Here, Eq.~(\ref{22}) contains oscillatory terms involving $\cos(2k\eta)$ and $\sin(2k\eta)$. Based on Eqs.~(\ref{var})  and (\ref{21}), the variance can be given by
\begin{eqnarray}
  \sigma_{\lambda \lambda'}^2 & = & \int \frac{\mathrm{d}^3 k}{(2 \pi)^3}
  P^{\lambda \lambda'} \left( \textbf{k} \right)\nonumber \\
  & = & \frac{1}{2} \delta^{\lambda \lambda'} \int \frac{\mathrm{d}^3 k}{(2
  \pi)^3} \int \frac{\mathrm{d}^3 p}{(2 \pi)^3} \left\{ \left( 1 - \left(
  \hat{\textbf{k}} \cdot \hat{\textbf{p}} \right)^2 \right)^2 \left(
  \frac{\left| \textbf{p} \right|}{k} \right)^4 P_{\Psi} \left( \left|
  \textbf{k} - \textbf{p} \right| \right) P_{\Psi} (p) \left( \frac{I_\text{A}^{(2)} +
  I_\text{B}^{(2)}}{2 (k \eta)^2} \right) \right\}~.\nonumber \\ \label{23}
\end{eqnarray}
The oscillatory terms vanish following the Riemann-Lebesgue lemma. The square of kernel function, $I_h^2$, reduces to ${(I_\text{A}^{(2)} + I_\text{B}^{(2)})}/{ (\sqrt{2} k \eta)^2}$, which is consistent with the setup of the envelope scheme. It indicates that the envelope of kernel function employed for calculating the spectrum can conserve the variance.
Based on Eqs.~(\ref{defIRD}) and (\ref{defBispectra}), we evaluate bispectrum of SIGWs in the form of
\begin{eqnarray}
  B^{\lambda_0 \lambda_1 \lambda_2} \left( \textbf{k}_1, \textbf{k}_2 \right)
  & = & 4 \int \frac{\mathrm{d}^3 p}{(2 \pi)^3} \Bigg\{ P_{\Psi} \left( \left|
  \textbf{k}_2 - \textbf{p} \right| \right) P_{\Psi} \left( \left|
  \textbf{k}_1 + \textbf{k}_2 - \textbf{p} \right| \right) P_{\Psi} (p)
  \mathbb{P}^{\lambda_0 \lambda_1 \lambda_2} \nonumber \\ && \times \left( \frac{\cos \left( \left|
    \textbf{k}_1 + \textbf{k}_2 \right| \eta \right) I_\text{A}^{(0)} + \sin \left( \left|
    \textbf{k}_1 + \textbf{k}_2 \right| \eta \right) I_\text{A}^{(0)} }{\left|
      \textbf{k}_1 + \textbf{k}_2 \right|\eta}\right) \nonumber \\ && \times \left( \frac{\cos
    (k_1 \eta) I_\text{A}^{(1)} + \sin \left( k_1 \eta \right) I_\text{A}^{(1)}}{k_1\eta} \right) \left(\frac{\cos (k_2
    \eta) I_\text{A}^{(2)} + \sin (k_2 \eta) I_\text{A}^{(2)}}{k_2\eta}\right) \Bigg\}\nonumber\\
  & = & \int \frac{\mathrm{d}^3 p}{(2 \pi)^3} \Bigg\{ P_{\Psi} \left( \left|
  \textbf{k}_2 - \textbf{p} \right| \right) P_{\Psi} \left( \left|
  \textbf{k}_1 + \textbf{k}_2 - \textbf{p} \right| \right) P_{\Psi} (p)
  \mathbb{P}^{\lambda_0 \lambda_1 \lambda_2} \nonumber \\ && \times \sum_{s, t = \pm} \Bigg( \cos 
  \left( \left( \left| \textbf{k}_1 + \textbf{k}_2 \right| + s   k_1 +
  t   k_2 \right) \eta \right) \nonumber \\ && \times \left( \frac{I_\text{A}^{(0)} I_\text{A}^{(1)} I_\text{A}^{(2)} - s
    t   I_\text{A}^{(0)} I_\text{B}^{(1)} I_\text{B}^{(2)} - t   I_\text{B}^{(0)} I_\text{A}^{(1)} I_\text{B}^{(2)} - s
      I_\text{B}^{(0)} I_\text{B}^{(1)} I_\text{A}^{(2)}}{\left|
        \textbf{k}_1 + \textbf{k}_2 \right| k_1 k_2 \eta^3} \right) \nonumber \\ &&  + \sin \left( \left(
  \left| \textbf{k}_1 + \textbf{k}_2 \right| + s   k_1 + t   k_2
  \right) \eta \right) \nonumber \\ && \times \left( \frac{t   I_\text{A}^{(0)} I_\text{A}^{(1)} I_\text{B}^{(2)} + s  
  I_\text{A}^{(0)} I_\text{B}^{(1)} I_\text{A}^{(2)} + I_\text{B}^{(0)} I_\text{A}^{(1)} I_\text{A}^{(2)} - s   t   I_\text{B}^{(0)} I_\text{B}^{(1)}
  I_\text{B}^{(2)}}{\left|
    \textbf{k}_1 + \textbf{k}_2 \right| k_1 k_2 \eta^3} \right) \Bigg) \Bigg\}~, \nonumber \\ \label{24}
\end{eqnarray}
where  $I^{(0)}_\text{A}\equiv I_\text{A} \left( \left| \textbf{k}_1 +
\textbf{k}_2 - \textbf{p} \right|, p, \left| \textbf{k}_1 + \textbf{k}_2
\right|, \eta \right)$, $I^{(1)}_\text{A}\equiv I_\text{A} \left( \left| \textbf{k}_1 + \textbf{k}_2 -
\textbf{p} \right|, \left| \textbf{k}_2 - \textbf{p} \right|, k_1, \eta
\right) $, and $I^{(2)}_\text{A}\equiv  I_\text{A} \left( \left| \textbf{k}_2 - \textbf{p} \right|, p, k_2, \eta
\right) $. 
Following Riemann-Lebesgue lemma in Eq.~(\ref{21}), the third moment in Eq.~(\ref{12}) tends to vanish. For a finite $\eta$, there are non-vanishing sub-leading order terms given as follows,
\begin{eqnarray}
  \mu^3_{\lambda_0 \lambda_1 \lambda_2} & = & \int \frac{\mathrm{d}^3 k_1}{(2
  \pi)^3} \int \frac{\mathrm{d}^3 k_2}{(2 \pi)^3} B_h^{\lambda_0 \lambda_1
  \lambda_2} \left( \textbf{k}_1, \textbf{k}_2 \right) \nonumber \\
  &\simeq & \sum_{s, t = \pm} \int \frac{\mathrm{d}^3 k_1}{(2
  \pi)^3} \int \frac{\mathrm{d}^3 k_2}{(2 \pi)^3}\int \frac{\mathrm{d}^3 p}{(2 \pi)^3} \Bigg\{ P_{\Psi} \left( \left|
    \textbf{k}_2 - \textbf{p} \right| \right) P_{\Psi} \left( \left|
    \textbf{k}_1 + \textbf{k}_2 - \textbf{p} \right| \right) P_{\Psi} (p)
    \mathbb{P}^{\lambda_0 \lambda_1 \lambda_2} \nonumber \\ && \times  \left( \frac{I_\text{A}^{(0)} I_\text{A}^{(1)} I_\text{A}^{(2)} - s
      t   I_\text{A}^{(0)} I_\text{B}^{(1)} I_\text{B}^{(2)} - t   I_\text{B}^{(0)} I_\text{A}^{(1)} I_\text{B}^{(2)} - s
        I_\text{B}^{(0)} I_\text{B}^{(1)} I_\text{A}^{(2)}}{\left|
          \textbf{k}_1 + \textbf{k}_2 \right| k_1 k_2 \eta^3} \right)  \nonumber \\ && \times
          \Theta\left[     \frac{1}{\eta}-   |\left| \textbf{k}_1 + \textbf{k}_2 \right| + s   k_1 + t   k_2|\right]  \Bigg\}~.    \label{25}
\end{eqnarray}
Here, one might notice that the momentum integrals involving the Heaviside step function yield an additional factor $\eta^{-1}$.
From Eq.~(\ref{25}), the third moment is dominated by the collinear-type and flattened-type bispectrum. It differs from the results obtained using the envelope scheme, in which the equilateral configurations dominate the bispectrum \cite{Espinosa:2018eve, Bartolo:2018evs, Bartolo:2018rku}. 
Finally, based on Eqs.~(\ref{23}), (\ref{25}), and (\ref{defSkew}), there is no ambiguity to calculate the skewness. 
Because the envelope scheme conserves variance and does not conserve third moment, it inevitably leads to a different value of skewness.
In other words, the envelope scheme does not conserve the skewness.

\ 

\subsection{Oscillation average scheme \label{appA}}

Addressing SIGWs generated during radiation-dominated era, we here introduce an oscillation average scheme that conserve both the variance and skewness. It is based on averaging the typical quantities as follows,
\begin{eqnarray}
  \mathcal{I}_\mathcal{C}\equiv\cos(w x) \mathcal{I}_\text{A}(x)~, &  \mathcal{I}_\mathcal{S}\equiv\sin(w x) \mathcal{I}_\text{B}(x)~, & \mathcal{I}_0\equiv \mathcal{I}_\text{C}(x)~,
\end{eqnarray}
where $w$ is finite, and the functions $I_\text{A,B,C}(x)$ do not depend on $w$ and trigonometric function of $x$. Following Riemann-Lebesgue lemma, we can set the oscillation averages of $\mathcal{I}_\mathcal{C}$, $\mathcal{I}_\mathcal{S}$, and $\mathcal{I}_0$  to be
\begin{subequations}
  \begin{eqnarray}
    \langle{\mathcal{I}}_\mathcal{C}\rangle_\text{osc} &\simeq& \left\{
      \begin{array}{cl}
        \mathcal{I}_\text{A}(x) ~&  \text{otherwise} \\
        0 ~& w x \gg  1  \\
      \end{array}~,
    \right.\\ 
    \langle{\mathcal{I}}_\mathcal{S}\rangle_\text{osc} &\simeq& 0~,\\
    \langle{\mathcal{I}}_0\rangle_\text{osc} &=& \mathcal{I}_\text{C}(x)~. 
  \end{eqnarray}
\end{subequations}
For simplicity, the oscillation average of $\mathcal{I}_\mathcal{C}$ can be symbolized as $\Delta_w[x]\mathcal{I}_\text{A}[x]$, where 
\begin{eqnarray}
  \Delta_w[x] &=& \left\{
    \begin{array}{cl}
      1 ~& \text{otherwise} \\
      0 ~& w x \gg  1  \\
    \end{array}~.
  \right.
\end{eqnarray}
The oscillation averages are employed for cancellation of high-frequency oscillation terms. We adopt a rough approximation in the regime of $wx \lesssim 1$.

For the sake of consistency, we utilize the oscillation average scheme for kernel functions in spectrum and bispectrum given in Eqs.~(\ref{22}) and (\ref{24}). First, the oscillation average of double kernel functions in spectrum can be given by
\begin{eqnarray}
  \langle I_h^2 \rangle_\text{osc} & \simeq & \frac{1}{(k\eta)^2}\left( \frac{1}{2}(I_\text{A}^2+I_\text{B}^2)+\frac{1}{2}\Delta_{k_\ast\eta}[2k/k_\ast](I_\text{A}^2-I_\text{B}^2) \right)~, \label{29}
\end{eqnarray}
where $I_h\equiv I_h\left( \left|\textbf{k} - \textbf{p} \right|, \left| \textbf{p} \right|, k, \eta\right)$, and $k_\ast$ is the pivot scale for detectors. In PTAs, the $k_\ast \eta$ is a quiet large number, and $k/k_\ast$ is around $\mathcal{O}(1)$. Thus, it leads to $\langle I_h^2 \rangle \simeq  ( (I_\text{A}^2+I_\text{B}^2))/(\sqrt{2} k\eta)^2$. Second, we utilize the oscillation average scheme for evaluating the triple kernel functions in bispectrum, namely,
\begin{eqnarray}
  \langle I_h^{(0)} I_h^{(1)} I_h^{(2)} \rangle_\text{osc} &\simeq& \frac{1}{4} \sum_{s, t = \pm}\Bigg(  \Delta_{k_\ast \eta} 
  \left[ \frac{ \left| \textbf{k}_1 + \textbf{k}_2 \right| + s   k_1 +
  t   k_2 }{k_\ast} \right] \nonumber \\ && \times  \left( \frac{I_\text{A}^{(0)} I_\text{A}^{(1)} I_\text{A}^{(2)} - s
    t   I_\text{A}^{(0)} I_\text{B}^{(1)} I_\text{B}^{(2)} - t   I_\text{B}^{(0)} I_\text{A}^{(1)} I_\text{B}^{(2)} - s
      I_\text{B}^{(0)} I_\text{B}^{(1)} I_\text{A}^{(2)}}{\left|
        \textbf{k}_1 + \textbf{k}_2 \right| k_1 k_2 \eta^3} \right)  \Bigg) ~, \nonumber \\ \label{30}
\end{eqnarray}
where the superscripts $(0)$, $(1)$, and $(2)$ have been introduced in Eq.~(\ref{24}). 
In PTAs, $k_\ast\eta$, $ \left| \textbf{k}_1 + \textbf{k}_2 \right|\eta$, $k_1\eta$, $k_2\eta$ are large but finite numbers because of finite observation time, while $( \left| \textbf{k}_1 + \textbf{k}_2 \right| + s   k_1 +t   k_2 )/k_\ast$ could tend to be zero. In this case, the bispectrum does not vanish when $\left| \textbf{k}_1 + \textbf{k}_2 \right| + s   k_1 +t   k_2 \lesssim \eta^{-1}$ as indicated in Eq.~(\ref{30}). Namely, both the colinear-type  and flattened-type non-Gaussianity could be non-vanishing. Additionally, for a single kernel function, one can obtain $\langle I_h \rangle_\text{osc}\simeq \Delta_{k_\ast \eta}[k/k_\ast]I_\text{A}/(k\eta)$, which indicates $\langle I_h \rangle_\text{osc}^2 \neq \langle I_h^2 \rangle_\text{osc}$.

By comparing the kernel functions in Eqs.~(\ref{29}) and (\ref{30}) with those in Eqs.~(\ref{23}) and (\ref{25}), our oscillation average scheme is shown to conserve both the variance and skewness of the SIGWs. 



\ 

\section{Bispectrum of SIGWs for a finite conformal time \label{III}} 

In this section, we will calculate the bispectrum $B^{\lambda_0 \lambda_1 \lambda_2} \left( \textbf{k}_1, \textbf{k}_2 \right)$ in collinear configurations and in the triangle-shape configurations. In addition to the SIGWs generated during the matter-dominated era, our primary focus is on employing the oscillation average scheme introduced in Sec.~\ref{IIIa} to calculate SIGWs generated during the radiation-dominated era. For the sake of practical setups on GW detectors like PTAs, we consider the conformal time $\eta$ to be a large but finite value.

\subsection{Bispectrum in the collinear configurations \label{III.A}}

The collinear configurations can be given by setting $\textbf{k}_1 = \alpha\textbf{k}_2$ for the bispectrum in Eq.~(\ref{defBispectra}), which reduces to
\begin{eqnarray}
  B^{\lambda_0 \lambda_1 \lambda_2} \left( \alpha \textbf{k}_2, \textbf{k}_2
  \right) & = & 4 \int \frac{{\rm d}^3 p}{(2 \pi)^3} \Big\{ P_{\Psi}\left(
  \left| \textbf{k}_2 - \textbf{p} \right| \right) P_{\Psi}\left( \left| (1 +
  \alpha) \textbf{k}_2 - \textbf{p} \right| \right) P_{\Psi}\left( \left|
  \textbf{p} \right| \right)  \mathbb{P}^{\lambda_0 \lambda_1
  \lambda_2} |_{\textbf{k}_1 = \alpha \textbf{k}_2} \nonumber\\ && \times I_h \left( \left|
  (1 + \alpha) \textbf{k}_2 - \textbf{p} \right|, p, \left| (1 + \alpha)
  \textbf{k}_2 \right|, \eta \right) I_h \left( \left| (1 + \alpha)
  \textbf{k}_2 - \textbf{p} \right|, \left| \textbf{k}_2 - \textbf{p} \right|,
  \alpha k_2, \eta \right) \nonumber\\ && \times I_h \left( \left| \textbf{k}_2 - \textbf{p}
  \right|, p, k_2, \eta \right) \Big\} \nonumber \\
  & = & \frac{4}{(2 \pi)^3} \int_0^{\infty} p^2 {\rm d} p \int_0^{\pi} \sin
  \theta {\rm d} \theta \Big\{ P_{\Psi}\left( \left| \textbf{k}_2 -
  \textbf{p} \right| \right) P_{\Psi}\left( \left| (1 + \alpha) \textbf{k}_2
  - \textbf{p} \right| \right) P_{\Psi}(p) \nonumber\\ && \times I_h \left( \left| (1 +
  \alpha) \textbf{k}_2 - \textbf{p} \right|, p, \left| (1 + y) \textbf{k}_2
  \right|, \eta \right) I_h \left( \left| (1 + \alpha) \textbf{k}_2 -
  \textbf{p} \right|, \left| \textbf{k}_2 - \textbf{p} \right|, \alpha k_2,
  \eta \right) \nonumber\\ && \times I_h \left( \left| \textbf{k}_2 - \textbf{p} \right|, p, k_2,
  \eta \right) \int_0^{2 \pi} {\rm d} \phi    \mathbb{P}^{\lambda_0
  \lambda_1 \lambda_2} |_{\textbf{k}_1 = \alpha \textbf{k}_2} \Big\}~. \label{biMD}
\end{eqnarray}
The $ \mathbb{P}^{\lambda_0 \lambda_1
\lambda_2} |_{\textbf{k}_1 = \alpha \textbf{k}_2} $ can be obtained by making use of the polarization tensor as follows,
\begin{eqnarray}
  e^+_{a   b} \left( \textbf{k}_2 \right)  \equiv  \frac{1}{\sqrt{2}}
  (e_{1, a} e_{1, b} - e_{2, a} e_{2, b})~, \hspace{0.5cm}
  e^{\times}_{a   b} \left( \textbf{k}_2 \right)  \equiv 
  \frac{1}{\sqrt{2}} (e_{1, a} e_{2, b} + e_{1, a} e_{2, b})~, \label{polTen1}
\end{eqnarray}
where $e_1$ and $e_2$ are the polarization vectors. In the Cartesian coordinate of
the $\textbf{p}$, we set $\hat{z} = \hat{\textbf{k}}_2$, $\hat{x}=e_1$ and $\hat{y}=e_2 $. The polar angle $\theta$ and azimuth
angle $\phi$ can be defined via $\hat{z} \cdot \textbf{p} = \hat{k} \cdot \textbf{p} = p \cos
\theta$, $\hat{x} \cdot \textbf{p} = p \sin \theta \cos \phi$ and $\hat{y} \cdot \textbf{p} = p
\sin \theta \cos \phi$. By making use of Eq.~(\ref{polTen1}) in the representation, one can obtain explicit expressions of the $ \mathbb{P}^{\lambda_0 \lambda_1 \lambda_2} |_{\textbf{k}_1 = \alpha \textbf{k}_2} $.

It is found that the bispectrum in Eq.~(\ref{biMD}) vanish, as given by the integrals over the azimuth angle $\phi$. For example, one can calculate the $+++$ component by associating with Eq.~(\ref{defPP}) in the collinear configurations, namely,
\begin{eqnarray}
 \int_0^{2 \pi} {\rm d} \phi    \mathbb{P}^{+ + +} |_{\textbf{k}_1 =
 \alpha \textbf{k}_2} = \int_0^{2 \pi} {\rm d} \phi \left\{ \frac{1}{2
 \sqrt{2} (1 + \alpha)^2 \alpha^2} \left( \frac{p \sin \theta}{k_2}
 \right)^6 \cos^3 (2 \phi) \right\} = 0 ~.
\end{eqnarray}
Above result is independent of curvature power spectrum $P_{\Psi}(k)$, and the kernel functions $I_h \left( \left| \textbf{k} - \textbf{p} \right|, p, k,\eta \right)$. In other words, it is independent of the initial conditions and dynamics of SIGWs. Hence, the vanishing bispectrum in the collinear configuration is a universal result. This result also suggests that bispectrum for SIGWs generated during radiation-dominated era vanish in the limit of $\eta \rightarrow\infty$. It challenges the detectability of SIGWs from the early time universe.

\

\subsection{Bispectrum in the triangular configurations \label{III.B}}

\begin{figure}[ht] \centering
  \includegraphics[width=0.5\linewidth]{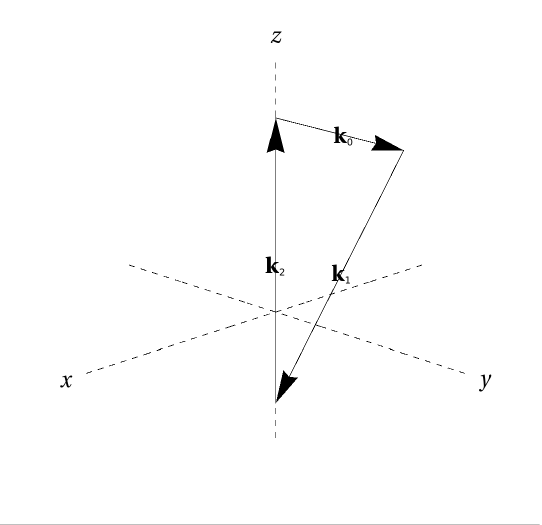}
  \caption{\label{F1} Schematic diagram for illustrating representation of momenta $\textbf{k}_0$, $\textbf{k}_1$, and $\textbf{k}_2$.}
\end{figure}

To obtain the triangular bispectrum, one should first provide the representation of the polarization tensors.
It is noted that the
polarization tensor is not as trivial as that in the collinear configurations, because these polarization tensors with respect to momenta $\textbf{k}_1$, $\textbf{k}_2$ and $\textbf{k}_0$ in Eq.~(\ref{defBh}) should be given, separately. 
Here, we establish the representation for polarization tensors as follows.
In the Cartesian coordinate system of the $\textbf{p}$ in Eq.~(\ref{defBispectra}), we have
$\hat{z} \cdot \textbf{p} = p \cos \theta$, $\hat{x} \cdot \textbf{p} = p \sin
\theta \cos \phi$ and $\hat{y} \cdot \textbf{p} = p \sin \theta \cos \phi$.
For simplicity, we set $\textbf{k}_2$ along the $z$-axis, namely, $\hat{\textbf{k}}_2=\hat{z}$. And the triangle formed by the momenta
$\textbf{k}_1$, $\textbf{k}_2$ and $\textbf{k}_0$ is set within the $x$-$z$
plane. In this setup, $\textbf{k}_1$ can be expressed as
\begin{eqnarray}
  \textbf{k}_1 & = & k_1 (\hat{x} \sin \theta_k + \hat{z} \cos \theta_k) ~,
\end{eqnarray}
where $\theta_k \equiv \arccos( \hat{\textbf{k}}_1 \cdot
\hat{\textbf{k}}_2 )$. The schematic diagram is shown in Fig.~\ref{F1}.
In this representation, the polarization vectors for momenta
$\textbf{k}_i$ are given by
\begin{subequations}
  \begin{eqnarray}
    e_1 \left( \textbf{k}_2 \right) & = & \hat{x} ~, \\
    e_1 \left( \textbf{k}_1 \right) & = & \hat{x} \cos \theta_k - \hat{z} \sin
    \theta_k ~, \\
    e_1 \left( \textbf{k}_0 \right) & = & \hat{x} \cos \theta_k' + \hat{z} \sin
    \theta_k' ~, \\
  e_2 \left( \textbf{k}_2 \right) &= &e_2 \left( \textbf{k}_1 \right) = e_2
  \left( \textbf{k}_0 \right) = \hat{y} ~,
\end{eqnarray}
\end{subequations}
where $\theta_k' \equiv \arccos ( \hat{\textbf{k}}_2 \cdot
\hat{\textbf{k}}_0 ) = \arccos ( \hat{\textbf{k}}_2 \cdot
( - \textbf{k}_1 - \textbf{k}_2 ) \left| \textbf{k}_1 +
\textbf{k}_2 \right|^{- 1} )$. Therefore, we can obtain the polarization tensors based on
\begin{subequations}
  \begin{eqnarray}
    e^+ \left( \textbf{k}_i \right) & \equiv & \frac{1}{\sqrt{2}} \left( e_1
    \left( \textbf{k}_i \right) \otimes e_1 \left( \textbf{k}_i \right) - e_2
    \left( \textbf{k}_i \right) \otimes e_2 \left( \textbf{k}_i \right) \right)
    ~, \\
    e^{\times} \left( \textbf{k}_i \right) & \equiv & \frac{1}{\sqrt{2}} \left(
    e_1 \left( \textbf{k}_i \right) \otimes e_2 \left( \textbf{k}_i \right) +
    e_1 \left( \textbf{k}_i \right) \otimes e_2 \left( \textbf{k}_i \right)
    \right) ~. 
  \end{eqnarray} \label{polTen2}
\end{subequations}
Utilizing the polariztion tensors in Eq.~(\ref{polTen2}), the explict expression of $\mathbb{P}^{\lambda_0
\lambda_1 \lambda_2}$ is obtained and presented in Appendix~{\ref{appC}}. 

In the subsequent parts, we will calculate the bispectrum for the matter-dominated era and the radiation-dominated era, separately.

\

\subsubsection{Bispectrum of SIGWs generated during matter-dominated era}

For the matter-dominated era, the kernel function given in Eq.~(\ref{defIMD}) is independent of momentum
$\textbf{p}$. Thus, the bispectrum in Eq.~(\ref{defBh}) can be rewritten as
\begin{eqnarray}
  B^{\lambda_0 \lambda_1 \lambda_2} \left( \textbf{k}_1, \textbf{k}_2 \right)
  & = & 4 I_h^{({\rm MD})} \left( \left| \textbf{k}_1 + \textbf{k}_2 \right|,
  \eta \right) I_h^{({\rm MD})} (k_1, \eta) I_h^{({\rm MD})} (k_2, \eta)
  \nonumber\\
  &  & \times \int \frac{{\rm d}^3 p}{(2 \pi)^3} \left\{ P_{\Psi}\left(
  \left| \textbf{k}_2 - \textbf{p} \right| \right) P_{\Psi}\left( \left|
  \textbf{k}_1 + \textbf{k}_2 - \textbf{p} \right| \right) P_{\Psi}(p)
  \mathbb{P}^{\lambda_0 \lambda_1 \lambda_2} \right\} ~. \nonumber \\  \label{defbiMD}
\end{eqnarray}
At the late time $\eta \rightarrow \infty$, the kernel
function $I_h^{({\rm MD})}$ tends to be a constant. It might suggest that the non-Gaussianity of SIGWs here could remain at the late time. We will further address it in Sec.~\ref{IV}. To compute the bispectrum in Eq.~(\ref{defbiMD}), we would utilize $\mathbb{P}^{\lambda_0
\lambda_1 \lambda_2}$ presented in Appendix~\ref{appC}. Additionally, we also adopt a phenomenological approach by considering the lognormal curvature spectrum as follows,
\begin{eqnarray}
  \mathcal{P}_{\Psi} (k) & = & \frac{A}{\sqrt{2 \pi} \Delta} e^{- \frac{1}{2
  \Delta^2} \left( \ln \left( \frac{k}{k_{\ast}} \right) \right)^2} ~. \label{lognormal}
\end{eqnarray}
The curvature power spectrum has a peak at $k_*$ with the width of $\Delta$.
As shown in Eq.~(\ref{var}), the dimensionless power spectrum is related to the power spectrum through $\mathcal{P}_{\Psi} (k) = k^3 (2 \pi^2)^{- 1} P_{\Psi}(k)$.

\begin{figure}[ht] \centering
  \includegraphics[width=1\linewidth]{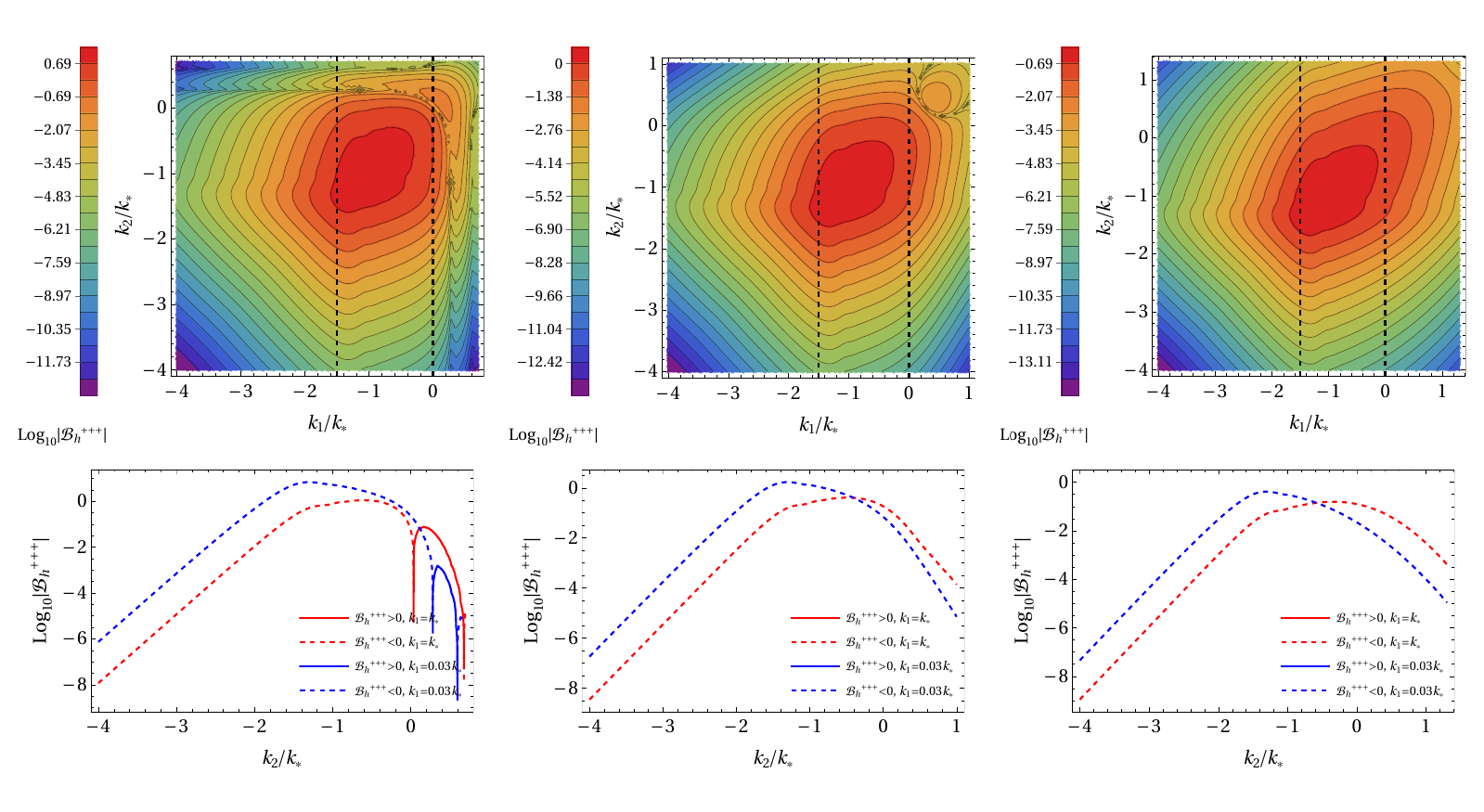}
  \caption{\label{F2} Top panel: $+++$ component of the dimensionless bispectrum as function of $k_1/k_*$ and $k_2/k_*$. Bottom panel: $+++$ component of the dimensionless bispectrum as function of $k_2/k_*$ for given $k_1/k_*$. In these plots, we have fixed $\theta_k=\pi/6$ and $k_*\eta=100$ and considered the width of the curvature power spectrum $\Delta=1/2$ (left panel), 1 (medium panel), and $2$ (right panel), respectively. Here, we set $A=1$.}
\end{figure}

\begin{figure}[ht] 
  \centering 
  \includegraphics[width=1\linewidth]{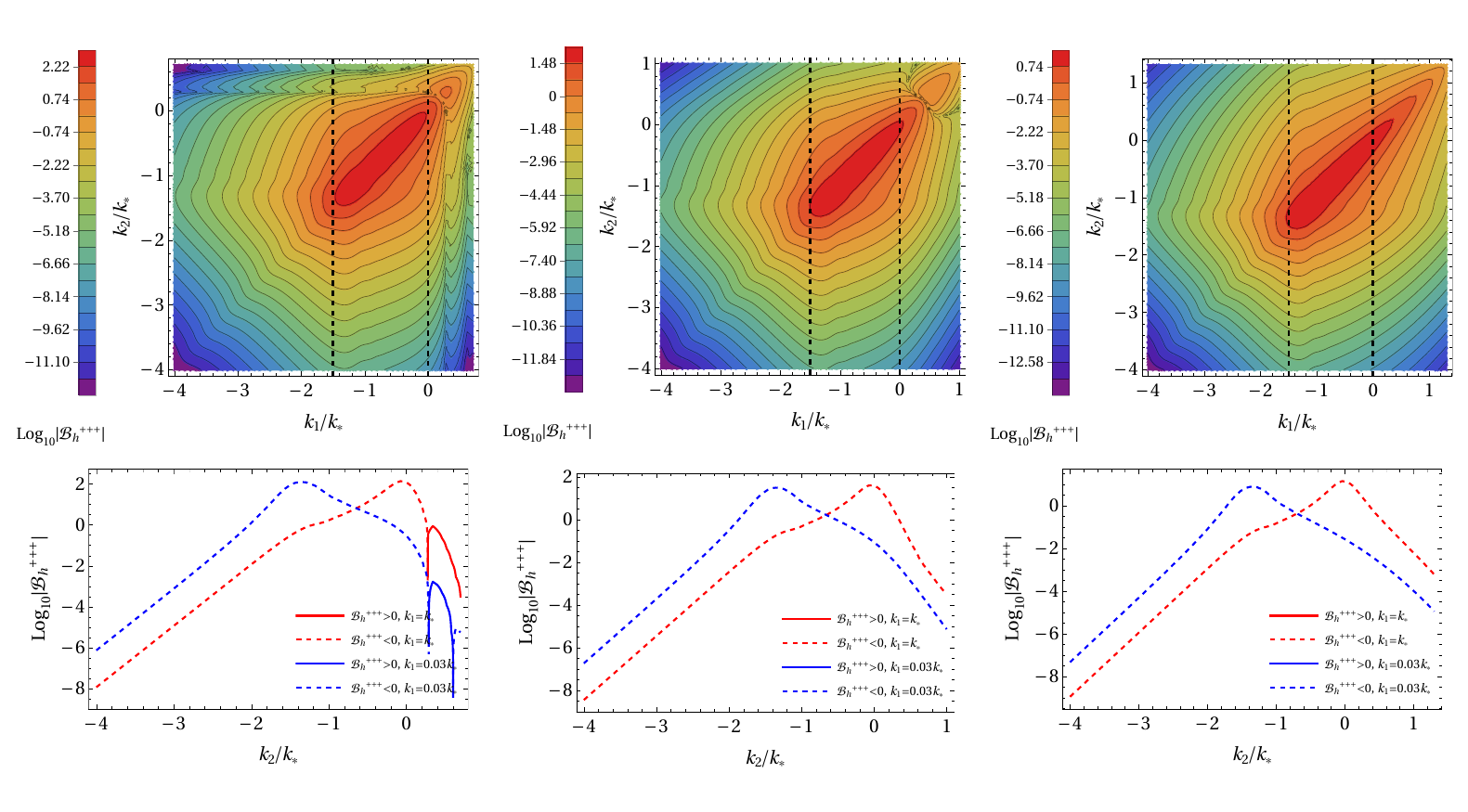}
  \caption{\label{F3}Top panel: $+++$ component of the dimensionless bispectrum as function of $k_1/k_*$ and $k_2/k_*$. Bottom panel: $+++$ component of the dimensionless bispectrum as function of $k_2/k_*$ for given $k_1/k_*$. In these plots, we have fixed $\theta_k=5\pi/6$ and $k_*\eta=100$, and considered the width of the curvature power spectrum $\Delta=1/2$ (left panel), $1$ (medium panel), and $2$ (right panel), respectively. Here, we set $A=1$.}
\end{figure}

Here, we calculate the bispectrum numerically, based on Monte-Carlo integration. It is found that the polarization components, $\times\times\times$, $++\times$, $+\times+$, and $\times++$ are much smaller than the rest of the components over several orders of magnitude. It is consistent with the result shown in the previous study that there are only four polarization components left to be non-vanishing \cite{Espinosa:2018eve}.  Associating the numerical results with Eq.~(\ref{defBh}), we find $B_h^{+++}(\textbf{k}_1,\textbf{k}_2)=B_h^{+++}(\textbf{k}_2,\textbf{k}_1)$, $B_h^{+\times\times}(\textbf{k}_1,\textbf{k}_2)=B_h^{+\times\times}(\textbf{k}_2,\textbf{k}_1)$, and $B_h^{++\times}(\textbf{k}_1,\textbf{k}_2)=B_h^{+\times+}(\textbf{k}_2,\textbf{k}_1)$. We also show the bispectrum as function of $k_1$ and $k_2$ for selected width $\Delta$ in Figs.~\ref{F2} and \ref{F3}. The bispectrum could be positive or negative, which are
denoted as solid and dashed curves in the bottom panels.
The maximum absolute value of the bispectrum amplitude decreases with width of the curvature power spectrum. It suggests that the peaked curvature power spectrum can enhance the bispectrum amplitude.
Comparing between Figs.~\ref{F2} and \ref{F3}, the absolute value of the bispectrum with $\theta_k = 5 \pi / 6$ has a smaller plateau than that for $\theta_k = \pi /6$.

\begin{figure}[ht] \centering
  \includegraphics[width=1\linewidth]{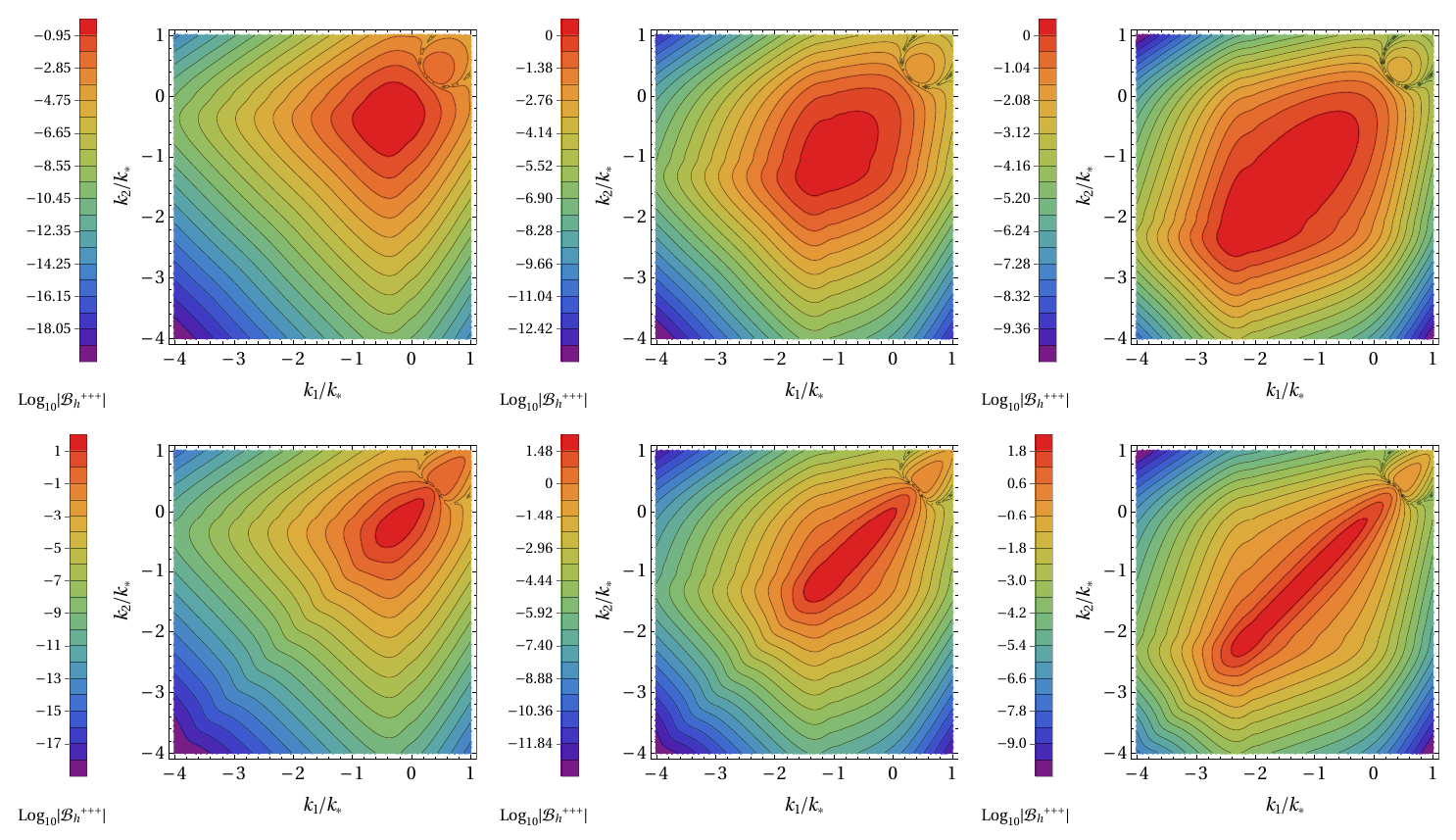}
  \caption{\label{F4} $+++$ component of the dimensionless bispectrum as function of $k_1/k_*$ and $k_2/k_*$ for $k_*\eta=10$ (left panels), $100$ (medium panels), and $1000$ (right panels), respectively. In theses plots, we have fixed the width $\Delta=1$, and $\theta_k=\pi/6$ (top panels) and $\theta_k=5\pi/6$ (bottom panels). Here, we set $A=1$.}
\end{figure}

\begin{figure}[ht] \centering
  \includegraphics[width=1\linewidth]{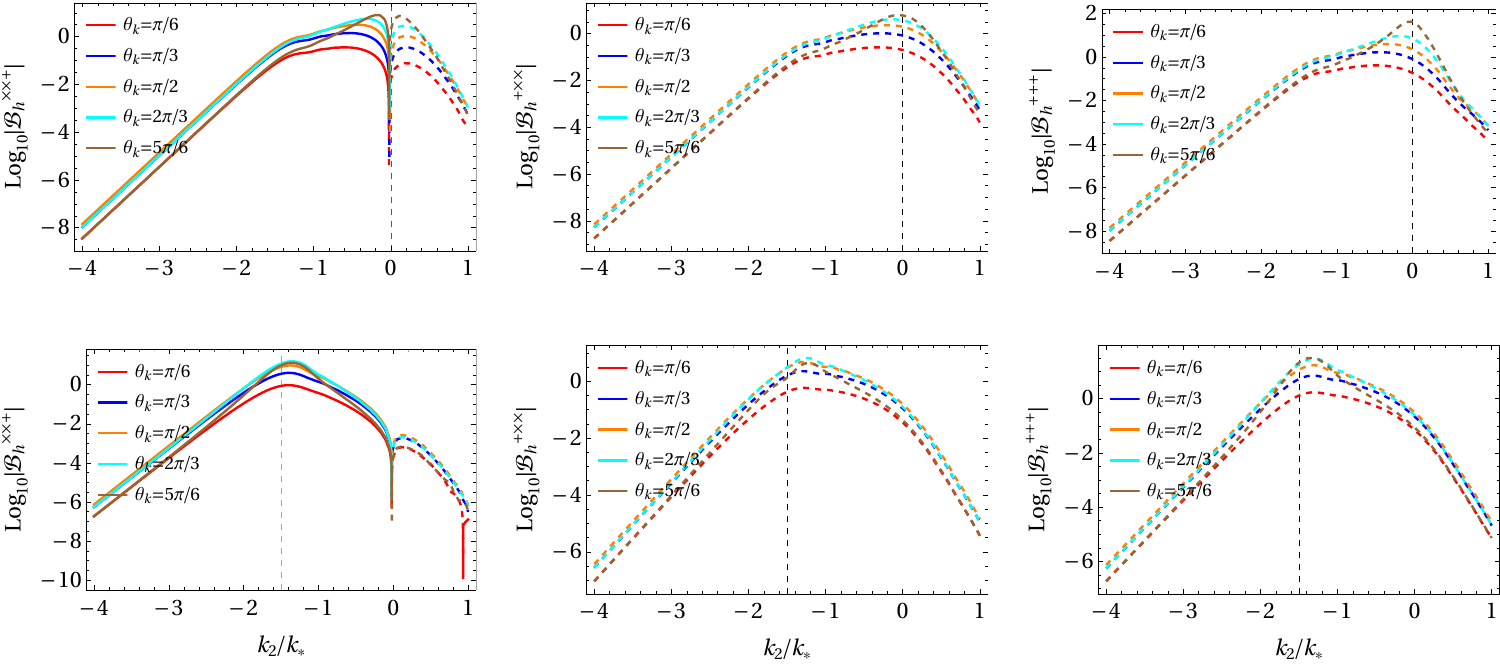}
  \caption{\label{F5} The dimensionless bispectrum as function of $k_2/k_*$ for selected $\theta_k$ with fixed $k_1/k_*=1$ (top panels) and $k_1/k_*=10^{-1.5}$ (bottom panels). In these plots, we consider the width $\Delta=1$, $k_*\eta=100$ and $A=1$}
\end{figure}

\begin{figure}[ht] \centering
  \includegraphics[width=1\linewidth]{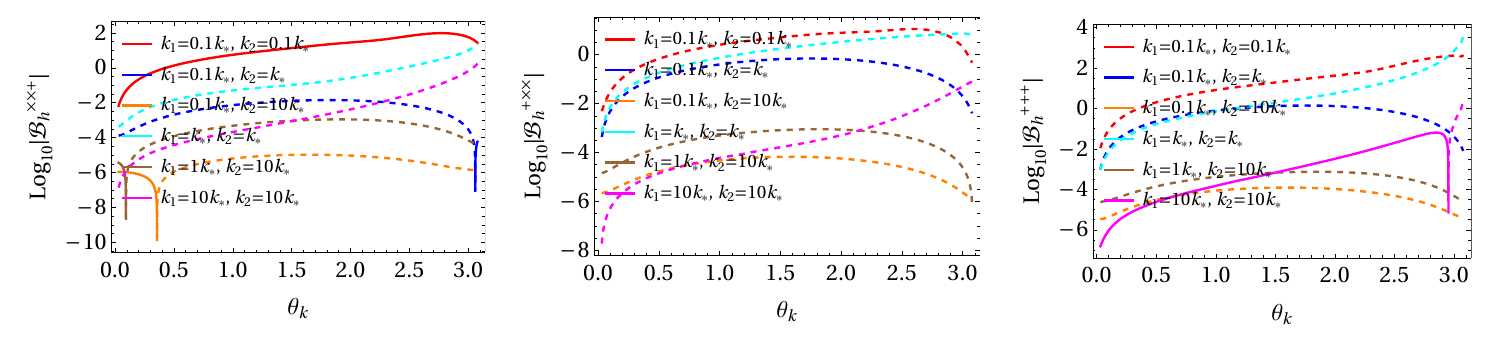}
  \caption{\label{F6} The dimensionless bispectrum as function of $\theta_k$ for selected $k_2/k_*$ and $k_1/k_*$. In these plots, we consider the width $\Delta=1$, $k_*\eta=100$ and $A=1$.}
\end{figure}

Fig.~\ref{F4} shows the bispectrum as function of $k_1$ and $k_1$ for selected $k_*\eta$. The plateau of the bispectrum increases with $k_*\eta$. 
Fig.~\ref{F5} shows the bispectrum as function of $k_2$ for selected $k_1$ and $\theta_k$.  The peaks of the bispectrum become narrower as $\theta_k$ increases.
To further show the relation between the bispectrum amplitude and the $\theta_k$, we present the bispectrum as function of
$\theta_k$ for given $k_1$ and $k_2$ in Fig.~\ref{F6}.
For $k_1$ and $k_2$ located on the plateau as shown in Fig.~\ref{F4}, the absolute value of bispectrum amplitudes monotonically increases with $\theta_k$. In other cases, the bispectrum amplitudes are suppressed as $\theta_k\rightarrow0$ or $\theta_k\rightarrow\pi$. The same conclusion holds for different polarization components, namely $\times\times+$, $+\times\times$, and $+++$.

\

\subsubsection{Bispectrum of SIGWs generated during radination-dominated era \label{III.B.2}}

In the radiation-dominated era, we are interested in the bispectrum at the late
time limit $\eta \rightarrow \infty$ because it might correspond to SGWB that we observe \cite{Kohri:2018awv}. In this case, the kernel function
at the late time was presented in Eq.~(\ref{defIRD}). By substituting the kernel
function into the bispectrum given in Eq.~(\ref{defBh}), it is found that the highly oscillatory behavior of bispectrum should be tackled for further computation. By making use of oscillation average introduced in Sec.~\ref{IIIa}, we have
\begin{eqnarray}
  B^{\lambda_0 \lambda_1 \lambda_2} \left( \textbf{k}_1, \textbf{k}_2 \right)
  & = & \Delta_{\eta} \left[ k_1 + k_2 - \left| \textbf{k}_1 + \textbf{k}_2
  \right| \right] B^{\lambda_0 \lambda_1 \lambda_2}_0 \left( \textbf{k}_1,
  \textbf{k}_2 \right) \nonumber\\ && + \Delta_{\eta} \left[ \left| \textbf{k}_1 +
  \textbf{k}_2 \right| - k_1 + k_2 \right] B^{\lambda_0 \lambda_1 \lambda_2}_+
  \left( \textbf{k}_1, \textbf{k}_2 \right) \nonumber\\ && + \Delta_{\eta} \left[ \left|
  \textbf{k}_1 + \textbf{k}_2 \right| + k_1 - k_2 \right] B^{\lambda_0
  \lambda_1 \lambda_2}_- \left( \textbf{k}_1, \textbf{k}_2 \right) ~,
  \label{BiRD1}
\end{eqnarray}
where the $\Delta_{\eta} [\ast]$ have been defined in Sec.~{\ref{appA}}, and 
\begin{eqnarray}
  B_*^{\lambda_0 \lambda_1 \lambda_2} \left( \textbf{k}_1, \textbf{k}_2
  \right)  =  \frac{1}{\left| \textbf{k}_1 + \textbf{k}_2 \right|  
  \left| \textbf{k}_1 \right|   \left| \textbf{k}_2 \right| \eta^3}
  \int \frac{{\rm d}^3 p}{(2 \pi)^3} \left\{ P_{\Psi}\left( \left|
  \textbf{k}_2 - \textbf{p} \right| \right) P_{\Psi}\left( \left|
  \textbf{k}_1 + \textbf{k}_2 - \textbf{p} \right| \right) P_{\Psi}(p)
  \mathbb{P}^{\lambda_0 \lambda_1 \lambda_2} I_* \right\} ~, \nonumber\\\label{33} \label{BiRD2}
\end{eqnarray}
and the $I_*$ with $*=0,\pm$ can be expressed in terms of the kernel functions in Eq.~(\ref{defIRD2}), namely
\begin{subequations}
  \begin{eqnarray}
    I_0 & \equiv & I_\text{A}^{(1)}  I_\text{A}^{(2)}  I_\text{A}^{(3)}   - I_\text{A}^{(1)} I_\text{B}^{(2)}  I_\text{B}^{(3)}    + I_\text{B}^{(1)}  I_\text{A}^{(2)}  I_\text{B}^{(3)}    + I_\text{B}^{(1)}  I_\text{B}^{(2)}  I_\text{A}^{(3)}  ~,  \\
    I_{\pm} & \equiv & I_\text{A}^{(1)}    I_\text{A}^{(2)}  I_\text{A}^{(3)}   + I_\text{A}^{(1)}  I_\text{B}^{(2)}  I_\text{B}^{(3)}    \mp I_\text{B}^{(1)}  I_\text{A}^{(2)}  I_\text{B}^{(3)}    \pm I_\text{B}^{(1)} I_\text{B}^{(2)}  I_\text{A}^{(3)}  ~.  
  \end{eqnarray} \label{BiRD3}
\end{subequations}
It is found that the bispectrum is of flattened-type non-Gaussianity for a large and finite $\eta$, because of $\lim_{\eta\rightarrow\infty} \Delta_{\eta}[*]=0$ for the triangular bispectrum. It might be understood that the oscillations can suppress the bispectrum amplitudes. 
It is different from the results in previous studies that the equilateral configuration dominates the bispectrum \cite{Espinosa:2018eve, Bartolo:2018evs, Bartolo:2018rku}. 

For illustration, we can let $k_1 + k_2 - \left| \textbf{k}_1 + \textbf{k}_2
\right| \equiv \epsilon / \eta$ and $0 < \epsilon \ll 1$, for example. In this scenario, the configuration for describing the flattened-type non-Gaussianity is presented in Tab.~\ref{T1}.
\begin{table}[ht]
  \centering
  \caption{The flattened bispectrum of SIGWs generated in radiation-dominated era}\label{T1}
\begin{ruledtabular}
  \begin{tabular}{c|cccc}
    Bispectra & Length relation & $\theta_{k,\rm min/max}$ & allowed $k_1$, $k_2$ & Bispectra shape\\ 
    \hline
    $B^{\lambda_0 \lambda_1 \lambda_2}_0$ & $k_1 + k_2 - \left| \textbf{k}_1 +
    \textbf{k}_2 \right| = \epsilon / \eta$ & $\sqrt{\left( \frac{k_2}{k_1} +
    1 \right) \frac{2 \epsilon}{k_2 \eta}}$ & / &
    \raisebox{0.0\height}{\includegraphics[width=0.2\linewidth]{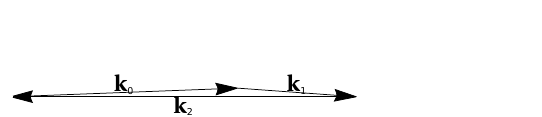}}\\
    $B^{\lambda_0 \lambda_1 \lambda_2}_{\pm}$ & $\left| \textbf{k}_1 +
    \textbf{k}_2 \right| \pm (k_2 - k_1) = \epsilon / \eta$ & $\pi -
    \sqrt{\left| \frac{k_2}{k_1} - 1 \right| \frac{2 \epsilon}{k_2 \eta}}$ & $\pm(k_1-k_2)>0$ &
    \raisebox{0.0\height}{\includegraphics[width=0.2\linewidth]{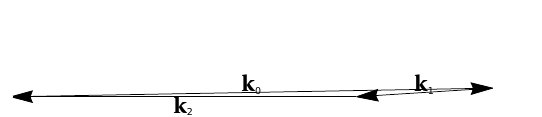}}
  \end{tabular}
\end{ruledtabular}  
\end{table}
In subsequent calculations, we prefer to quantify the flattened non-Gaussianity with a single parameter $\theta_k$. 
Thus we can futher evaluate Eq.~(\ref{33}) by expanding the expression at the $\theta_k \rightarrow 0$ for $B_0^{\lambda_0 \lambda_1 \lambda_2}$, namely,
\begin{eqnarray}
  B_0^{\lambda_0 \lambda_1 \lambda_2} \left( \textbf{k}_1, \textbf{k}_2
  \right) & \simeq & \frac{\theta_k^2}{(k_1 + k_2)   k_1 k_2 \eta^3}
  \int \frac{{\rm d}^3 p}{(2 \pi)^3} \Big\{ P_{\Psi}\left( \left|
  \textbf{k}_2 - \textbf{p} \right| \right) P_{\Psi}\left( \left| (k_1 + k_2)
  \hat{\textbf{k}}_2 - \textbf{p} \right| \right) P_{\Psi}(p)
  \mathbb{B}_1 \nonumber\\
  && + P_{\Psi}\left( \left| \textbf{k}_2 - \textbf{p} \right|
  \right) P_{\Psi}' \left( \left| \left( k_1 + k_2 \right)
  \hat{\textbf{k}}_2 - \textbf{p} \right| \right) P_{\Psi}(p)
  \bar{\mathbb{B}}_1 \nonumber\\
  && + P_{\Psi}\left( \left| \textbf{k}_2 - \textbf{p}
  \right| \right) P_{\Psi}'' \left( \left| (k_1 + k_2) \hat{\textbf{k}}_2
  - \textbf{p} \right| \right) P_{\Psi}(p) \tilde{\mathbb{B}}_1 \Big\}
  ~,  \label{36} 
\end{eqnarray}
where the $P'_{\Psi} (k)$ and $P_{\Psi}'' (k)$ are the derivative with respect
to  $k$, and
\begin{subequations}
  \begin{eqnarray}
    \mathbb{B}_0 & = & \frac{1}{2} \left. \frac{\partial^2 \mathbb{P}^{\lambda_0
    \lambda_1 \lambda_2}}{\partial \theta_k^2} \right|_{\theta_k = 0}   
    I_0 |_{\theta_k = 0} - \frac{k_1 p \cos \phi \sin \phi}{\left| (k_1 + k_2)
    \hat{\textbf{k}}_2 - \textbf{p} \right|} \left. \frac{\partial
    \mathbb{P}^{\lambda_0 \lambda_1 \lambda_2}}{\partial \theta_k}
    \right|_{\theta_k = 0} \left. \frac{\partial I_0}{\partial \left|
    \textbf{k}_1 + \textbf{k}_2 - \textbf{p} \right|} \right|_{\theta_k = 0} \nonumber \\ 
    && +
    \frac{k_1^2 p^2 \cos^2 \phi \sin^2 \theta    \mathbb{P}^{\lambda_0
    \lambda_1 \lambda_2} |_{\theta_k = 0}}{2 \left| (k_1 + k_2)
    \hat{\textbf{k}}_2 - \textbf{p} \right|^3}   \Bigg( \left| (k_1 +
    k_2) \hat{\textbf{k}}_2 - \textbf{p} \right| \left. \frac{\partial^2
    I_0}{\partial \left| \textbf{k}_1 + \textbf{k}_2 - \textbf{p} \right|^2}
    \right|_{\theta_k = 0} \nonumber\\ &&  - \left. \frac{\partial I_0}{\partial \left|
    \textbf{k}_1 + \textbf{k}_2 - \textbf{p} \right|} \right|_{\theta_k = 0}
    \Bigg) ~,
     \\
    \bar{\mathbb{B}}_0 & = & - \frac{k_1 p \cos \phi \sin \theta}{\left| (k_1 +
    k_2) \hat{\textbf{k}}_2 - \textbf{p} \right|} \left. \frac{\partial
    \mathbb{P}^{\lambda_0 \lambda_1 \lambda_2}}{\partial \theta_k}
    \right|_{\theta_k = 0}    I_0 |_{\theta_k = 0} \nonumber\\
    && + \frac{k_1^2 p^2
    \cos^2 \phi \sin^2 \theta    \mathbb{P}^{\lambda_0 \lambda_1
    \lambda_2} |_{\theta_k = 0}}{2 \left| (k_1 + k_2) \hat{\textbf{k}}_2 -
    \textbf{p} \right|^3}   \left( 2 \left| (k_1 + k_2)
    \hat{\textbf{k}}_2 - \textbf{p} \right| \left. \frac{\partial
    I_0}{\partial \left| \textbf{k}_1 + \textbf{k}_2 - \textbf{p} \right|}
    \right|_{\theta_k = 0} -    I_0 |_{\theta_k = 0} \right)~, \nonumber \\ \\ 
    \tilde{\mathbb{B}}_0 & = & \frac{k_1^2 p^2 \cos^2 \phi \sin^2 \theta
     \mathbb{P}^{\lambda_0 \lambda_1 \lambda_2} |_{\theta_k =
    0}    I_0 |_{\theta_k = 0}}{2 \left| (k_1 + k_2)
    \hat{\textbf{k}}_2 - \textbf{p} \right|^2}~.
  \end{eqnarray}
\end{subequations}
Similarly, we expand $B_{\pm}^{\lambda_0 \lambda_1 \lambda_2}$ at the
$\theta_k = \pi$, namely,
\begin{eqnarray}
  B_{\pm}^{\lambda_0 \lambda_1 \lambda_2} \left( \textbf{k}_1, \textbf{k}_2
  \right) & \simeq & \frac{(\theta_k - \pi)^2}{| k_2 - k_1 |   k_1 k_2
  \eta^3} \int \frac{{\rm d}^3 p}{(2 \pi)^3} \Big\{ P_{\Psi}\left( \left|
  \textbf{k}_2 - \textbf{p} \right| \right) P_{\Psi}\left( \left| (k_2 - k_1)
  \hat{\textbf{k}}_2 - \textbf{p} \right| \right) P_{\Psi}(p)
  \mathbb{B}_i \nonumber\\
  && + P_{\Psi}\left( \left| \textbf{k}_2 - \textbf{p} \right|
  \right) P_{\Psi}' \left( \left| \left( k_2 - k_1 \right)
  \hat{\textbf{k}}_2 - \textbf{p} \right| \right) P_{\Psi}(p)
  \bar{\mathbb{B}}_i \nonumber\\
  && + P_{\Psi}\left( \left| \textbf{k}_2 - \textbf{p}
  \right| \right) P_{\Psi}'' \left( \left| (k_2 - k_1) \hat{\textbf{k}}_2
  - \textbf{p} \right| \right) P_{\Psi}(p) \tilde{\mathbb{B}}_i \Big\}
  ~, \label{38}
\end{eqnarray}
where 
\begin{subequations}
  \begin{eqnarray}
    \mathbb{B}_{\pm} & = & \frac{1}{2} \left. \frac{\partial^2
    \mathbb{P}^{\lambda_0 \lambda_1 \lambda_2}}{\partial \theta_k^2}
    \right|_{\theta_k = \pi}    I_{\pm} |_{\theta_k = \pi} + \frac{k_1 p
    \cos \phi \sin \phi}{\left| (k_2 - k_1) \hat{\textbf{k}}_2 - \textbf{p}
    \right|} \left. \frac{\partial \mathbb{P}^{\lambda_0 \lambda_1
    \lambda_2}}{\partial \theta_k} \right|_{\theta_k = \pi} \left.
    \frac{\partial I_{\pm}}{\partial \left| \textbf{k}_1 + \textbf{k}_2 -
    \textbf{p} \right|} \right|_{\theta_k = \pi} \nonumber \\ 
    && + \frac{k_1^2 p^2 \cos^2 \phi
    \sin^2 \theta    \mathbb{P}^{\lambda_0 \lambda_1 \lambda_2}
    |_{\theta_k = 0}}{2 \left| (k_2 - k_1) \hat{\textbf{k}}_2 - \textbf{p}
    \right|^3}   \Bigg( \left| (k_2 - k_1) \hat{\textbf{k}}_2 -
    \textbf{p} \right| \left. \frac{\partial^2 I_{\pm}}{\partial \left|
    \textbf{k}_1 + \textbf{k}_2 - \textbf{p} \right|^2} \right|_{\theta_k = \pi}
    \nonumber \\ 
    && - \left. \frac{\partial I_{\pm}}{\partial \left| \textbf{k}_1 + \textbf{k}_2
    - \textbf{p} \right|} \right|_{\theta_k = \pi} \Bigg) ~,\\
    \bar{\mathbb{B}}_{\pm} & = & \frac{k_1 p \cos \phi \sin \theta}{\left| (k_2
    - k_1) \hat{\textbf{k}}_2 - \textbf{p} \right|} \left. \frac{\partial
    \mathbb{P}^{\lambda_0 \lambda_1 \lambda_2}}{\partial \theta_k}
    \right|_{\theta_k = \pi}    I_{\pm} |_{\theta_k = \pi} \nonumber \\ 
    && + \frac{k_1^2
    p^2 \cos^2 \phi \sin^2 \theta    \mathbb{P}^{\lambda_0 \lambda_1
    \lambda_2} |_{\theta_k = \pi}}{2 \left| (k_2 - k_1) \hat{\textbf{k}}_2 -
    \textbf{p} \right|^3}   \Bigg( 2 \left| (k_2 - k_1)
    \hat{\textbf{k}}_2  - \textbf{p} \right| \left. \frac{\partial
    I_{\pm}}{\partial \left| \textbf{k}_1 + \textbf{k}_2 - \textbf{p} \right|}
    \right|_{\theta_k = \pi} -    I_{\pm} |_{\theta_k = \pi} \Bigg)
    ~, \nonumber \\ \\
    \tilde{\mathbb{B}}_{\pm} & = & \frac{k_1^2 p^2 \cos^2 \phi \sin^2 \theta
     \mathbb{P}^{\lambda_0 \lambda_1 \lambda_2} |_{\theta_k =
    \pi}    I_{\pm} |_{\theta_k = \pi}}{2 \left| (k_2 - k_1)
    \hat{\textbf{k}}_2 - \textbf{p} \right|^2} ~. 
  \end{eqnarray}
\end{subequations}
Due to the integrals over coordinate $\phi$, the leading orders of the
bispectra are shown to be proportional to the $\theta_k^2$ in Eq.~(\ref{36}) and $(\pi-\theta_k)^2$ in Eq.~(\ref{38}). However, it is not as simple as expected. It is found that the expansion at
$\theta_k \rightarrow 0$ or $\pi$ introduce singularities in the terms of
$\partial^2 I_{*} / \partial \left| \textbf{k}_1 + \textbf{k}_2 -
\textbf{p} \right|_{\theta_k = 0, \pi}$, thereby leading to the integrals to
be devergent. To obtain correct results, regularization scheme should be
employed as presented in Appendix~\ref{appB}. In this context, it is noted that the regularized singular terms would dominate the bispectrum amplitudes and result in the bispectrum proportional to $\theta_k^{\beta}$ or $(\pi-\theta_k)^{\beta}$, where $\beta < 2$. This feature is extensively illustrated in Appendix~\ref{appB}. Here, in this sense, the leading orders of the expansions in Eqs.~(\ref{36}) and (\ref{38}) can reduce to
\begin{subequations}
  \begin{eqnarray}
    B_0^{\lambda_0 \lambda_1 \lambda_2} \left( \textbf{k}_1, \textbf{k}_2
    \right) & \simeq & \frac{\theta_k^2}{(k_1 + k_2)   k_1 k_2 \eta^3}
    \int \frac{{\rm d}^3 p}{(2 \pi)^3} \Bigg\{ P_{\Psi}\left( \left|
    \textbf{k}_2 - \textbf{p} \right| \right) P_{\Psi}\left( \left| (k_1 + k_2)
    \hat{\textbf{k}}_2 - \textbf{p} \right| \right) P_{\Psi}(p)
    \bar{\mathbb{P}}^{\lambda_0 \lambda_1 \lambda_2}_0 \nonumber\\
    && \times \frac{p^8 \sin^8
    \theta}{64 \sqrt{2} k_2^2 (k_1 + k_2)^2 ((k_1 + k_2)^2 + p^2 - 2 (k_1 + k_2)
    p \cos \theta)} \left. \frac{\partial^2 I_0}{\partial \left|
    \textbf{k}_1 + \textbf{k}_2 - \textbf{p} \right|^2} \right|_{\theta_k = 0}
    \Bigg\} ~, \nonumber\\
    \\
    B_{\pm}^{\lambda_0 \lambda_1 \lambda_2} \left( \textbf{k}_1, \textbf{k}_2
    \right) & \simeq & \frac{(\theta_k - \pi)^2}{| k_2 - k_1 |   k_1 k_2
    \eta^3} \int \frac{{\rm d}^3 p}{(2 \pi)^3} \Bigg\{ P_{\Psi}\left( \left|
    \textbf{k}_2 - \textbf{p} \right| \right) P_{\Psi}\left( \left| (k_2 - k_1)
    \hat{\textbf{k}}_2 - \textbf{p} \right| \right) P_{\Psi}(p)
    \bar{\mathbb{P}}^{\lambda_0 \lambda_1 \lambda_2}_{\pm} \nonumber\\
    && \times \frac{p^8 \sin^8
    \theta}{64 \sqrt{2} k_2^2 (k_2 - k_1)^2 ((k_2 - k_1)^2 + p^2 - 2 (k_2 - k_1)
    p \cos \theta)} \left. \frac{\partial^2 I_{\pm}}{\partial \left|
    \textbf{k}_1 + \textbf{k}_2 - \textbf{p} \right|^2} \right|_{\theta_k = \pi}
    \Bigg\} ~, \nonumber\\
  \end{eqnarray} \label{40}
\end{subequations}
where $\bar{\mathbb{P}}^{\lambda_0 \lambda_1 \lambda_2}_{\ast}$ is defined in
Tab.~\ref{T2}. The $\bar{\mathbb{P}}^{\lambda_0 \lambda_1 \lambda_2}$ with constant components indicates that the polarization components differ only in bispectrum amplitudes. Based on the regularization scheme presented in Appendix~\ref{appB}, we introduce regularization parameters $\varepsilon_0$ and $\varepsilon_\pi$. It can result in $\sqrt{\varepsilon_0} B_0^{\lambda_0 \lambda_1 \lambda_2}$ and $\sqrt{\varepsilon_\pi} B_\pm^{\lambda_0 \lambda_1 \lambda_2}$ yielding finite values. Here, we ansatz $\varepsilon_0=\varepsilon_\pi\equiv\varepsilon$ for the sake of simplicity. 
\begin{table}[ht] 
  \centering 
  \caption{The values of $\bar{\mathbb{P}}_\ast$ in Eqs.~(\ref{40})}\label{T2}
  \begin{ruledtabular}
    \begin{tabular}{c|ccc}
    {$\lambda_0 \lambda_1 \lambda_2$} & $\bar{\mathbb{P}}_0$ &
    $\bar{\mathbb{P}}_+$ & $\bar{\mathbb{P}}_-$\\
    \hline
    $^{\times \times +}$ & $- 1$ & $- 1$ & $ 1$\\
    $^{\times + \times}$ & $- 1$ & $ 1$ & $- 1$\\
    $^{+ \times \times}$ & $1$ & $- 1$ & $- 1$\\
    $^{+ + +}$ & 3 & 3 & 3\\
    others & 0 & 0 & 0
  \end{tabular}
\end{ruledtabular}
\end{table}
Associating the results in Tab.~\ref{T2} with Eq.~(\ref{defBh}), we can also  obtain
\begin{subequations}
  \begin{eqnarray}
    B_h^{\times\times+}(k_1,k_2,\theta_k\rightarrow0) = B_h^{\times+\times}(k_2,k_1,\theta_k\rightarrow0) &=& -\frac{1}{3}(B_0^{+++}(k_1,k_2)+B_0^{+++}(k_2,k_1))~, \nonumber \\ \\
    B_h^{+++}(k_1,k_2,\theta_k\rightarrow0) = 3B_h^{\times++}(k_1,k_2,\theta_k\rightarrow0) &=& B_0^{+++}(k_1,k_2)+B_0^{+++}(k_2,k_1)~, 
  \end{eqnarray}
  and
  \begin{eqnarray}
    B_h^{\times\times+}(k_1,k_2,\theta_k\rightarrow\pi) = B_h^{\times+\times}(k_2,k_1,\theta_k\rightarrow\pi) &=& \frac{1}{3}(B_\pm^{+++}(k_1,k_2)+B_\pm^{+++}(k_2,k_1))\nonumber\\
    &&\times(\Theta(k_2-k_1) - \Theta(k_1-k_2))~, \nonumber \\ \\
    B_h^{+++}(k_1,k_2,\theta_k\rightarrow\pi) = 3B_h^{\times++}(k_1,k_2,\theta_k\rightarrow\pi) &=& B_\pm^{+++}(k_1,k_2)+B_\pm^{+++}(k_2,k_1)~, \nonumber \\ 
  \end{eqnarray} \label{BhfromB}
\end{subequations}
where the $\Theta(x)$ is Heaviside step function.
It shows that the different polarization components of the bispectrum $B_h^{\lambda_0\lambda_1\lambda_2}$ can all be derived from two independent quantities, i.e., $B_0^{+++}$ and $B_\pm^{+++}$. 

\begin{figure}[ht] \centering
  \includegraphics[width=1\linewidth]{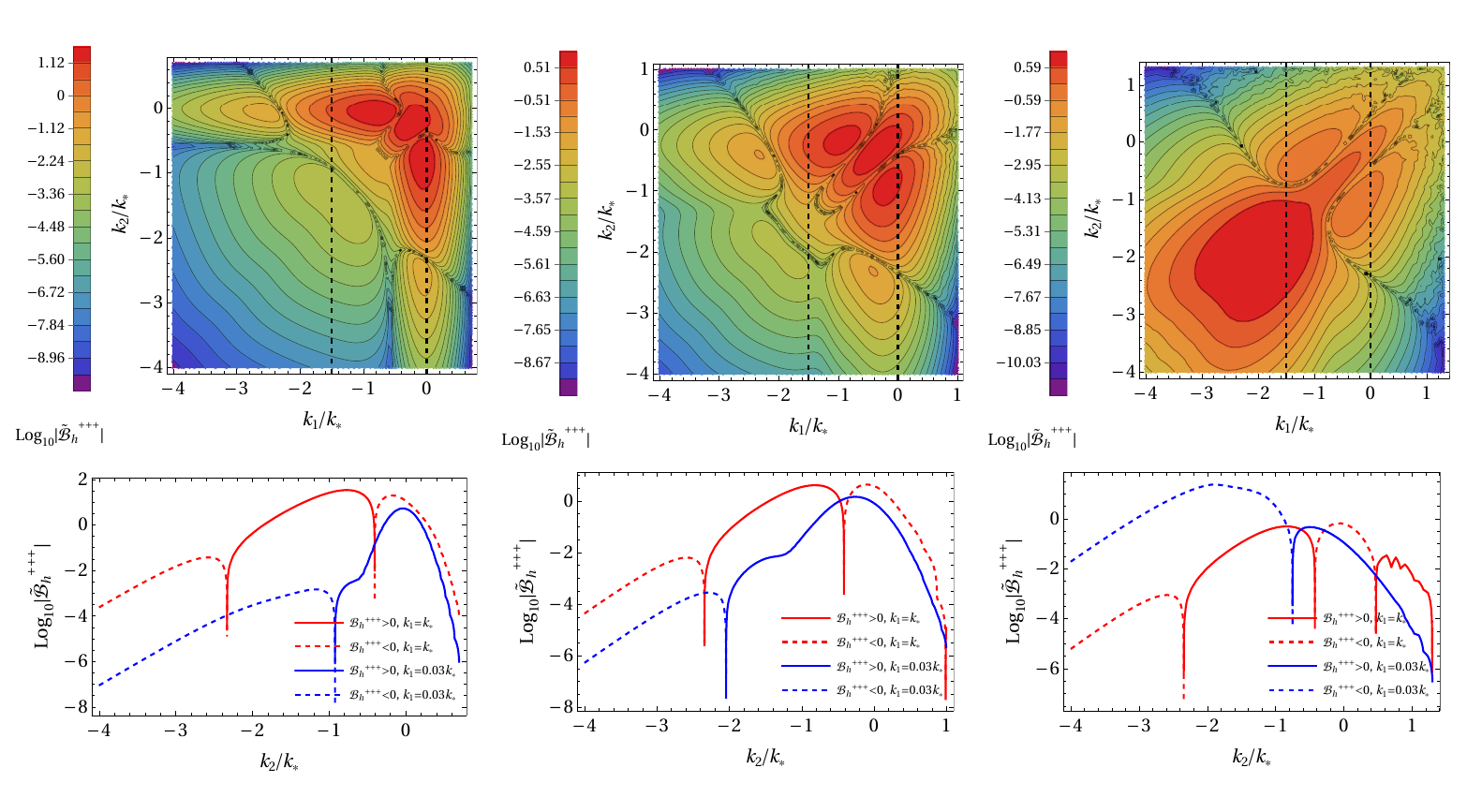}
  \caption{\label{F7} The dimensionless bispectrum at $\theta_k\rightarrow0$  as function of $k_1 / k_{\ast}$ and $k_2 / k_{\ast}$
  for $\Delta=1/2$ (left panel), 1 (medium panel), $2$ (right panel). Here, we have $\tilde{\mathcal{B}}_h^{+++}\equiv(k_*\eta)^3\theta_k^{-2}\sqrt{\varepsilon}{\mathcal{B}}_h^{+++}$. Here, we set $A=1$.}
\end{figure}

\begin{figure}[ht] \centering
  \includegraphics[width=1\linewidth]{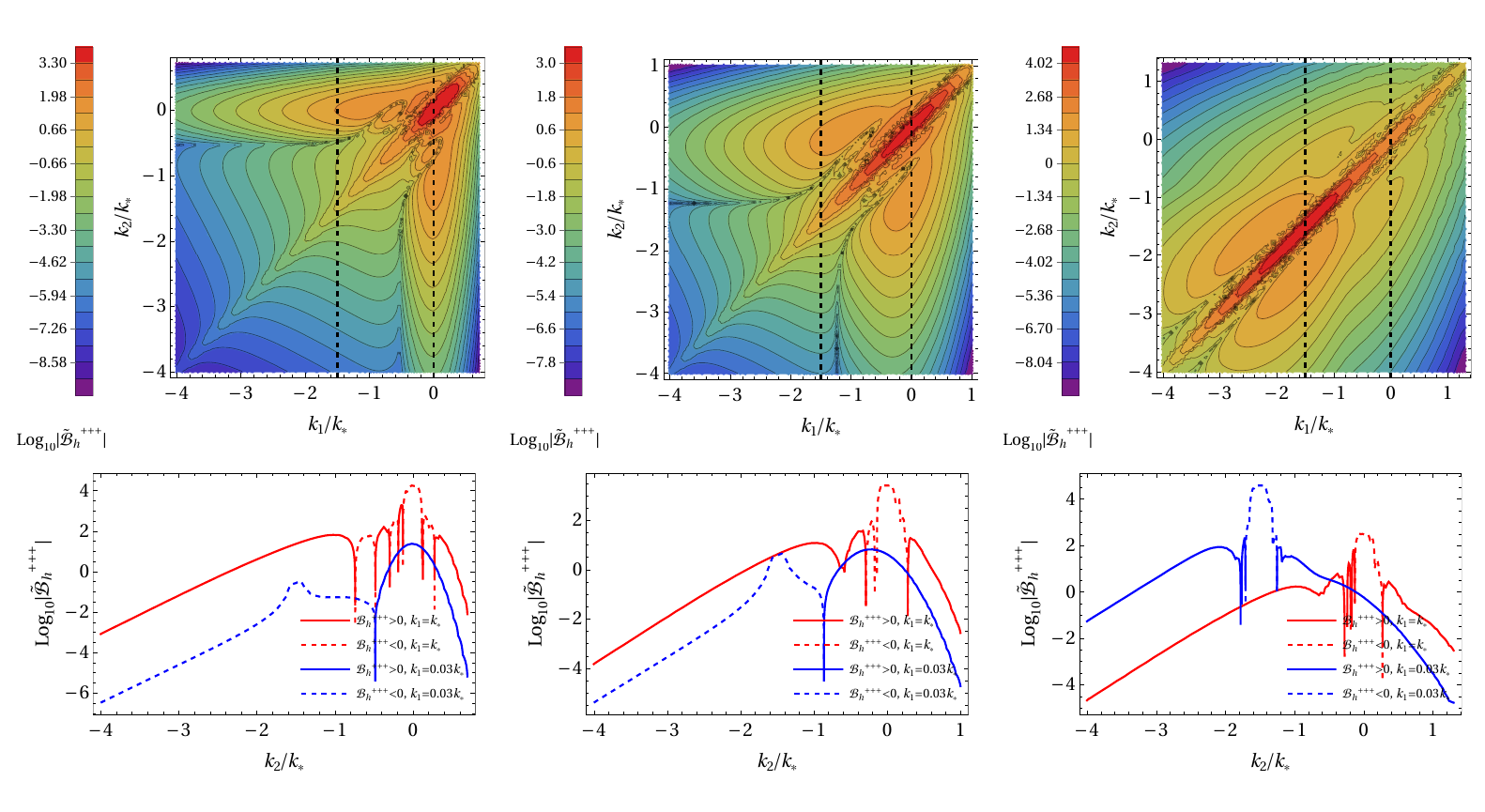}
  \caption{\label{F8} The dimensionless bispectrum at $\theta_k\rightarrow\pi$  as function of $k_1 / k_{\ast}$ and $k_2 / k_{\ast}$
  for $\Delta=1/2$ (left panel), 1 (medium panel), $2$ (right panel). Here, we have $\tilde{\mathcal{B}}_h^{+++}\equiv(k_*\eta)^3\theta_k^{-2}\sqrt{\varepsilon}{\mathcal{B}}_h^{+++}$. Here, we set $A=1$.}
\end{figure}

In the phenomenological approach, we also consider the
log-normal curvature power spectrum as given in Eq.~(\ref{lognormal}). Figs.~\ref{F7} and \ref{F8} show the 
bispectrum as function of $k_1$ and $k_2$ for selected $\Delta$ with $\theta_k\rightarrow0$ and $\theta_k\rightarrow\pi$, respectively. We also calculated it numerically, based on Monte-Carlo integration. Because of the
constant $\bar{\mathbb{P}}^{\lambda_0 \lambda_1 \lambda_2}_{\ast}$ as
mentioned above, we only plot the $+ + +$ polarization components.
The bispectrum amplitude is shown to be enhanced for a narrower width $\Delta$.
Compared between Fig.~\ref{F7} and Fig.~\ref{F8},
the maximum bispectrum amplitudes for $\theta_k \simeq \pi$ are much larger than that of $\theta_k \rightarrow 0$. 

\

\section{Skewness of the scalar-induced gravitational waves\label{IV}}

As mentioned in Sec.~{\ref{II}}, the degree of non-Gaussianity of SIGWs can be quantified by the skewness. In this section, we wish to examine whether the non-Gaussianity of SIGWs is significant. To be specific, examine whether we have a skewness $|\Gamma| \ll 1$.
The skewness will be calculated by making use of the variances in the matter-dominated era and radiation-dominated era, separately.
Since we have different polarization components for the bispectrum, one can rewrite the skewness in Eq.~(\ref{defSkew}) with polarization indices, namely, 
\begin{eqnarray}
  \Gamma_{\lambda_0\lambda_1\lambda_2}=\frac{\mu^3_{\lambda_0\lambda_1\lambda_2}}{\bar{\sigma}^3}~,
\end{eqnarray}
where the average variance is given by $\bar{\sigma}=(\sigma^{++}+\sigma^{\times\times})/2$ based on Eq.~(\ref{var}) and the unpolarized power spectrum given in Eq.~(\ref{spectrah}).

\begin{figure}[ht] \centering
  \includegraphics[width=0.8\linewidth]{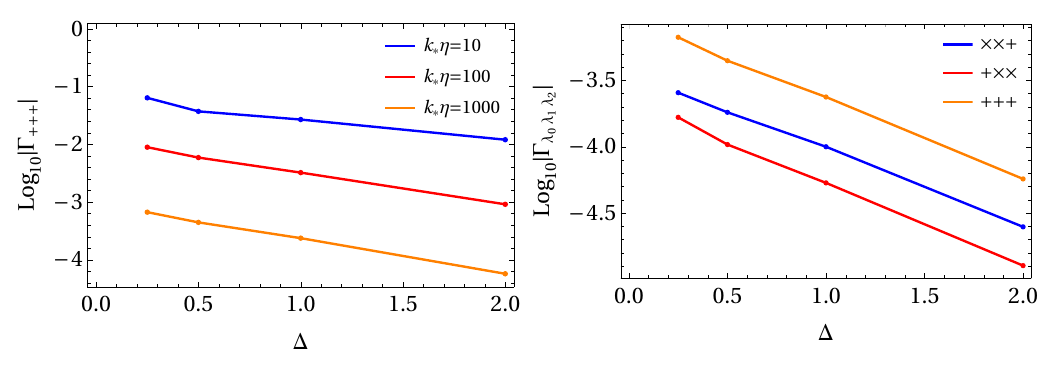}
  \caption{\label{F9} The skewness for SIGWs generated during matter-dominated era as function of width $\Delta$ of the curvature power spectrum for selected $k_*\eta$ (left panel) and different polarizations (right panels).}
\end{figure}

 The skewness for SIGWs generated during matter-dominated era as function of width $\Delta$ is presented in Fig.~\ref{F9}. It is calculated, numerically.
It is found that the skewness decreases with the width $\Delta$, indicating that the peaked curvature power spectrum can result in the enhancements of the non-Guassianity of SIGWs. For SIGWs generated from the exit of the inflationary era $\eta=0$ to the $\eta\propto k_*^{-1}$, the skewness is shown to be suppressed as the  $k_*\eta$ increases. Specifically, we have the skewness $\propto (k_*\eta)^{-1}$, approximately. It indicates that the non-Guassianity is mostly generated at the time when the perturbations enter the horizon. As SIGWs evolve to the late time with a large $k_*\eta$, the superposition of the waves would suppress the non-Gaussianity.  In the right panel of Fig.~\ref{F9}, different polarization components of the skewness have similar behavior with respect to the $\Delta$. The $+++$ component of the skewness has the largest amplitude, thereby indicating the largest non-Gaussianity. According to the results in Fig.~\ref{F9}, the skewness is tended to less than $1$.

\begin{figure}[ht] \centering
  \includegraphics[width=0.55\linewidth]{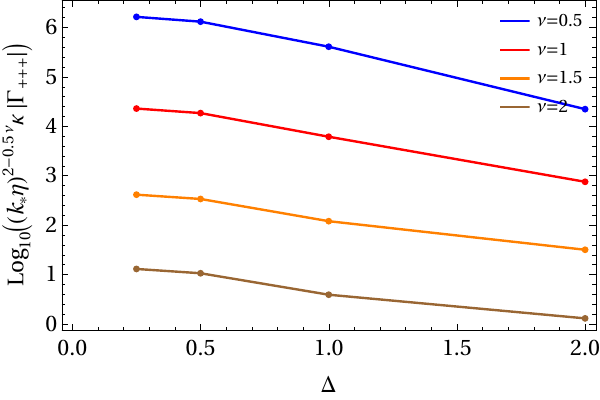}
  \caption{The skewness for SIGWs generated during radiation-dominated era as function of width $\Delta$ of the curvature power spectrum for selected $\nu$. We have let $\kappa_0=\kappa_\pi\equiv\kappa$ in the plot.\label{F10}}
\end{figure}

In radiation-dominated era, the third moments of SIGWs in Eq.~(\ref{trimon}) reduce to 
\begin{eqnarray}
  \mu_{\lambda_0 \lambda_1 \lambda_2}^3 & = & \frac{1}{2} \int_{-
  \infty}^{\infty}  {\rm d}\ln   k_1 \int_{- \infty}^{\infty}  {\rm d}\ln
    k_2 \Bigg\{ \int_0^{\theta_{k, 0}}  {\rm d} \theta_k \{ \theta_k
   \mathcal{B}_h^{\lambda_0 \lambda_1 \lambda_2} |_{\theta_k
  \rightarrow 0} \} + \int_{\pi - \theta_{k, \pi}}^{\pi}  {\rm d} \theta_k \{
  (\pi - \theta_k)  \mathcal{B}_h^{\lambda_0 \lambda_1 \lambda_2}
  |_{\theta_k \rightarrow \pi} \} \Bigg\} \nonumber\\
  & = & \frac{(k_{\ast} \eta)^{\frac{\nu}{2} - 5}}{2 (4 - \nu)} \int_{-
  \infty}^{\infty}  {\rm d} \ln   k_1 \int_{- \infty}^{\infty}  {\rm d} \ln
    k_2 \Bigg\{ \kappa_0^{- 1} (k_{\ast} \eta)^3 \theta_k^{- 2}
  \sqrt{\varepsilon_0} \mathcal{B}_h^{\lambda_0 \lambda_1 \lambda_2}
  |_{\theta_{k \rightarrow 0}} \left( \left( \frac{k_2}{k_1} + 1 \right)
  \frac{2 k_{\ast}}{k_2} \right)^{2 - \frac{\nu}{2}} \nonumber\\
  &  & + \kappa_{\pi}^{- 1} (k_{\ast} \eta)^3 (\pi - \theta_k)^{- 2}
  \sqrt{\varepsilon_{\pi}} \mathcal{B}_h^{\lambda_0 \lambda_1 \lambda_2}
  |_{\theta_k \rightarrow \pi} \left( \left| \frac{k_2}{k_1} - 1 \right|
  \frac{2 k_{\ast}}{k_2} \right)^{2 - \frac{\nu}{2}} \Bigg\}~, \label{3momRD}
\end{eqnarray} 
where the $\theta_{k,0}$ and $\theta_{k,\pi}$ are given based on the $\theta_{k,\rm min/max}$ in Tab.~\ref{T1}, namely,
\begin{eqnarray}
  \theta_{k, 0}  \equiv  \sqrt{\left( \frac{k_2}{k_1} + 1 \right)
  \frac{2}{k_2 \eta}}~, \hspace{0.5cm}
  \theta_{k, \pi}  \equiv  \sqrt{\left| \frac{k_2}{k_1} - 1 \right|
  \frac{2}{k_2 \eta}}~.
\end{eqnarray}
By employing the regularization scheme, the integral for the bispectrum
$\sqrt{\varepsilon}  \mathcal{B}_h^{\lambda_0 \lambda_1 \lambda_2}
|_{\theta_k \rightarrow 0, \pi}$ is finite. We ansatz $\sqrt{\varepsilon_0} = \kappa_0 \theta_k^{\nu}$ and
$\sqrt{\varepsilon_{\pi}} = \kappa_{\pi} (\pi - \theta_k)^{\nu}$, and $0 < \nu < 2$ in Eq.~(\ref{3momRD}). The upper bound of $\nu$ is determined by the requirement for the collinear configuration that the bispectrum vanish for $\theta_k=\pi-\theta_k = 0$. Because of the poor numerical precision of the integrations, we could not obtain the actual value of $\nu$. Additional discussion on this aspect is provided in Appendix~\ref{appB}, where we present an example indicating $\nu\simeq 1.4$. Here, we show the skewness as function of the $\Delta$ with the undetermined parameter $\nu$ in Fig.~\ref{F10}. We only consider the $+++$ polarization components as the suggestion of Eqs.~(\ref{BhfromB}).
It also indicates that the non-Gaussianity of SIGWs is enhanced due to the peaked curvature power spectrum.  The conclusion remains robust because the monotonicity of function $|\Gamma(\Delta)|$ is independent of the choice of $\nu$. Besides, the skewness of SIGWS also decreases with the $k_*\eta$. We can obtain the analytical result of the skewness $\propto (k_*\eta)^{\frac{\nu}{2}-2}$. Because of $0<\nu<2$, the non-Gaussinity for SIGWs generated during radiation-dominated era decays more rapidly with $\eta$ compared to that for the matter-dominated era. 
Therefore, the skewness is expected to be much less than $1$ due to the substantial value of $k_*\eta$ at the late time.

\

\section{Conclusions and Discussions\label{V}}

This paper investigated the intrinsic non-Gaussianity of SIGWs influenced by formation of primordial black holes. In a phenomenological approach, we considered the primordial curvature spectrum modeled as a lognormal one. The bispectrum and skewness of SIGWs generated during the matter-dominated era and the radiation-dominated era were calculated, separately. 
The bispectrum vanishes in the collinear configuration, which is shown to be independent of both the initial conditions and the dynamics of SIGWs. Based on our proposed oscillation average scheme, the bispectrum of SIGWs generated during the radiation-dominated era is dominated by the flattened configuration. And there are four polarization components $\times\times+$, $\times+\times$, $+\times\times$ and $+++$ left to be non-vanishing. 
Utilizing the skewness for quantifying the non-Gaussianity, it was found that the curvature power spectrum with a  narrower width can result in an enhancement of the   non-Gaussianity. The conclusion holds for both the SIGWs generated during the radiation-dominated era and the matter-dominated era. 

The decay of skewness over time indicates that the accumulation or superposition of waves possesses the capacity to suppress the non-Gaussiananity of SIGWs. 
The skewness for radiation-dominated era decays more rapidly compared to that for matter-dominated era. This can be attributed to the fact that the oscillations of SIGWs in the radiation-dominated era also serve to suppress the non-Gaussianity. 

We introduced alternative oscillation average scheme that can conserve the skewness of SIGWs. It ensure us to correctly examine whether a large curvature perturbation induced by formation of primordial black hole can affect the intrinsic non-Gaussianity.
The most interesting conclusion of this study lies in the peaked curvature power spectrum can result in an enhancement of the non-Guassianity of SIGWs. It motivates further studies on the non-Gaussinanity of SIGWs with the curvature power spectrum modeled as a delta function. In fact, we have obtained the preliminary results as presented in Appendix~\ref{appD} and found it difficult to understand the nature cut-off of the bispectrum. Perhaps, it is expected to be addressed in future studies.  



\ 

{\it Acknowledgments. }The author thanks Prof.~Sai Wang for useful discussions.
This work is supported by the National Natural Science Foundation of China under grants No.~12305073 and No.~12347101.

\

\ 

\ 

\appendix

\

\section{Regularization \label{appB}}

In this study, we encounter a situation that function $f (x, \lambda)$ has a
finite integral over $x$, while when we expand it with respect to the
parameter $\lambda$, the resulting integral diverges.
For illustration, we consider a simple example with a function in the form of 
\begin{eqnarray}
  f (x, \lambda) & \equiv & \frac{\sin x}{x + \lambda} ~,
\end{eqnarray}
its integral over $x$ can be analytically given by
\begin{eqnarray}
  \int_0^1 f (x, \lambda) {\rm d} x & = & ({\rm Ci} (\lambda) - {\rm Ci} (1 +
  \lambda)) \sin \lambda - ({\rm Si} (\lambda) - {\rm Si} (1 + \lambda))
  \cos \lambda \nonumber\\
  & = & {\rm Si} (1) + (\gamma - 1 - {\rm Ci} (1) + {\rm Si} (1) + \log
  \lambda) \lambda +\mathcal{O} (\lambda^2) \nonumber\\
  & = & {\rm Si} (1) + \lambda \ln \lambda +\mathcal{O} (\lambda)~, \label{B2}
\end{eqnarray}
where $\gamma$ is Euler's constant, and we have expanded the integral for a
small $\lambda$ in the second equal sign. On the other side, we expanded the function $f (x,
\lambda)$ first, namely,
\begin{eqnarray}
  f (x, \lambda) & = & \frac{\sin x}{x} - \frac{\sin x}{x^2} \lambda
  +\mathcal{O} (\lambda^2) ~.
\end{eqnarray}
It shows that the leading term is consistent with that in Eq.~(\ref{B2}),
\begin{eqnarray}
  \int_0^1 {\rm d} x \left\{ \frac{\sin x}{x} \right\} & = & {\rm Si} (1)~.
\end{eqnarray}
However, one might find it difficult to obtain consistent subleading terms in
Eq.~(\ref{B2}), because the expansion introduces a singularity in $\sin x /
x^2$, making the integral diverge.

Here, we should employ regularization to handle the singularity. By introducing a small
parameter $\varepsilon$ to the singular term to smooth out the singularity, thereby obtaining the integral as follows
\begin{eqnarray}
  \int_0^1 {\rm d} \lambda \left\{ - \frac{\sin x}{x^2 + \varepsilon}
  \right\} & = & - \frac{1}{2 \sqrt{\varepsilon}} \Big( \left( 2 {\rm Ci}
  \sqrt{\varepsilon} - {\rm Ci} \left( 1 - i \sqrt{\varepsilon} \right) -
  {\rm Ci} \left( 1 + i \sqrt{\varepsilon} \right) \right) \sinh
  \sqrt{\varepsilon} \nonumber\\ && + \left( - 2 {\rm Si} \sqrt{\varepsilon} + i  
  {\rm Si} \left( 1 - i \sqrt{\varepsilon} \right) - i   {\rm Si}
  \left( 1 + i \sqrt{\varepsilon} \right) \right) \cosh \sqrt{\varepsilon}
  \Big) \nonumber\\
  & = & \frac{1}{2} \log \varepsilon +\mathcal{O} (\varepsilon) ~. \label{B5}
\end{eqnarray}
It shows that the integral is finite. Finally, matching the result Eq.~(\ref{B5}) with the subleading order terms in Eq.~(\ref{B2}), we obtain
\begin{eqnarray}
  \varepsilon & = & \lambda^2 ~.
\end{eqnarray}
It shows that the subleading order in the expansion is not proportional to the
$\lambda$, but actually the $\lambda \ln \lambda$. The divergence of the
integration for the $\sin x / x^2$ is caused by the $\ln \lambda$ at singularity $\lambda \rightarrow 0$. 
Additionally, it indicates that the integration of singular terms at the order $\mathcal{O} (\lambda^n)$ can yield an outcome of order $\mathcal{O} (\lambda^m)$, where $m < n$. This feature was found and has been used to simplify our calculation in Sec.~{\ref{III}}.

As mentioned in Sec.~\ref{III.B.2}, the singularities are introduced due to the expansion with respect
to small $\theta_k$. To be specific, the singularities exist in the
terms of $\partial I_{\ast} / \partial \left| \textbf{k }_1 + \textbf{k}_2 -
\textbf{p} \right|$ and $\partial^2 I_{\ast} / \partial \left| \textbf{k}_1 +
\textbf{k}_2 - \textbf{p} \right|^2$ in Eqs.~(\ref{36}) and (\ref{38}). The former term does not
result in divergence of the integration if Cauchy principal value for the integral is
employed, while the divergence from the latter term is inevitable. Besides, as the singular term would dominate
the outcome of the integral, it can be used for simplification in
Eq.~(\ref{40}). Unlike the aforementioned simple example, obtaining the analytical outcome of the integral is not feasible in this case. The small regularization parameter, $\varepsilon$, should be numerically matched with the expansion parameter.

\begin{figure}
  \includegraphics[width=1\linewidth]{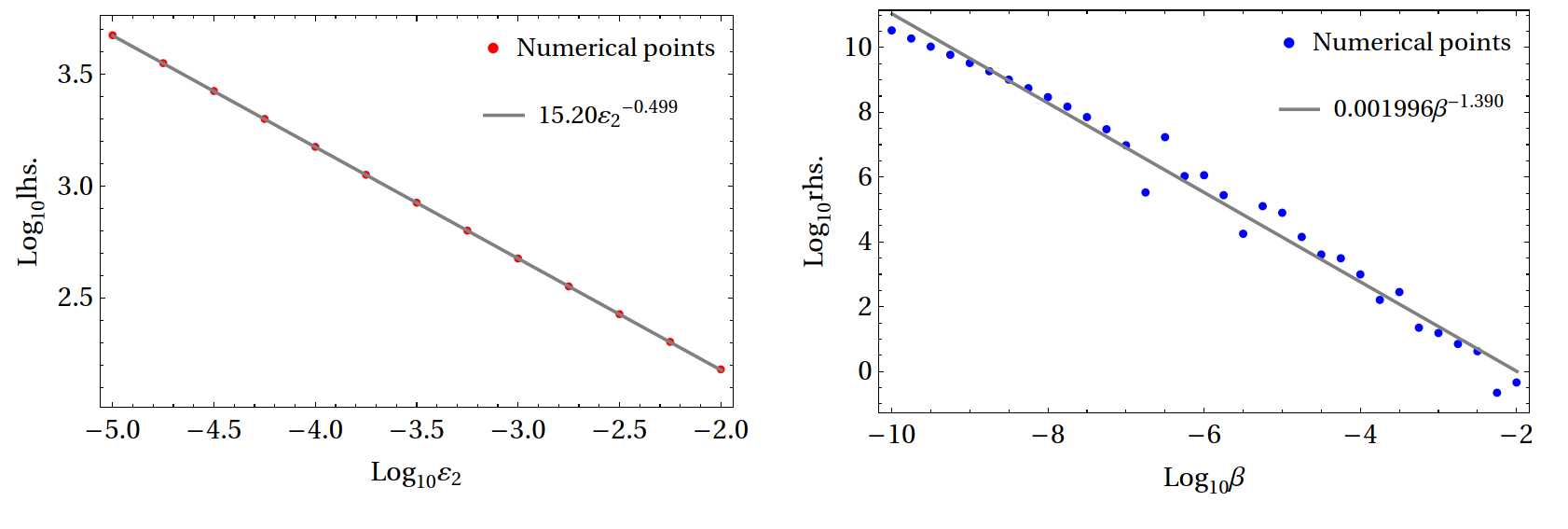}
  \caption{\label{AF1}Left panel: numerical results and the fits for the left-hand side of Eq.~(\ref{B13}) with respect to regularization parameter $\varepsilon$. Right panel: numerical results and the fits for the right-hand side of Eq.~(\ref{B13}) with respect to the parameter $\beta$.}
\end{figure}

Because the singularities in
Eq.~(\ref{40}) are from the expansion of the kernel functions $I_\text{B} ( | \textbf{k} - \textbf{p} |, \textbf{p}, k, \eta )$,
we here consider an example with $k=\eta=1$ as follows
\begin{eqnarray}
  {\rm Func} (\beta) & = & \int \frac{{\rm d}^3 p}{(2 \pi)^3} \left\{ e^{- \ln
  \left( \frac{\left| \textbf{k} - \textbf{p} \right|}{k} \right)^2 - \ln
  \left( \frac{p}{k} \right)^2} I_\text{B} \left( \left| \textbf{k} - \textbf{p}
  \right| + \beta, p, 1, 1 \right) \right\} \nonumber \\
  & = & \frac{1}{(2 \pi)^2} \int_0^{\infty} {\rm d} u \int_{| 1 -  v
  |}^{1 +  v} {\rm d}  v \{ e^{-   (\ln   u)^2 - (\ln
   v)^2} u  v I_\text{B} (u + \beta,  v, 1, 1) \} ~,
\end{eqnarray}
where we have utilized variable substitution $u = \left| \textbf{k} -
\textbf{p} \right|$ and $ v = p$ in the second equal sign. For the small
$\beta$, we wish to obtain the expansion of ${\rm Func}(\beta)$ with respect to the $\beta$. Firstly,
we can expand the $I_\text{B}$ with respect to small $\beta$, namely
\begin{eqnarray}
  {\rm Func} (\beta) & = & {\rm Func}_0 + \beta {\rm Func}_1 +
  \frac{\beta^2}{2} {\rm Func}_2 +\mathcal{O} (3) ~,\label{B8}
\end{eqnarray}
where we should calculate the integrations as follows,
\begin{subequations}
  \begin{eqnarray}
    {\rm Func}_0 & \equiv & \frac{1}{(2 \pi)^2} \int_0^{\infty} {\rm d} u \int_{| 1 -
     v |}^{1 +  v} {\rm d}  v \{ e^{-   (\ln  
    u)^2 - (\ln  v)^2} u  v I_\text{B} (u,  v, 1, 1) \} ~,\\
    {\rm Func}_1 & \equiv & \frac{1}{(2 \pi)^2} \int_0^{\infty} {\rm d} u \int_{| 1 -
     v |}^{1 +  v} {\rm d}  v \{ e^{-   (\ln  
    u)^2 - (\ln  v)^2} u  v \partial_u I_\text{B} (u,  v, 1, 1) \} ~, \\
    {\rm Func}_2 & \equiv & \frac{1}{(2 \pi)^2} \int_0^{\infty} {\rm d} u \int_{| 1 -
     v |}^{1 +  v} {\rm d}  v \{ e^{-   (\ln  
    u)^2 - (\ln  v)^2} u  v \partial_u^2 I_\text{B} (u,  v, 1, 1) \} ~.
  \end{eqnarray}
\end{subequations}
Because there are singularities in the derivative of $I_\text{B}$, we introduce the
regularization parameters $\varepsilon_1$ and $\varepsilon_2$, namely,
\begin{eqnarray}
  \partial_u I_\text{B} (u,  v, 1, 1)_{\varepsilon_1} & = & \frac{27}{4 u^4
   v^3} \Bigg( \frac{4 u  v (- u^6 + u^4 ( v^2 - 3) + 3
  ( v^2 - 3)^3 - 3 u^2 ( v^2 - 3) ( v^2 + 5))}{((u - v)^2 -
  3) ((u + v)^2 - 3) + \varepsilon_1 {\rm sign} (((u - v)^2 - 3) ((u + v)^2 -
  3))}  \nonumber \\
  &  & - (9 + u^2 - 3  v^2) (u^2 +  v^2 - 3) \ln \left| \frac{(u
  - v)^2 - 3}{(u + v)^2 - 3} \right| \Bigg)~,\\
  \partial_u^2 I_\text{B} (u,  v, 1, 1)_{\varepsilon_2} & = & \frac{27}{u^5
   v^3} \Bigg( \frac{4 u v}{((u - v)^2 -
  3)^2 ((u + v)^2 - 3)^2+\varepsilon_2} \Big(u^8 \left(v^2+3\right)+2 u^6 \left(v^2-3\right) \left(v^2+9\right) \nonumber \\ && -2 u^4
  \left(v^2-3\right) \left(5 v^4+3 v^2+54\right)+10 u^2 \left(v^2-3\right)^3
  \left(v^2+3\right)-3 \left(v^2-3\right)^5 \Big) \nonumber \\ && +
  \left(v^2-3\right) \left(u^2+3 v^2-9\right) \log \left|\frac{(u+v)^2-3}{(u-v)^2-3}\right| \Bigg)~. \label{B11}
\end{eqnarray}
Here, the Cauchy principal value for the ${\rm Func}_1$ can be given by
\begin{eqnarray}
   PV  {\rm Func_1} & = & \frac{1}{(2\pi)^2}\lim_{\varepsilon_1
  \rightarrow 0^+} \int_0^{\infty} {\rm d} u \int_{| 1 -  v |}^{1 +
   v} {\rm d}  v \{ e^{-   (\ln   u)^2 - (\ln
   v)^2} u  v \partial_u I_\text{B} (u,  v, 1, 1)_{\varepsilon_1}
  \} ~.
\end{eqnarray}
One can numerically check that the integral on the right-hand side of the above equation is finite. Given $\varepsilon_1$ being small enough, the value of $\varepsilon_1$ would
not affect the outcome of the integral. The divergence of ${\rm Func}_2$ is
inevitable. By introducing the parameter $\varepsilon_2$ as shown in Eq.~(\ref{B11}), we can
calculate the integral numerically. On the left-hand side of
Fig.~{\ref{AF1}}, we obtain ${\rm Func}_2 \propto \varepsilon_2^{- 1 / 2}$. Finally, we can relate the parameter $\varepsilon_2$ and $\beta$ based
on
\begin{eqnarray}
  \frac{1}{(2 \pi)^2} \int_0^{\infty} {\rm d} u \int_{| 1 -  v |}^{1 +
   v} {\rm d}  v \{ e^{-   (\ln   u)^2 - (\ln
   v)^2} u  v \partial_u^2 I_\text{B} (u,  v, 1, 1)_{\varepsilon_2}
  \} & = & \frac{2 ({\rm Func} (\beta) - {\rm Func}_0 - \beta
  PV{\rm Func}_1)}{\beta^2}~. \nonumber\\ \label{B13}
\end{eqnarray}
The numerical outcomes on the right-hand side of Eq.~(\ref{B13}) is shown on the right-hand side
of Fig.~{\ref{AF1}}. And we can obtain
\begin{equation}
  \varepsilon_2 \simeq (619.4 \beta)^{2.788}~. \label{B14}
\end{equation}
By making use of the results in Eq.~(\ref{B14}), we can rewrite the Eq.~(\ref{B8}) as
\begin{eqnarray}
  {\rm Func} (\beta) & = & {\rm Func}_0 + \beta PV{\rm Func}_1 + 6.4 \times
  10^{- 5} \beta^{0.6} \left( \sqrt{\varepsilon_2} {\rm Func}_2 \right)
  +\mathcal{O} (2) ~,
\end{eqnarray} 
where the integral $PV{\rm Func}_1$ and $\sqrt{\varepsilon_2} {\rm Func}_2$
are finite. Here, the subleading order of the expansion is given by
${\rm Func}_2$, instead of the temrs $PV \rm Func_1$.

Here, we did not suggest that the regularization parameter $\varepsilon_0$ and $\varepsilon_\pi$ in Sec.~\ref{III.B.2} are proportional to $\theta_k^{2.788}$. We indeed found that Eq.~(\ref{40}) is proportional to $\sqrt{\varepsilon_*}^{-1}$ with the numerical integrals. However, the reliable numerical outcome (like the right-hand side of Fig.~\ref{AF1}) for  $\theta_k\rightarrow0$ or $\theta_k\rightarrow\pi$ is difficult to obtain. Hence, we have to posit the ansatz $\sqrt{\varepsilon_0} = \kappa_0 \theta_k^{\nu}$ and $\sqrt{\varepsilon_{\pi}} = \kappa_{\pi} (\pi - \theta_k)^{\nu}$ with the undetermined parameters $\kappa_*$ and $\nu$.

\section{Explicit expressions of $\mathbb{P}^{\lambda_0\lambda_1\lambda_2}$ \label{appC}} 

The $\mathbb{P}^{\lambda_0\lambda_1\lambda_2}$ for $\lambda_*=\times,+$ used in Sec.~\ref{III} is presented as follows,
\begin{eqnarray}
\mathbb{P}^{\times\times\times}&=& \frac{1}{\sqrt{2} k_1^2 k_2^2 (k_1^2+2 k_1 k_2 \cos \theta_k+k_2^2)^{3/2}}\Big( 
  -4 p^5 \sin ^4\theta \sin ^3\phi  \cos \phi  (\sin \theta \cos \phi  (k_1 \cos \theta_k+k_2)\nonumber\\ &&
   -k_1 \cos \theta \sin \theta_k) 
  (\sin \theta_k
  (k_2-p \cos \theta)+p \sin \theta \cos \theta_k \cos \phi )~,\\
\mathbb{P}^{\times\times+}&=& \frac{1}{ \sqrt{2} k_1^2 k_2^2 (k_1^2+2 k_1 k_2 \cos \theta_k+k_2^2)^{3/2}}\Big( 
  -2 p^5 \sin ^4\theta \sin ^2\phi  \cos (2 \phi ) (\sin \theta \cos \phi (k_1 \cos \theta_k+k_2)\nonumber\\ &&
  -k_1 \cos \theta \sin \theta_k) (\sin \theta_k
  (k_2-p \cos \theta)+p \sin \theta \cos \theta_k \cos \phi )~,
\end{eqnarray}
\begin{eqnarray}
\mathbb{P}^{\times+\times}&=& \frac{1}{ \sqrt{2} k_1^2 k_2^2 (k_1^2+2 k_1 k_2 \cos \theta_k+k_2^2)^{3/2}}\Big( 
  -2 p^4 \sin ^3\theta \sin ^2\phi  \cos \phi  (\sin \theta \cos \phi  (k_1 \cos \theta_k+k_2)\nonumber\\ &&
  -k_1 \cos \theta \sin \theta_k)  (\cos ^2\theta_k
   (p^2 \sin ^2\theta \cos ^2\phi -k_2^2 )+k_2^2-2 p \cos \theta  (-k_2 \cos ^2\theta_k+k_2\nonumber\\ &&
   +p \sin \theta \sin \theta_k \cos \theta_k
  \cos \phi  )+k_2 p \sin \theta \sin (2 \theta_k) \cos \phi \nonumber\\ &&
  +p^2 \cos ^2\theta \sin ^2\theta_k-p^2 \sin ^2\theta \sin ^2\phi  )~,\\
\mathbb{P}^{\times++}&=& \frac{1}{ \sqrt{2} k_1^2 k_2^2 (k_1^2+2 k_1 k_2 \cos \theta_k+k_2^2)^{3/2}}\Big( 
  -p^4 \sin ^3\theta \sin \phi  \cos (2 \phi ) (\sin \theta \cos \phi  (k_1 \cos \theta_k+k_2)\nonumber\\ &&
   -k_1 \cos \theta \sin \theta_k)  (\cos ^2\theta_k
   (p^2 \sin ^2\theta \cos ^2\phi -k_2^2 )+k_2^2-2 p \cos \theta  (-k_2 \cos ^2\theta_k+k_2 \nonumber\\ &&
    +p \sin \theta \sin \theta_k \cos \theta_k
  \cos \phi  )+k_2 p \sin \theta \sin (2 \theta_k) \cos \phi +p^2 \cos ^2\theta \sin ^2\theta_k\nonumber\\ &&
   -p^2 \sin ^2\theta \sin ^2\phi  )
\Big)~,\end{eqnarray}
\begin{eqnarray}
\mathbb{P}^{+\times\times}&=& \frac{1}{2 \sqrt{2} k_1^2 k_2^2 \left(k_1^2+2 k_1 k_2 \cos \theta_k+k_2^2\right)^3}\Big( 
  -4 p^5 \sin ^3\theta \sin ^2\phi  \cos \phi   (-\sin ^2\theta \cos ^2\phi  (k_1 \cos \theta_k\nonumber\\ &&+k_2)^2  (k_1^2
  +2 k_1 k_2 \cos \theta_k+k_2^2 )+\sin ^2\theta \sin ^2\phi   (k_1^2+2 k_1 k_2 \cos \theta_k+k_2^2 )^2\nonumber\\ &&+k_1^2 \sin \theta \cos \theta \sin (2 \theta_k)
   (k_1^2
   +3 k_2^2 ) \cos \phi +k_1 k_2 \sin (2 \theta ) \sin \theta_k \cos \phi   (k_1^2 \cos (2 \theta_k)\nonumber\\ &&+2
  k_1^2+k_2^2 )
   -k_1^2  \cos ^2\theta  \sin ^2\theta_k  (k_1^2+2 k_1 k_2 \cos \theta_k+k_2^2 ) ) (\sin \theta_k (k_2-p \cos \theta)\nonumber\\ &&+p \sin \theta \cos \theta_k \cos \phi )~,\\
\mathbb{P}^{+\times+}&=& \frac{1}{2 \sqrt{2} k_1^2 k_2^2 \left(k_1^2+2 k_1 k_2 \cos \theta_k+k_2^2\right)^3}\Big( 
  -2 p^5 \sin ^3\theta \sin \phi  \cos (2 \phi )  (-\sin ^2\theta \cos ^2\phi  (k_1 \cos \theta_k\nonumber\\ &&
   +k_2)^2  (k_1^2+2 k_1 k_2 \cos \theta_k+k_2^2 )+\sin ^2\theta \sin ^2\phi   (k_1^2+2 k_1 k_2 \cos \theta_k+k_2^2 )^2\nonumber\\ &&
   +k_1^2 \sin \theta \cos \theta \sin (2 \theta_k)
   (k_1^2+3 k_2^2 ) \cos \phi +k_1 k_2 \sin (2 \theta ) \sin \theta_k \cos \phi   (k_1^2 \cos (2 \theta_k)+2
  k_1^2+k_2^2 )\nonumber\\ &&
  -k_1^2  \cos ^2\theta  \sin ^2\theta_k  (k_1^2+2 k_1 k_2 \cos \theta_k+k_2^2 ) ) (\sin \theta_k (k_2-p \cos \theta)\nonumber\\ &&
  +p \sin \theta \cos \theta_k \cos \phi )
\Big)~,\end{eqnarray}
\begin{eqnarray}
\mathbb{P}^{++\times}&=& \frac{1}{2 \sqrt{2} k_1^2 k_2^2 \left(k_1^2+2 k_1 k_2 \cos \theta_k+k_2^2\right)^3}\Big( 
  2 p^4 \sin ^2\theta \sin \phi  \cos \phi   (\sin ^2\theta \cos ^2\phi  (k_1 \cos \theta_k\nonumber\\ &&
  +k_2)^2  (k_1^2+2 k_1 k_2 \cos \theta_k+k_2^2 )+ (k_1^2+2 k_1 k_2 \cos \theta_k+k_2^2 )  (k_1^2 \cos ^2\theta \sin ^2\theta_k\nonumber\\ &&
  -\sin ^2\theta \sin ^2\phi 
   (k_1^2+2 k_1 k_2 \cos \theta_k+k_2^2 ) )-k_1 \sin (2 \theta ) \sin \theta_k \cos \phi   (k_1 \cos \theta_k
   (k_1^2+3 k_2^2 )\nonumber\\ &&
   +k_2  (k_1^2 \cos (2 \theta_k)+2 k_1^2+k_2^2 ) ) )  (\cos ^2\theta_k  (p^2 \sin ^2\theta \cos
  ^2\phi -k_2^2 )+k_2^2\nonumber\\ &&
  -2 p \cos \theta  (k_2 \sin ^2\theta_k+p \sin \theta \sin \theta_k \cos \theta_k \cos \phi  )+k_2 p
  \sin \theta \sin (2 \theta_k) \cos \phi \nonumber\\ &&
  +p^2 \cos ^2\theta \sin ^2\theta_k-p^2 \sin ^2\theta \sin ^2\phi  )~,\\
\mathbb{P}^{+++}&=& \frac{1}{2 \sqrt{2} k_1^2 k_2^2 \left(k_1^2+2 k_1 k_2 \cos \theta_k+k_2^2\right)^3}\Big( 
  p^4 \sin ^2\theta \cos (2 \phi )  (\sin ^2\theta \cos ^2\phi  (k_1 \cos \theta_k+k_2)^2  (k_1^2\nonumber\\ &&
  +2 k_1 k_2 \cos \theta_k+k_2^2 )+ (k_1^2+2 k_1 k_2 \cos \theta_k+k_2^2 )  (k_1^2 \cos ^2\theta \sin ^2\theta_k-\sin ^2\theta \sin ^2\phi 
   (k_1^2\nonumber\\ &&
   +2 k_1 k_2 \cos \theta_k+k_2^2 ) )-k_1 \sin (2 \theta ) \sin \theta_k \cos \phi   (k_1 \cos \theta_k
   (k_1^2+3 k_2^2 )+k_2  (k_1^2 \cos (2 \theta_k)\nonumber\\ &&
   +2 k_1^2+k_2^2 ) ) )  (\cos ^2\theta_k  (p^2 \sin ^2\theta \cos
  ^2\phi -k_2^2 )+k_2^2-2 p \cos \theta  (k_2 \sin ^2\theta_k\nonumber\\ &&
  +p \sin \theta \sin \theta_k \cos \theta_k \cos \phi  )+k_2 p
  \sin \theta \sin (2 \theta_k) \cos \phi +p^2 \cos ^2\theta \sin ^2\theta_k\nonumber\\ &&
  -p^2 \sin ^2\theta \sin ^2\phi  )
\Big)~.
\end{eqnarray}

\section{SIGWs induced by a peaked curvature power spectrum \label{appD}}
Utilizing the variable substitution that can transform the momentum
$\textbf{p}$ into three dimensionless quantities $u,  v, w$, namely,
\[  v = \frac{p}{k_2}, \hspace{0.25cm} u = \frac{\left| \textbf{k}_2 -
   \textbf{p} \right|}{k_2}, \hspace{0.25cm} w = \frac{\left| \textbf{k}_1 +
   \textbf{k}_2 - \textbf{p} \right|}{k_2} . \]
we can evaluate the bispectrum in Eq.~(\ref{BiRD2}) as
\begin{eqnarray}
  B_{\ast}^{\lambda_0 \lambda_1 \lambda_2} \left( \textbf{k}_1, \textbf{k}_2
  \right) & = & \frac{1}{(2 \pi)^3 \left| \textbf{k}_1 + \textbf{k}_2 \right|
    \left| \textbf{k}_1 \right|   \left| \textbf{k}_2 \right|
  \eta^3} \int_0^{\infty} \int_{| 1 -  v |}^{1 +  v}
  \int_{w_-}^{w_+} \mathcal{D} ( v, u, k_1, k_2, \theta_k) {\rm d}
   v {\rm d} u {\rm d} w \nonumber\\
  &  & \left\{ P_{\Phi} (k_2 u) P_{\Phi} (k_2 w) P_{\Phi} (k_2  v)
  I_{\ast} \left( k_1, k_2, \left| \textbf{k}_1 + \textbf{k}_2 \right|, v, u,
  w \right) \mathbb{P}^{\lambda_0 \lambda_1 \lambda_3} |_{u, w, v} \right\}~, \nonumber \\
\end{eqnarray}
where $*=0,\pm$, the $I_* \left( k_1, k_2, \left| \textbf{k}_1 + \textbf{k}_2 \right|, v, u,
w \right)$ is given in Eq.~(\ref{BiRD3}), and
\begin{eqnarray}
  \mathcal{D}( v, u, w, k_1, k_2, \theta_k) & = & 4 u  v w   k_2^5 \Big( - k_1^4 - 2 (1 + u^2
  -  v^2) \cos \theta_k k_1^3 k_2 \nonumber\\ && - \left( u^4 + (1 -  v^2)^2 - 2
  w^2 - 2 u^2 \left(  v^2 - 2 \cos^2 \theta_k \right) \right) k_1^2
  k_2^2 \nonumber\\ &&  - 2 (1 + u^2 -  v^2) (u - w) (u + w) \cos \theta_k   k_1
  k_2^3 + (u^2 - w^2)^2 k_2^4 \Big)^{- \frac{1}{2}}~, \nonumber \\ \\
  w_{\pm}( v, u, k_1, k_2, \theta_k) & = & \bigg( u^2 + \left( \frac{k_1}{k_2} \right)^2  +
  \frac{k_1}{k_2} \Big( \cos \theta_k (1 + u^2 -  v^2) \nonumber\\ && \pm \sin
  \theta_k \sqrt{- ( u^2-1)^2 + 2 (1 + u^2)  v^2 -  v^4}
  \Big) \bigg)^{\frac{1}{2}}~. 
\end{eqnarray}
Considering a peaked curvature power spectrum $\mathcal{P}_\Psi=A k_* \delta(k-k_*)$, we obtain
\begin{eqnarray}
  B_{\star}^{\lambda_0 \lambda_1 \lambda_2} \left( \textbf{k}_1, \textbf{k}_2
  \right) & = & \frac{A^3 \pi^3}{k_{\ast}^6 \left| \textbf{k}_1 +
  \textbf{k}_2 \right| k_1 k_2^4 \eta^3} \mathcal{D} \left(
  \frac{k_{\ast}}{k_2}, \frac{k_{\ast}}{k_2}, \frac{k_{\ast}}{k_2}, k_1, k_2, \theta_k \right)
  \Theta \left( 2 \min \left[ 1, \frac{k_2 \sin \theta_k}{\left| \textbf{k}_1
  + \textbf{k}_2 \right|} \right] - \frac{k_2}{k_{\ast}} \right) \nonumber \\
  &  & \times I_{\star} \left( k_1, k_2, \left| \textbf{k}_1 + \textbf{k}_2
  \right|, \frac{k_{\ast}}{k_2}, \frac{k_{\ast}}{k_2}, \frac{k_{\ast}}{k_2}
  \right) \mathbb{P}^{\lambda_0 \lambda_1 \lambda_3} |_{u = w = v = k_{\ast} /
  k_2}~. \label{D4}
\end{eqnarray}
where $\star=0,\pm$  and $\Theta(x)$ is Heaviside step functions.  

Associating the condition of the flattened non-Gaussianity presented in Tab.~\ref{T1} with the Heaviside step functions in Eq.~(\ref{D4}), it is found that the bispectrum shows to be non-vanishing on large scale $k_1, k_2\lesssim\mathcal{\mathcal{O}(\sqrt{\epsilon/\eta})}$. It might suggest that the GW detectors on the current frequency band can not detect the bispectrum for the SIGWs induced by the $\delta$-peaked curvature power spectrum.

\bibliography{cite} 

\begin{thebibliography}{65}%
\makeatletter
\providecommand \@ifxundefined [1]{%
 \@ifx{#1\undefined}
}%
\providecommand \@ifnum [1]{%
 \ifnum #1\expandafter \@firstoftwo
 \else \expandafter \@secondoftwo
 \fi
}%
\providecommand \@ifx [1]{%
 \ifx #1\expandafter \@firstoftwo
 \else \expandafter \@secondoftwo
 \fi
}%
\providecommand \natexlab [1]{#1}%
\providecommand \enquote  [1]{``#1''}%
\providecommand \bibnamefont  [1]{#1}%
\providecommand \bibfnamefont [1]{#1}%
\providecommand \citenamefont [1]{#1}%
\providecommand \href@noop [0]{\@secondoftwo}%
\providecommand \href [0]{\begingroup \@sanitize@url \@href}%
\providecommand \@href[1]{\@@startlink{#1}\@@href}%
\providecommand \@@href[1]{\endgroup#1\@@endlink}%
\providecommand \@sanitize@url [0]{\catcode `\\12\catcode `\$12\catcode `\&12\catcode `\#12\catcode `\^12\catcode `\_12\catcode `\%12\relax}%
\providecommand \@@startlink[1]{}%
\providecommand \@@endlink[0]{}%
\providecommand \url  [0]{\begingroup\@sanitize@url \@url }%
\providecommand \@url [1]{\endgroup\@href {#1}{\urlprefix }}%
\providecommand \urlprefix  [0]{URL }%
\providecommand \Eprint [0]{\href }%
\providecommand \doibase [0]{http://dx.doi.org/}%
\providecommand \selectlanguage [0]{\@gobble}%
\providecommand \bibinfo  [0]{\@secondoftwo}%
\providecommand \bibfield  [0]{\@secondoftwo}%
\providecommand \translation [1]{[#1]}%
\providecommand \BibitemOpen [0]{}%
\providecommand \bibitemStop [0]{}%
\providecommand \bibitemNoStop [0]{.\EOS\space}%
\providecommand \EOS [0]{\spacefactor3000\relax}%
\providecommand \BibitemShut  [1]{\csname bibitem#1\endcsname}%
\let\auto@bib@innerbib\@empty
\bibitem [{\citenamefont {Hellings}\ and\ \citenamefont {Downs}(1983)}]{Hellings:1983fr}%
  \BibitemOpen
  \bibfield  {author} {\bibinfo {author} {\bibfnamefont {R.~w.}\ \bibnamefont {Hellings}}\ and\ \bibinfo {author} {\bibfnamefont {G.~s.}\ \bibnamefont {Downs}},\ }\href {\doibase 10.1086/183954} {\bibfield  {journal} {\bibinfo  {journal} {Astrophys. J. Lett.}\ }\textbf {\bibinfo {volume} {265}},\ \bibinfo {pages} {L39} (\bibinfo {year} {1983})}\BibitemShut {NoStop}%
\bibitem [{\citenamefont {Agazie}\ \emph {et~al.}(2023{\natexlab{a}})\citenamefont {Agazie} \emph {et~al.}}]{NANOGrav:2023gor}%
  \BibitemOpen
  \bibfield  {author} {\bibinfo {author} {\bibfnamefont {G.}~\bibnamefont {Agazie}} \emph {et~al.} (\bibinfo {collaboration} {NANOGrav}),\ }\href {\doibase 10.3847/2041-8213/acdac6} {\bibfield  {journal} {\bibinfo  {journal} {Astrophys. J. Lett.}\ }\textbf {\bibinfo {volume} {951}},\ \bibinfo {pages} {L8} (\bibinfo {year} {2023}{\natexlab{a}})},\ \Eprint {http://arxiv.org/abs/2306.16213} {arXiv:2306.16213 [astro-ph.HE]} \BibitemShut {NoStop}%
\bibitem [{\citenamefont {Antoniadis}\ \emph {et~al.}(2023)\citenamefont {Antoniadis} \emph {et~al.}}]{EPTA:2023fyk}%
  \BibitemOpen
  \bibfield  {author} {\bibinfo {author} {\bibfnamefont {J.}~\bibnamefont {Antoniadis}} \emph {et~al.} (\bibinfo {collaboration} {EPTA, InPTA:}),\ }\href {\doibase 10.1051/0004-6361/202346844} {\bibfield  {journal} {\bibinfo  {journal} {Astron. Astrophys.}\ }\textbf {\bibinfo {volume} {678}},\ \bibinfo {pages} {A50} (\bibinfo {year} {2023})},\ \Eprint {http://arxiv.org/abs/2306.16214} {arXiv:2306.16214 [astro-ph.HE]} \BibitemShut {NoStop}%
\bibitem [{\citenamefont {Reardon}\ \emph {et~al.}(2023)\citenamefont {Reardon} \emph {et~al.}}]{Reardon:2023gzh}%
  \BibitemOpen
  \bibfield  {author} {\bibinfo {author} {\bibfnamefont {D.~J.}\ \bibnamefont {Reardon}} \emph {et~al.},\ }\href {\doibase 10.3847/2041-8213/acdd02} {\bibfield  {journal} {\bibinfo  {journal} {Astrophys. J. Lett.}\ }\textbf {\bibinfo {volume} {951}},\ \bibinfo {pages} {L6} (\bibinfo {year} {2023})},\ \Eprint {http://arxiv.org/abs/2306.16215} {arXiv:2306.16215 [astro-ph.HE]} \BibitemShut {NoStop}%
\bibitem [{\citenamefont {Xu}\ \emph {et~al.}(2023)\citenamefont {Xu} \emph {et~al.}}]{Xu:2023wog}%
  \BibitemOpen
  \bibfield  {author} {\bibinfo {author} {\bibfnamefont {H.}~\bibnamefont {Xu}} \emph {et~al.},\ }\href {\doibase 10.1088/1674-4527/acdfa5} {\bibfield  {journal} {\bibinfo  {journal} {Res. Astron. Astrophys.}\ }\textbf {\bibinfo {volume} {23}},\ \bibinfo {pages} {075024} (\bibinfo {year} {2023})},\ \Eprint {http://arxiv.org/abs/2306.16216} {arXiv:2306.16216 [astro-ph.HE]} \BibitemShut {NoStop}%
\bibitem [{\citenamefont {Afzal}\ \emph {et~al.}(2023)\citenamefont {Afzal} \emph {et~al.}}]{NANOGrav:2023hvm}%
  \BibitemOpen
  \bibfield  {author} {\bibinfo {author} {\bibfnamefont {A.}~\bibnamefont {Afzal}} \emph {et~al.} (\bibinfo {collaboration} {NANOGrav}),\ }\href {\doibase 10.3847/2041-8213/acdc91} {\bibfield  {journal} {\bibinfo  {journal} {Astrophys. J. Lett.}\ }\textbf {\bibinfo {volume} {951}},\ \bibinfo {pages} {L11} (\bibinfo {year} {2023})},\ \Eprint {http://arxiv.org/abs/2306.16219} {arXiv:2306.16219 [astro-ph.HE]} \BibitemShut {NoStop}%
\bibitem [{\citenamefont {Antoniadis}\ \emph {et~al.}(2024{\natexlab{a}})\citenamefont {Antoniadis} \emph {et~al.}}]{EPTA:2023xxk}%
  \BibitemOpen
  \bibfield  {author} {\bibinfo {author} {\bibfnamefont {J.}~\bibnamefont {Antoniadis}} \emph {et~al.} (\bibinfo {collaboration} {EPTA, InPTA}),\ }\href {\doibase 10.1051/0004-6361/202347433} {\bibfield  {journal} {\bibinfo  {journal} {Astron. Astrophys.}\ }\textbf {\bibinfo {volume} {685}},\ \bibinfo {pages} {A94} (\bibinfo {year} {2024}{\natexlab{a}})},\ \Eprint {http://arxiv.org/abs/2306.16227} {arXiv:2306.16227 [astro-ph.CO]} \BibitemShut {NoStop}%
\bibitem [{\citenamefont {Agazie}\ \emph {et~al.}(2023{\natexlab{b}})\citenamefont {Agazie} \emph {et~al.}}]{NANOGrav:2023pdq}%
  \BibitemOpen
  \bibfield  {author} {\bibinfo {author} {\bibfnamefont {G.}~\bibnamefont {Agazie}} \emph {et~al.} (\bibinfo {collaboration} {NANOGrav}),\ }\href {\doibase 10.3847/2041-8213/ace18a} {\bibfield  {journal} {\bibinfo  {journal} {Astrophys. J. Lett.}\ }\textbf {\bibinfo {volume} {951}},\ \bibinfo {pages} {L50} (\bibinfo {year} {2023}{\natexlab{b}})},\ \Eprint {http://arxiv.org/abs/2306.16222} {arXiv:2306.16222 [astro-ph.HE]} \BibitemShut {NoStop}%
\bibitem [{\citenamefont {Agazie}\ \emph {et~al.}(2023{\natexlab{c}})\citenamefont {Agazie} \emph {et~al.}}]{NANOGrav:2023hfp}%
  \BibitemOpen
  \bibfield  {author} {\bibinfo {author} {\bibfnamefont {G.}~\bibnamefont {Agazie}} \emph {et~al.} (\bibinfo {collaboration} {NANOGrav}),\ }\href {\doibase 10.3847/2041-8213/ace18b} {\bibfield  {journal} {\bibinfo  {journal} {Astrophys. J. Lett.}\ }\textbf {\bibinfo {volume} {952}},\ \bibinfo {pages} {L37} (\bibinfo {year} {2023}{\natexlab{c}})},\ \Eprint {http://arxiv.org/abs/2306.16220} {arXiv:2306.16220 [astro-ph.HE]} \BibitemShut {NoStop}%
\bibitem [{\citenamefont {Antoniadis}\ \emph {et~al.}(2024{\natexlab{b}})\citenamefont {Antoniadis} \emph {et~al.}}]{EPTA:2023gyr}%
  \BibitemOpen
  \bibfield  {author} {\bibinfo {author} {\bibfnamefont {J.}~\bibnamefont {Antoniadis}} \emph {et~al.} (\bibinfo {collaboration} {EPTA, InPTA}),\ }\href {\doibase 10.1051/0004-6361/202348568} {\bibfield  {journal} {\bibinfo  {journal} {Astron. Astrophys.}\ }\textbf {\bibinfo {volume} {690}},\ \bibinfo {pages} {A118} (\bibinfo {year} {2024}{\natexlab{b}})},\ \Eprint {http://arxiv.org/abs/2306.16226} {arXiv:2306.16226 [astro-ph.HE]} \BibitemShut {NoStop}%
\bibitem [{\citenamefont {Ananda}\ \emph {et~al.}(2007)\citenamefont {Ananda}, \citenamefont {Clarkson},\ and\ \citenamefont {Wands}}]{Ananda:2006af}%
  \BibitemOpen
  \bibfield  {author} {\bibinfo {author} {\bibfnamefont {K.~N.}\ \bibnamefont {Ananda}}, \bibinfo {author} {\bibfnamefont {C.}~\bibnamefont {Clarkson}}, \ and\ \bibinfo {author} {\bibfnamefont {D.}~\bibnamefont {Wands}},\ }\href {\doibase 10.1103/PhysRevD.75.123518} {\bibfield  {journal} {\bibinfo  {journal} {Phys. Rev. D}\ }\textbf {\bibinfo {volume} {75}},\ \bibinfo {pages} {123518} (\bibinfo {year} {2007})},\ \Eprint {http://arxiv.org/abs/gr-qc/0612013} {arXiv:gr-qc/0612013} \BibitemShut {NoStop}%
\bibitem [{\citenamefont {Baumann}\ \emph {et~al.}(2007)\citenamefont {Baumann}, \citenamefont {Steinhardt}, \citenamefont {Takahashi},\ and\ \citenamefont {Ichiki}}]{Baumann:2007zm}%
  \BibitemOpen
  \bibfield  {author} {\bibinfo {author} {\bibfnamefont {D.}~\bibnamefont {Baumann}}, \bibinfo {author} {\bibfnamefont {P.~J.}\ \bibnamefont {Steinhardt}}, \bibinfo {author} {\bibfnamefont {K.}~\bibnamefont {Takahashi}}, \ and\ \bibinfo {author} {\bibfnamefont {K.}~\bibnamefont {Ichiki}},\ }\href {\doibase 10.1103/PhysRevD.76.084019} {\bibfield  {journal} {\bibinfo  {journal} {Phys. Rev. D}\ }\textbf {\bibinfo {volume} {76}},\ \bibinfo {pages} {084019} (\bibinfo {year} {2007})},\ \Eprint {http://arxiv.org/abs/hep-th/0703290} {arXiv:hep-th/0703290} \BibitemShut {NoStop}%
\bibitem [{\citenamefont {Espinosa}\ \emph {et~al.}(2018)\citenamefont {Espinosa}, \citenamefont {Racco},\ and\ \citenamefont {Riotto}}]{Espinosa:2018eve}%
  \BibitemOpen
  \bibfield  {author} {\bibinfo {author} {\bibfnamefont {J.~R.}\ \bibnamefont {Espinosa}}, \bibinfo {author} {\bibfnamefont {D.}~\bibnamefont {Racco}}, \ and\ \bibinfo {author} {\bibfnamefont {A.}~\bibnamefont {Riotto}},\ }\href {\doibase 10.1088/1475-7516/2018/09/012} {\bibfield  {journal} {\bibinfo  {journal} {JCAP}\ }\textbf {\bibinfo {volume} {09}},\ \bibinfo {pages} {012} (\bibinfo {year} {2018})},\ \Eprint {http://arxiv.org/abs/1804.07732} {arXiv:1804.07732 [hep-ph]} \BibitemShut {NoStop}%
\bibitem [{\citenamefont {Kohri}\ and\ \citenamefont {Terada}(2018)}]{Kohri:2018awv}%
  \BibitemOpen
  \bibfield  {author} {\bibinfo {author} {\bibfnamefont {K.}~\bibnamefont {Kohri}}\ and\ \bibinfo {author} {\bibfnamefont {T.}~\bibnamefont {Terada}},\ }\href {\doibase 10.1103/PhysRevD.97.123532} {\bibfield  {journal} {\bibinfo  {journal} {Phys. Rev. D}\ }\textbf {\bibinfo {volume} {97}},\ \bibinfo {pages} {123532} (\bibinfo {year} {2018})},\ \Eprint {http://arxiv.org/abs/1804.08577} {arXiv:1804.08577 [gr-qc]} \BibitemShut {NoStop}%
\bibitem [{\citenamefont {Harigaya}\ \emph {et~al.}(2023)\citenamefont {Harigaya}, \citenamefont {Inomata},\ and\ \citenamefont {Terada}}]{Harigaya:2023pmw}%
  \BibitemOpen
  \bibfield  {author} {\bibinfo {author} {\bibfnamefont {K.}~\bibnamefont {Harigaya}}, \bibinfo {author} {\bibfnamefont {K.}~\bibnamefont {Inomata}}, \ and\ \bibinfo {author} {\bibfnamefont {T.}~\bibnamefont {Terada}},\ }\href {\doibase 10.1103/PhysRevD.108.123538} {\bibfield  {journal} {\bibinfo  {journal} {Phys. Rev. D}\ }\textbf {\bibinfo {volume} {108}},\ \bibinfo {pages} {123538} (\bibinfo {year} {2023})},\ \Eprint {http://arxiv.org/abs/2309.00228} {arXiv:2309.00228 [astro-ph.CO]} \BibitemShut {NoStop}%
\bibitem [{\citenamefont {Guth}\ and\ \citenamefont {Pi}(1982)}]{Guth:1982ec}%
  \BibitemOpen
  \bibfield  {author} {\bibinfo {author} {\bibfnamefont {A.~H.}\ \bibnamefont {Guth}}\ and\ \bibinfo {author} {\bibfnamefont {S.~Y.}\ \bibnamefont {Pi}},\ }\href {\doibase 10.1103/PhysRevLett.49.1110} {\bibfield  {journal} {\bibinfo  {journal} {Phys. Rev. Lett.}\ }\textbf {\bibinfo {volume} {49}},\ \bibinfo {pages} {1110} (\bibinfo {year} {1982})}\BibitemShut {NoStop}%
\bibitem [{\citenamefont {Starobinsky}(1982)}]{Starobinsky:1982ee}%
  \BibitemOpen
  \bibfield  {author} {\bibinfo {author} {\bibfnamefont {A.~A.}\ \bibnamefont {Starobinsky}},\ }\href {\doibase 10.1016/0370-2693(82)90541-X} {\bibfield  {journal} {\bibinfo  {journal} {Phys. Lett. B}\ }\textbf {\bibinfo {volume} {117}},\ \bibinfo {pages} {175} (\bibinfo {year} {1982})}\BibitemShut {NoStop}%
\bibitem [{\citenamefont {Grishchuk}(1974)}]{Grishchuk:1974ny}%
  \BibitemOpen
  \bibfield  {author} {\bibinfo {author} {\bibfnamefont {L.~P.}\ \bibnamefont {Grishchuk}},\ }\href@noop {} {\bibfield  {journal} {\bibinfo  {journal} {Zh. Eksp. Teor. Fiz.}\ }\textbf {\bibinfo {volume} {67}},\ \bibinfo {pages} {825} (\bibinfo {year} {1974})}\BibitemShut {NoStop}%
\bibitem [{\citenamefont {Starobinsky}(1979)}]{Starobinsky:1979ty}%
  \BibitemOpen
  \bibfield  {author} {\bibinfo {author} {\bibfnamefont {A.~A.}\ \bibnamefont {Starobinsky}},\ }\href@noop {} {\bibfield  {journal} {\bibinfo  {journal} {JETP Lett.}\ }\textbf {\bibinfo {volume} {30}},\ \bibinfo {pages} {682} (\bibinfo {year} {1979})}\BibitemShut {NoStop}%
\bibitem [{\citenamefont {Thomas}\ \emph {et~al.}(2023)\citenamefont {Thomas}, \citenamefont {Thomas},\ and\ \citenamefont {Joy}}]{Thomas:2023poh}%
  \BibitemOpen
  \bibfield  {author} {\bibinfo {author} {\bibfnamefont {R.}~\bibnamefont {Thomas}}, \bibinfo {author} {\bibfnamefont {J.}~\bibnamefont {Thomas}}, \ and\ \bibinfo {author} {\bibfnamefont {M.}~\bibnamefont {Joy}},\ }\href {\doibase 10.1016/j.dark.2023.101313} {\bibfield  {journal} {\bibinfo  {journal} {Phys. Dark Univ.}\ }\textbf {\bibinfo {volume} {42}},\ \bibinfo {pages} {101313} (\bibinfo {year} {2023})}\BibitemShut {NoStop}%
\bibitem [{\citenamefont {Cai}(2023)}]{Cai:2022lec}%
  \BibitemOpen
  \bibfield  {author} {\bibinfo {author} {\bibfnamefont {Y.}~\bibnamefont {Cai}},\ }\href {\doibase 10.1103/PhysRevD.107.063512} {\bibfield  {journal} {\bibinfo  {journal} {Phys. Rev. D}\ }\textbf {\bibinfo {volume} {107}},\ \bibinfo {pages} {063512} (\bibinfo {year} {2023})},\ \Eprint {http://arxiv.org/abs/2212.10893} {arXiv:2212.10893 [gr-qc]} \BibitemShut {NoStop}%
\bibitem [{\citenamefont {Peng}\ \emph {et~al.}(2022)\citenamefont {Peng}, \citenamefont {Zeng}, \citenamefont {Fu},\ and\ \citenamefont {Guo}}]{Peng:2022ttg}%
  \BibitemOpen
  \bibfield  {author} {\bibinfo {author} {\bibfnamefont {Z.-Z.}\ \bibnamefont {Peng}}, \bibinfo {author} {\bibfnamefont {Z.-M.}\ \bibnamefont {Zeng}}, \bibinfo {author} {\bibfnamefont {C.}~\bibnamefont {Fu}}, \ and\ \bibinfo {author} {\bibfnamefont {Z.-K.}\ \bibnamefont {Guo}},\ }\href {\doibase 10.1103/PhysRevD.106.124044} {\bibfield  {journal} {\bibinfo  {journal} {Phys. Rev. D}\ }\textbf {\bibinfo {volume} {106}},\ \bibinfo {pages} {124044} (\bibinfo {year} {2022})},\ \Eprint {http://arxiv.org/abs/2209.10374} {arXiv:2209.10374 [gr-qc]} \BibitemShut {NoStop}%
\bibitem [{\citenamefont {Cai}\ and\ \citenamefont {Piao}(2022)}]{Cai:2022nqv}%
  \BibitemOpen
  \bibfield  {author} {\bibinfo {author} {\bibfnamefont {Y.}~\bibnamefont {Cai}}\ and\ \bibinfo {author} {\bibfnamefont {Y.-S.}\ \bibnamefont {Piao}},\ }\href {\doibase 10.1007/JHEP06(2022)067} {\bibfield  {journal} {\bibinfo  {journal} {JHEP}\ }\textbf {\bibinfo {volume} {06}},\ \bibinfo {pages} {067} (\bibinfo {year} {2022})},\ \Eprint {http://arxiv.org/abs/2201.04552} {arXiv:2201.04552 [gr-qc]} \BibitemShut {NoStop}%
\bibitem [{\citenamefont {Fumagalli}\ \emph {et~al.}(2022)\citenamefont {Fumagalli}, \citenamefont {Palma}, \citenamefont {Renaux-Petel}, \citenamefont {Sypsas}, \citenamefont {Witkowski},\ and\ \citenamefont {Zenteno}}]{Fumagalli:2021mpc}%
  \BibitemOpen
  \bibfield  {author} {\bibinfo {author} {\bibfnamefont {J.}~\bibnamefont {Fumagalli}}, \bibinfo {author} {\bibfnamefont {G.~A.}\ \bibnamefont {Palma}}, \bibinfo {author} {\bibfnamefont {S.}~\bibnamefont {Renaux-Petel}}, \bibinfo {author} {\bibfnamefont {S.}~\bibnamefont {Sypsas}}, \bibinfo {author} {\bibfnamefont {L.~T.}\ \bibnamefont {Witkowski}}, \ and\ \bibinfo {author} {\bibfnamefont {C.}~\bibnamefont {Zenteno}},\ }\href {\doibase 10.1007/JHEP03(2022)196} {\bibfield  {journal} {\bibinfo  {journal} {JHEP}\ }\textbf {\bibinfo {volume} {03}},\ \bibinfo {pages} {196} (\bibinfo {year} {2022})},\ \Eprint {http://arxiv.org/abs/2111.14664} {arXiv:2111.14664 [astro-ph.CO]} \BibitemShut {NoStop}%
\bibitem [{\citenamefont {Cai}\ and\ \citenamefont {Piao}(2021)}]{Cai:2020qpu}%
  \BibitemOpen
  \bibfield  {author} {\bibinfo {author} {\bibfnamefont {Y.}~\bibnamefont {Cai}}\ and\ \bibinfo {author} {\bibfnamefont {Y.-S.}\ \bibnamefont {Piao}},\ }\href {\doibase 10.1103/PhysRevD.103.083521} {\bibfield  {journal} {\bibinfo  {journal} {Phys. Rev. D}\ }\textbf {\bibinfo {volume} {103}},\ \bibinfo {pages} {083521} (\bibinfo {year} {2021})},\ \Eprint {http://arxiv.org/abs/2012.11304} {arXiv:2012.11304 [gr-qc]} \BibitemShut {NoStop}%
\bibitem [{\citenamefont {Ito}\ \emph {et~al.}(2021)\citenamefont {Ito}, \citenamefont {Soda},\ and\ \citenamefont {Yamaguchi}}]{Ito:2020neq}%
  \BibitemOpen
  \bibfield  {author} {\bibinfo {author} {\bibfnamefont {A.}~\bibnamefont {Ito}}, \bibinfo {author} {\bibfnamefont {J.}~\bibnamefont {Soda}}, \ and\ \bibinfo {author} {\bibfnamefont {M.}~\bibnamefont {Yamaguchi}},\ }\href {\doibase 10.1088/1475-7516/2021/03/033} {\bibfield  {journal} {\bibinfo  {journal} {JCAP}\ }\textbf {\bibinfo {volume} {03}},\ \bibinfo {pages} {033} (\bibinfo {year} {2021})},\ \Eprint {http://arxiv.org/abs/2009.03611} {arXiv:2009.03611 [astro-ph.CO]} \BibitemShut {NoStop}%
\bibitem [{\citenamefont {Maldacena}(2003)}]{Maldacena:2002vr}%
  \BibitemOpen
  \bibfield  {author} {\bibinfo {author} {\bibfnamefont {J.~M.}\ \bibnamefont {Maldacena}},\ }\href {\doibase 10.1088/1126-6708/2003/05/013} {\bibfield  {journal} {\bibinfo  {journal} {JHEP}\ }\textbf {\bibinfo {volume} {05}},\ \bibinfo {pages} {013} (\bibinfo {year} {2003})},\ \Eprint {http://arxiv.org/abs/astro-ph/0210603} {arXiv:astro-ph/0210603} \BibitemShut {NoStop}%
\bibitem [{\citenamefont {Bartolo}\ \emph {et~al.}(2004)\citenamefont {Bartolo}, \citenamefont {Komatsu}, \citenamefont {Matarrese},\ and\ \citenamefont {Riotto}}]{Bartolo:2004if}%
  \BibitemOpen
  \bibfield  {author} {\bibinfo {author} {\bibfnamefont {N.}~\bibnamefont {Bartolo}}, \bibinfo {author} {\bibfnamefont {E.}~\bibnamefont {Komatsu}}, \bibinfo {author} {\bibfnamefont {S.}~\bibnamefont {Matarrese}}, \ and\ \bibinfo {author} {\bibfnamefont {A.}~\bibnamefont {Riotto}},\ }\href {\doibase 10.1016/j.physrep.2004.08.022} {\bibfield  {journal} {\bibinfo  {journal} {Phys. Rept.}\ }\textbf {\bibinfo {volume} {402}},\ \bibinfo {pages} {103} (\bibinfo {year} {2004})},\ \Eprint {http://arxiv.org/abs/astro-ph/0406398} {arXiv:astro-ph/0406398} \BibitemShut {NoStop}%
\bibitem [{\citenamefont {Bartolo}\ \emph {et~al.}(2021)\citenamefont {Bartolo}, \citenamefont {Caloni}, \citenamefont {Orlando},\ and\ \citenamefont {Ricciardone}}]{Bartolo:2020gsh}%
  \BibitemOpen
  \bibfield  {author} {\bibinfo {author} {\bibfnamefont {N.}~\bibnamefont {Bartolo}}, \bibinfo {author} {\bibfnamefont {L.}~\bibnamefont {Caloni}}, \bibinfo {author} {\bibfnamefont {G.}~\bibnamefont {Orlando}}, \ and\ \bibinfo {author} {\bibfnamefont {A.}~\bibnamefont {Ricciardone}},\ }\href {\doibase 10.1088/1475-7516/2021/03/073} {\bibfield  {journal} {\bibinfo  {journal} {JCAP}\ }\textbf {\bibinfo {volume} {03}},\ \bibinfo {pages} {073} (\bibinfo {year} {2021})},\ \Eprint {http://arxiv.org/abs/2008.01715} {arXiv:2008.01715 [astro-ph.CO]} \BibitemShut {NoStop}%
\bibitem [{\citenamefont {Aoki}\ \emph {et~al.}(2021)\citenamefont {Aoki}, \citenamefont {Gorji}, \citenamefont {Mizuno},\ and\ \citenamefont {Mukohyama}}]{Aoki:2020ila}%
  \BibitemOpen
  \bibfield  {author} {\bibinfo {author} {\bibfnamefont {K.}~\bibnamefont {Aoki}}, \bibinfo {author} {\bibfnamefont {M.~A.}\ \bibnamefont {Gorji}}, \bibinfo {author} {\bibfnamefont {S.}~\bibnamefont {Mizuno}}, \ and\ \bibinfo {author} {\bibfnamefont {S.}~\bibnamefont {Mukohyama}},\ }\href {\doibase 10.1088/1475-7516/2021/01/054} {\bibfield  {journal} {\bibinfo  {journal} {JCAP}\ }\textbf {\bibinfo {volume} {01}},\ \bibinfo {pages} {054} (\bibinfo {year} {2021})},\ \Eprint {http://arxiv.org/abs/2010.03973} {arXiv:2010.03973 [gr-qc]} \BibitemShut {NoStop}%
\bibitem [{\citenamefont {Adshead}\ and\ \citenamefont {Lim}(2010)}]{Adshead:2009bz}%
  \BibitemOpen
  \bibfield  {author} {\bibinfo {author} {\bibfnamefont {P.}~\bibnamefont {Adshead}}\ and\ \bibinfo {author} {\bibfnamefont {E.~A.}\ \bibnamefont {Lim}},\ }\href {\doibase 10.1103/PhysRevD.82.024023} {\bibfield  {journal} {\bibinfo  {journal} {Phys. Rev. D}\ }\textbf {\bibinfo {volume} {82}},\ \bibinfo {pages} {024023} (\bibinfo {year} {2010})},\ \Eprint {http://arxiv.org/abs/0912.1615} {arXiv:0912.1615 [astro-ph.CO]} \BibitemShut {NoStop}%
\bibitem [{\citenamefont {Cai}\ \emph {et~al.}(2019)\citenamefont {Cai}, \citenamefont {Pi},\ and\ \citenamefont {Sasaki}}]{Cai:2018dig}%
  \BibitemOpen
  \bibfield  {author} {\bibinfo {author} {\bibfnamefont {R.-g.}\ \bibnamefont {Cai}}, \bibinfo {author} {\bibfnamefont {S.}~\bibnamefont {Pi}}, \ and\ \bibinfo {author} {\bibfnamefont {M.}~\bibnamefont {Sasaki}},\ }\href {\doibase 10.1103/PhysRevLett.122.201101} {\bibfield  {journal} {\bibinfo  {journal} {Phys. Rev. Lett.}\ }\textbf {\bibinfo {volume} {122}},\ \bibinfo {pages} {201101} (\bibinfo {year} {2019})},\ \Eprint {http://arxiv.org/abs/1810.11000} {arXiv:1810.11000 [astro-ph.CO]} \BibitemShut {NoStop}%
\bibitem [{\citenamefont {Inomata}\ \emph {et~al.}(2021)\citenamefont {Inomata}, \citenamefont {Kawasaki}, \citenamefont {Mukaida},\ and\ \citenamefont {Yanagida}}]{Inomata:2020xad}%
  \BibitemOpen
  \bibfield  {author} {\bibinfo {author} {\bibfnamefont {K.}~\bibnamefont {Inomata}}, \bibinfo {author} {\bibfnamefont {M.}~\bibnamefont {Kawasaki}}, \bibinfo {author} {\bibfnamefont {K.}~\bibnamefont {Mukaida}}, \ and\ \bibinfo {author} {\bibfnamefont {T.~T.}\ \bibnamefont {Yanagida}},\ }\href {\doibase 10.1103/PhysRevLett.126.131301} {\bibfield  {journal} {\bibinfo  {journal} {Phys. Rev. Lett.}\ }\textbf {\bibinfo {volume} {126}},\ \bibinfo {pages} {131301} (\bibinfo {year} {2021})},\ \Eprint {http://arxiv.org/abs/2011.01270} {arXiv:2011.01270 [astro-ph.CO]} \BibitemShut {NoStop}%
\bibitem [{\citenamefont {Atal}\ and\ \citenamefont {Dom\`enech}(2021)}]{Atal:2021jyo}%
  \BibitemOpen
  \bibfield  {author} {\bibinfo {author} {\bibfnamefont {V.}~\bibnamefont {Atal}}\ and\ \bibinfo {author} {\bibfnamefont {G.}~\bibnamefont {Dom\`enech}},\ }\href {\doibase 10.1088/1475-7516/2021/06/001} {\bibfield  {journal} {\bibinfo  {journal} {JCAP}\ }\textbf {\bibinfo {volume} {06}},\ \bibinfo {pages} {001} (\bibinfo {year} {2021})},\ \bibinfo {note} {[Erratum: JCAP 10, E01 (2023)]},\ \Eprint {http://arxiv.org/abs/2103.01056} {arXiv:2103.01056 [astro-ph.CO]} \BibitemShut {NoStop}%
\bibitem [{\citenamefont {Yuan}\ and\ \citenamefont {Huang}(2021)}]{Yuan:2020iwf}%
  \BibitemOpen
  \bibfield  {author} {\bibinfo {author} {\bibfnamefont {C.}~\bibnamefont {Yuan}}\ and\ \bibinfo {author} {\bibfnamefont {Q.-G.}\ \bibnamefont {Huang}},\ }\href {\doibase 10.1016/j.physletb.2021.136606} {\bibfield  {journal} {\bibinfo  {journal} {Phys. Lett. B}\ }\textbf {\bibinfo {volume} {821}},\ \bibinfo {pages} {136606} (\bibinfo {year} {2021})},\ \Eprint {http://arxiv.org/abs/2007.10686} {arXiv:2007.10686 [astro-ph.CO]} \BibitemShut {NoStop}%
\bibitem [{\citenamefont {Adshead}\ \emph {et~al.}(2021)\citenamefont {Adshead}, \citenamefont {Lozanov},\ and\ \citenamefont {Weiner}}]{Adshead:2021hnm}%
  \BibitemOpen
  \bibfield  {author} {\bibinfo {author} {\bibfnamefont {P.}~\bibnamefont {Adshead}}, \bibinfo {author} {\bibfnamefont {K.~D.}\ \bibnamefont {Lozanov}}, \ and\ \bibinfo {author} {\bibfnamefont {Z.~J.}\ \bibnamefont {Weiner}},\ }\href {\doibase 10.1088/1475-7516/2021/10/080} {\bibfield  {journal} {\bibinfo  {journal} {JCAP}\ }\textbf {\bibinfo {volume} {10}},\ \bibinfo {pages} {080} (\bibinfo {year} {2021})},\ \Eprint {http://arxiv.org/abs/2105.01659} {arXiv:2105.01659 [astro-ph.CO]} \BibitemShut {NoStop}%
\bibitem [{\citenamefont {Ragavendra}(2022)}]{Ragavendra:2021qdu}%
  \BibitemOpen
  \bibfield  {author} {\bibinfo {author} {\bibfnamefont {H.~V.}\ \bibnamefont {Ragavendra}},\ }\href {\doibase 10.1103/PhysRevD.105.063533} {\bibfield  {journal} {\bibinfo  {journal} {Phys. Rev. D}\ }\textbf {\bibinfo {volume} {105}},\ \bibinfo {pages} {063533} (\bibinfo {year} {2022})},\ \Eprint {http://arxiv.org/abs/2108.04193} {arXiv:2108.04193 [astro-ph.CO]} \BibitemShut {NoStop}%
\bibitem [{\citenamefont {Rezazadeh}\ \emph {et~al.}(2022)\citenamefont {Rezazadeh}, \citenamefont {Teimoori}, \citenamefont {Karimi},\ and\ \citenamefont {Karami}}]{Rezazadeh:2021clf}%
  \BibitemOpen
  \bibfield  {author} {\bibinfo {author} {\bibfnamefont {K.}~\bibnamefont {Rezazadeh}}, \bibinfo {author} {\bibfnamefont {Z.}~\bibnamefont {Teimoori}}, \bibinfo {author} {\bibfnamefont {S.}~\bibnamefont {Karimi}}, \ and\ \bibinfo {author} {\bibfnamefont {K.}~\bibnamefont {Karami}},\ }\href {\doibase 10.1140/epjc/s10052-022-10735-w} {\bibfield  {journal} {\bibinfo  {journal} {Eur. Phys. J. C}\ }\textbf {\bibinfo {volume} {82}},\ \bibinfo {pages} {758} (\bibinfo {year} {2022})},\ \Eprint {http://arxiv.org/abs/2110.01482} {arXiv:2110.01482 [gr-qc]} \BibitemShut {NoStop}%
\bibitem [{\citenamefont {Zhang}(2022)}]{Zhang:2021rqs}%
  \BibitemOpen
  \bibfield  {author} {\bibinfo {author} {\bibfnamefont {F.}~\bibnamefont {Zhang}},\ }\href {\doibase 10.1103/PhysRevD.105.063539} {\bibfield  {journal} {\bibinfo  {journal} {Phys. Rev. D}\ }\textbf {\bibinfo {volume} {105}},\ \bibinfo {pages} {063539} (\bibinfo {year} {2022})},\ \Eprint {http://arxiv.org/abs/2112.10516} {arXiv:2112.10516 [gr-qc]} \BibitemShut {NoStop}%
\bibitem [{\citenamefont {Lin}\ \emph {et~al.}(2023)\citenamefont {Lin}, \citenamefont {Gao}, \citenamefont {Gong}, \citenamefont {Lu}, \citenamefont {Wang},\ and\ \citenamefont {Zhang}}]{Lin:2021vwc}%
  \BibitemOpen
  \bibfield  {author} {\bibinfo {author} {\bibfnamefont {J.}~\bibnamefont {Lin}}, \bibinfo {author} {\bibfnamefont {S.}~\bibnamefont {Gao}}, \bibinfo {author} {\bibfnamefont {Y.}~\bibnamefont {Gong}}, \bibinfo {author} {\bibfnamefont {Y.}~\bibnamefont {Lu}}, \bibinfo {author} {\bibfnamefont {Z.}~\bibnamefont {Wang}}, \ and\ \bibinfo {author} {\bibfnamefont {F.}~\bibnamefont {Zhang}},\ }\href {\doibase 10.1103/PhysRevD.107.043517} {\bibfield  {journal} {\bibinfo  {journal} {Phys. Rev. D}\ }\textbf {\bibinfo {volume} {107}},\ \bibinfo {pages} {043517} (\bibinfo {year} {2023})},\ \Eprint {http://arxiv.org/abs/2111.01362} {arXiv:2111.01362 [gr-qc]} \BibitemShut {NoStop}%
\bibitem [{\citenamefont {Meng}\ \emph {et~al.}(2022)\citenamefont {Meng}, \citenamefont {Yuan},\ and\ \citenamefont {Huang}}]{Meng:2022ixx}%
  \BibitemOpen
  \bibfield  {author} {\bibinfo {author} {\bibfnamefont {D.-S.}\ \bibnamefont {Meng}}, \bibinfo {author} {\bibfnamefont {C.}~\bibnamefont {Yuan}}, \ and\ \bibinfo {author} {\bibfnamefont {Q.-g.}\ \bibnamefont {Huang}},\ }\href {\doibase 10.1103/PhysRevD.106.063508} {\bibfield  {journal} {\bibinfo  {journal} {Phys. Rev. D}\ }\textbf {\bibinfo {volume} {106}},\ \bibinfo {pages} {063508} (\bibinfo {year} {2022})},\ \Eprint {http://arxiv.org/abs/2207.07668} {arXiv:2207.07668 [astro-ph.CO]} \BibitemShut {NoStop}%
\bibitem [{\citenamefont {Chen}\ \emph {et~al.}(2022)\citenamefont {Chen}, \citenamefont {Yu},\ and\ \citenamefont {Wu}}]{Chen:2022dqr}%
  \BibitemOpen
  \bibfield  {author} {\bibinfo {author} {\bibfnamefont {L.-Y.}\ \bibnamefont {Chen}}, \bibinfo {author} {\bibfnamefont {H.}~\bibnamefont {Yu}}, \ and\ \bibinfo {author} {\bibfnamefont {P.}~\bibnamefont {Wu}},\ }\href {\doibase 10.1103/PhysRevD.106.063537} {\bibfield  {journal} {\bibinfo  {journal} {Phys. Rev. D}\ }\textbf {\bibinfo {volume} {106}},\ \bibinfo {pages} {063537} (\bibinfo {year} {2022})},\ \Eprint {http://arxiv.org/abs/2210.05201} {arXiv:2210.05201 [gr-qc]} \BibitemShut {NoStop}%
\bibitem [{\citenamefont {Abe}\ \emph {et~al.}(2023)\citenamefont {Abe}, \citenamefont {Inui}, \citenamefont {Tada},\ and\ \citenamefont {Yokoyama}}]{Abe:2022xur}%
  \BibitemOpen
  \bibfield  {author} {\bibinfo {author} {\bibfnamefont {K.~T.}\ \bibnamefont {Abe}}, \bibinfo {author} {\bibfnamefont {R.}~\bibnamefont {Inui}}, \bibinfo {author} {\bibfnamefont {Y.}~\bibnamefont {Tada}}, \ and\ \bibinfo {author} {\bibfnamefont {S.}~\bibnamefont {Yokoyama}},\ }\href {\doibase 10.1088/1475-7516/2023/05/044} {\bibfield  {journal} {\bibinfo  {journal} {JCAP}\ }\textbf {\bibinfo {volume} {05}},\ \bibinfo {pages} {044} (\bibinfo {year} {2023})},\ \Eprint {http://arxiv.org/abs/2209.13891} {arXiv:2209.13891 [astro-ph.CO]} \BibitemShut {NoStop}%
\bibitem [{\citenamefont {Chang}\ \emph {et~al.}(2024)\citenamefont {Chang}, \citenamefont {Kuang}, \citenamefont {Wu}, \citenamefont {Zhou},\ and\ \citenamefont {Zhu}}]{Chang:2023aba}%
  \BibitemOpen
  \bibfield  {author} {\bibinfo {author} {\bibfnamefont {Z.}~\bibnamefont {Chang}}, \bibinfo {author} {\bibfnamefont {Y.-T.}\ \bibnamefont {Kuang}}, \bibinfo {author} {\bibfnamefont {D.}~\bibnamefont {Wu}}, \bibinfo {author} {\bibfnamefont {J.-Z.}\ \bibnamefont {Zhou}}, \ and\ \bibinfo {author} {\bibfnamefont {Q.-H.}\ \bibnamefont {Zhu}},\ }\href {\doibase 10.1103/PhysRevD.109.L041303} {\bibfield  {journal} {\bibinfo  {journal} {Phys. Rev. D}\ }\textbf {\bibinfo {volume} {109}},\ \bibinfo {pages} {L041303} (\bibinfo {year} {2024})},\ \Eprint {http://arxiv.org/abs/2311.05102} {arXiv:2311.05102 [astro-ph.CO]} \BibitemShut {NoStop}%
\bibitem [{\citenamefont {Garcia-Saenz}\ \emph {et~al.}(2023)\citenamefont {Garcia-Saenz}, \citenamefont {Pinol}, \citenamefont {Renaux-Petel},\ and\ \citenamefont {Werth}}]{Garcia-Saenz:2022tzu}%
  \BibitemOpen
  \bibfield  {author} {\bibinfo {author} {\bibfnamefont {S.}~\bibnamefont {Garcia-Saenz}}, \bibinfo {author} {\bibfnamefont {L.}~\bibnamefont {Pinol}}, \bibinfo {author} {\bibfnamefont {S.}~\bibnamefont {Renaux-Petel}}, \ and\ \bibinfo {author} {\bibfnamefont {D.}~\bibnamefont {Werth}},\ }\href {\doibase 10.1088/1475-7516/2023/03/057} {\bibfield  {journal} {\bibinfo  {journal} {JCAP}\ }\textbf {\bibinfo {volume} {03}},\ \bibinfo {pages} {057} (\bibinfo {year} {2023})},\ \Eprint {http://arxiv.org/abs/2207.14267} {arXiv:2207.14267 [astro-ph.CO]} \BibitemShut {NoStop}%
\bibitem [{\citenamefont {Bartolo}\ \emph {et~al.}(2019{\natexlab{a}})\citenamefont {Bartolo}, \citenamefont {Bertacca}, \citenamefont {Matarrese}, \citenamefont {Peloso}, \citenamefont {Ricciardone}, \citenamefont {Riotto},\ and\ \citenamefont {Tasinato}}]{Bartolo:2019oiq}%
  \BibitemOpen
  \bibfield  {author} {\bibinfo {author} {\bibfnamefont {N.}~\bibnamefont {Bartolo}}, \bibinfo {author} {\bibfnamefont {D.}~\bibnamefont {Bertacca}}, \bibinfo {author} {\bibfnamefont {S.}~\bibnamefont {Matarrese}}, \bibinfo {author} {\bibfnamefont {M.}~\bibnamefont {Peloso}}, \bibinfo {author} {\bibfnamefont {A.}~\bibnamefont {Ricciardone}}, \bibinfo {author} {\bibfnamefont {A.}~\bibnamefont {Riotto}}, \ and\ \bibinfo {author} {\bibfnamefont {G.}~\bibnamefont {Tasinato}},\ }\href {\doibase 10.1103/PhysRevD.100.121501} {\bibfield  {journal} {\bibinfo  {journal} {Phys. Rev. D}\ }\textbf {\bibinfo {volume} {100}},\ \bibinfo {pages} {121501} (\bibinfo {year} {2019}{\natexlab{a}})},\ \Eprint {http://arxiv.org/abs/1908.00527} {arXiv:1908.00527 [astro-ph.CO]} \BibitemShut {NoStop}%
\bibitem [{\citenamefont {Bartolo}\ \emph {et~al.}(2020)\citenamefont {Bartolo}, \citenamefont {Bertacca}, \citenamefont {De~Luca}, \citenamefont {Franciolini}, \citenamefont {Matarrese}, \citenamefont {Peloso}, \citenamefont {Ricciardone}, \citenamefont {Riotto},\ and\ \citenamefont {Tasinato}}]{Bartolo:2019zvb}%
  \BibitemOpen
  \bibfield  {author} {\bibinfo {author} {\bibfnamefont {N.}~\bibnamefont {Bartolo}}, \bibinfo {author} {\bibfnamefont {D.}~\bibnamefont {Bertacca}}, \bibinfo {author} {\bibfnamefont {V.}~\bibnamefont {De~Luca}}, \bibinfo {author} {\bibfnamefont {G.}~\bibnamefont {Franciolini}}, \bibinfo {author} {\bibfnamefont {S.}~\bibnamefont {Matarrese}}, \bibinfo {author} {\bibfnamefont {M.}~\bibnamefont {Peloso}}, \bibinfo {author} {\bibfnamefont {A.}~\bibnamefont {Ricciardone}}, \bibinfo {author} {\bibfnamefont {A.}~\bibnamefont {Riotto}}, \ and\ \bibinfo {author} {\bibfnamefont {G.}~\bibnamefont {Tasinato}},\ }\href {\doibase 10.1088/1475-7516/2020/02/028} {\bibfield  {journal} {\bibinfo  {journal} {JCAP}\ }\textbf {\bibinfo {volume} {02}},\ \bibinfo {pages} {028} (\bibinfo {year} {2020})},\ \Eprint {http://arxiv.org/abs/1909.12619} {arXiv:1909.12619 [astro-ph.CO]} \BibitemShut {NoStop}%
\bibitem [{\citenamefont {Li}\ \emph {et~al.}(2023)\citenamefont {Li}, \citenamefont {Wang}, \citenamefont {Zhao},\ and\ \citenamefont {Kohri}}]{Li:2023qua}%
  \BibitemOpen
  \bibfield  {author} {\bibinfo {author} {\bibfnamefont {J.-P.}\ \bibnamefont {Li}}, \bibinfo {author} {\bibfnamefont {S.}~\bibnamefont {Wang}}, \bibinfo {author} {\bibfnamefont {Z.-C.}\ \bibnamefont {Zhao}}, \ and\ \bibinfo {author} {\bibfnamefont {K.}~\bibnamefont {Kohri}},\ }\href {\doibase 10.1088/1475-7516/2023/10/056} {\bibfield  {journal} {\bibinfo  {journal} {JCAP}\ }\textbf {\bibinfo {volume} {10}},\ \bibinfo {pages} {056} (\bibinfo {year} {2023})},\ \Eprint {http://arxiv.org/abs/2305.19950} {arXiv:2305.19950 [astro-ph.CO]} \BibitemShut {NoStop}%
\bibitem [{\citenamefont {Li}\ \emph {et~al.}(2024)\citenamefont {Li}, \citenamefont {Wang}, \citenamefont {Zhao},\ and\ \citenamefont {Kohri}}]{Li:2023xtl}%
  \BibitemOpen
  \bibfield  {author} {\bibinfo {author} {\bibfnamefont {J.-P.}\ \bibnamefont {Li}}, \bibinfo {author} {\bibfnamefont {S.}~\bibnamefont {Wang}}, \bibinfo {author} {\bibfnamefont {Z.-C.}\ \bibnamefont {Zhao}}, \ and\ \bibinfo {author} {\bibfnamefont {K.}~\bibnamefont {Kohri}},\ }\href {\doibase 10.1088/1475-7516/2024/06/039} {\bibfield  {journal} {\bibinfo  {journal} {JCAP}\ }\textbf {\bibinfo {volume} {06}},\ \bibinfo {pages} {039} (\bibinfo {year} {2024})},\ \Eprint {http://arxiv.org/abs/2309.07792} {arXiv:2309.07792 [astro-ph.CO]} \BibitemShut {NoStop}%
\bibitem [{\citenamefont {Bartolo}\ \emph {et~al.}(2019{\natexlab{b}})\citenamefont {Bartolo}, \citenamefont {De~Luca}, \citenamefont {Franciolini}, \citenamefont {Lewis}, \citenamefont {Peloso},\ and\ \citenamefont {Riotto}}]{Bartolo:2018evs}%
  \BibitemOpen
  \bibfield  {author} {\bibinfo {author} {\bibfnamefont {N.}~\bibnamefont {Bartolo}}, \bibinfo {author} {\bibfnamefont {V.}~\bibnamefont {De~Luca}}, \bibinfo {author} {\bibfnamefont {G.}~\bibnamefont {Franciolini}}, \bibinfo {author} {\bibfnamefont {A.}~\bibnamefont {Lewis}}, \bibinfo {author} {\bibfnamefont {M.}~\bibnamefont {Peloso}}, \ and\ \bibinfo {author} {\bibfnamefont {A.}~\bibnamefont {Riotto}},\ }\href {\doibase 10.1103/PhysRevLett.122.211301} {\bibfield  {journal} {\bibinfo  {journal} {Phys. Rev. Lett.}\ }\textbf {\bibinfo {volume} {122}},\ \bibinfo {pages} {211301} (\bibinfo {year} {2019}{\natexlab{b}})},\ \Eprint {http://arxiv.org/abs/1810.12218} {arXiv:1810.12218 [astro-ph.CO]} \BibitemShut {NoStop}%
\bibitem [{\citenamefont {Bartolo}\ \emph {et~al.}(2019{\natexlab{c}})\citenamefont {Bartolo}, \citenamefont {De~Luca}, \citenamefont {Franciolini}, \citenamefont {Peloso}, \citenamefont {Racco},\ and\ \citenamefont {Riotto}}]{Bartolo:2018rku}%
  \BibitemOpen
  \bibfield  {author} {\bibinfo {author} {\bibfnamefont {N.}~\bibnamefont {Bartolo}}, \bibinfo {author} {\bibfnamefont {V.}~\bibnamefont {De~Luca}}, \bibinfo {author} {\bibfnamefont {G.}~\bibnamefont {Franciolini}}, \bibinfo {author} {\bibfnamefont {M.}~\bibnamefont {Peloso}}, \bibinfo {author} {\bibfnamefont {D.}~\bibnamefont {Racco}}, \ and\ \bibinfo {author} {\bibfnamefont {A.}~\bibnamefont {Riotto}},\ }\href {\doibase 10.1103/PhysRevD.99.103521} {\bibfield  {journal} {\bibinfo  {journal} {Phys. Rev. D}\ }\textbf {\bibinfo {volume} {99}},\ \bibinfo {pages} {103521} (\bibinfo {year} {2019}{\natexlab{c}})},\ \Eprint {http://arxiv.org/abs/1810.12224} {arXiv:1810.12224 [astro-ph.CO]} \BibitemShut {NoStop}%
\bibitem [{\citenamefont {Bartolo}\ \emph {et~al.}(2018)\citenamefont {Bartolo}, \citenamefont {Domcke}, \citenamefont {Figueroa}, \citenamefont {Garc\'\i{}a-Bellido}, \citenamefont {Peloso}, \citenamefont {Pieroni}, \citenamefont {Ricciardone}, \citenamefont {Sakellariadou}, \citenamefont {Sorbo},\ and\ \citenamefont {Tasinato}}]{Bartolo:2018qqn}%
  \BibitemOpen
  \bibfield  {author} {\bibinfo {author} {\bibfnamefont {N.}~\bibnamefont {Bartolo}}, \bibinfo {author} {\bibfnamefont {V.}~\bibnamefont {Domcke}}, \bibinfo {author} {\bibfnamefont {D.~G.}\ \bibnamefont {Figueroa}}, \bibinfo {author} {\bibfnamefont {J.}~\bibnamefont {Garc\'\i{}a-Bellido}}, \bibinfo {author} {\bibfnamefont {M.}~\bibnamefont {Peloso}}, \bibinfo {author} {\bibfnamefont {M.}~\bibnamefont {Pieroni}}, \bibinfo {author} {\bibfnamefont {A.}~\bibnamefont {Ricciardone}}, \bibinfo {author} {\bibfnamefont {M.}~\bibnamefont {Sakellariadou}}, \bibinfo {author} {\bibfnamefont {L.}~\bibnamefont {Sorbo}}, \ and\ \bibinfo {author} {\bibfnamefont {G.}~\bibnamefont {Tasinato}},\ }\href {\doibase 10.1088/1475-7516/2018/11/034} {\bibfield  {journal} {\bibinfo  {journal} {JCAP}\ }\textbf {\bibinfo {volume} {11}},\ \bibinfo {pages} {034} (\bibinfo {year} {2018})},\ \Eprint {http://arxiv.org/abs/1806.02819} {arXiv:1806.02819 [astro-ph.CO]} \BibitemShut {NoStop}%
\bibitem [{\citenamefont {Tsuneto}\ \emph {et~al.}(2019)\citenamefont {Tsuneto}, \citenamefont {Ito}, \citenamefont {Noumi},\ and\ \citenamefont {Soda}}]{Tsuneto:2018tif}%
  \BibitemOpen
  \bibfield  {author} {\bibinfo {author} {\bibfnamefont {M.}~\bibnamefont {Tsuneto}}, \bibinfo {author} {\bibfnamefont {A.}~\bibnamefont {Ito}}, \bibinfo {author} {\bibfnamefont {T.}~\bibnamefont {Noumi}}, \ and\ \bibinfo {author} {\bibfnamefont {J.}~\bibnamefont {Soda}},\ }\href {\doibase 10.1088/1475-7516/2019/03/032} {\bibfield  {journal} {\bibinfo  {journal} {JCAP}\ }\textbf {\bibinfo {volume} {03}},\ \bibinfo {pages} {032} (\bibinfo {year} {2019})},\ \Eprint {http://arxiv.org/abs/1812.10615} {arXiv:1812.10615 [gr-qc]} \BibitemShut {NoStop}%
\bibitem [{\citenamefont {Powell}\ and\ \citenamefont {Tasinato}(2020)}]{Powell:2019kid}%
  \BibitemOpen
  \bibfield  {author} {\bibinfo {author} {\bibfnamefont {C.}~\bibnamefont {Powell}}\ and\ \bibinfo {author} {\bibfnamefont {G.}~\bibnamefont {Tasinato}},\ }\href {\doibase 10.1088/1475-7516/2020/01/017} {\bibfield  {journal} {\bibinfo  {journal} {JCAP}\ }\textbf {\bibinfo {volume} {01}},\ \bibinfo {pages} {017} (\bibinfo {year} {2020})},\ \Eprint {http://arxiv.org/abs/1910.04758} {arXiv:1910.04758 [gr-qc]} \BibitemShut {NoStop}%
\bibitem [{\citenamefont {Tasinato}(2022)}]{Tasinato:2022xyq}%
  \BibitemOpen
  \bibfield  {author} {\bibinfo {author} {\bibfnamefont {G.}~\bibnamefont {Tasinato}},\ }\href {\doibase 10.1103/PhysRevD.105.083506} {\bibfield  {journal} {\bibinfo  {journal} {Phys. Rev. D}\ }\textbf {\bibinfo {volume} {105}},\ \bibinfo {pages} {083506} (\bibinfo {year} {2022})},\ \Eprint {http://arxiv.org/abs/2203.15440} {arXiv:2203.15440 [gr-qc]} \BibitemShut {NoStop}%
\bibitem [{\citenamefont {Zhu}(2023)}]{Zhu:2022bwf}%
  \BibitemOpen
  \bibfield  {author} {\bibinfo {author} {\bibfnamefont {Q.-H.}\ \bibnamefont {Zhu}},\ }\href {\doibase 10.1103/PhysRevD.107.103519} {\bibfield  {journal} {\bibinfo  {journal} {Phys. Rev. D}\ }\textbf {\bibinfo {volume} {107}},\ \bibinfo {pages} {103519} (\bibinfo {year} {2023})},\ \Eprint {http://arxiv.org/abs/2301.00311} {arXiv:2301.00311 [gr-qc]} \BibitemShut {NoStop}%
\bibitem [{\citenamefont {Sasaki}\ \emph {et~al.}(2018)\citenamefont {Sasaki}, \citenamefont {Suyama}, \citenamefont {Tanaka},\ and\ \citenamefont {Yokoyama}}]{Sasaki:2018dmp}%
  \BibitemOpen
  \bibfield  {author} {\bibinfo {author} {\bibfnamefont {M.}~\bibnamefont {Sasaki}}, \bibinfo {author} {\bibfnamefont {T.}~\bibnamefont {Suyama}}, \bibinfo {author} {\bibfnamefont {T.}~\bibnamefont {Tanaka}}, \ and\ \bibinfo {author} {\bibfnamefont {S.}~\bibnamefont {Yokoyama}},\ }\href {\doibase 10.1088/1361-6382/aaa7b4} {\bibfield  {journal} {\bibinfo  {journal} {Class. Quant. Grav.}\ }\textbf {\bibinfo {volume} {35}},\ \bibinfo {pages} {063001} (\bibinfo {year} {2018})},\ \Eprint {http://arxiv.org/abs/1801.05235} {arXiv:1801.05235 [astro-ph.CO]} \BibitemShut {NoStop}%
\bibitem [{\citenamefont {Dom\`enech}(2021)}]{Domenech:2021ztg}%
  \BibitemOpen
  \bibfield  {author} {\bibinfo {author} {\bibfnamefont {G.}~\bibnamefont {Dom\`enech}},\ }\href {\doibase 10.3390/universe7110398} {\bibfield  {journal} {\bibinfo  {journal} {Universe}\ }\textbf {\bibinfo {volume} {7}},\ \bibinfo {pages} {398} (\bibinfo {year} {2021})},\ \Eprint {http://arxiv.org/abs/2109.01398} {arXiv:2109.01398 [gr-qc]} \BibitemShut {NoStop}%
\bibitem [{\citenamefont {Franciolini}\ \emph {et~al.}(2023)\citenamefont {Franciolini}, \citenamefont {Iovino}, \citenamefont {Vaskonen},\ and\ \citenamefont {Veermae}}]{Franciolini:2023pbf}%
  \BibitemOpen
  \bibfield  {author} {\bibinfo {author} {\bibfnamefont {G.}~\bibnamefont {Franciolini}}, \bibinfo {author} {\bibfnamefont {A.}~\bibnamefont {Iovino}, \bibfnamefont {Junior.}}, \bibinfo {author} {\bibfnamefont {V.}~\bibnamefont {Vaskonen}}, \ and\ \bibinfo {author} {\bibfnamefont {H.}~\bibnamefont {Veermae}},\ }\href {\doibase 10.1103/PhysRevLett.131.201401} {\bibfield  {journal} {\bibinfo  {journal} {Phys. Rev. Lett.}\ }\textbf {\bibinfo {volume} {131}},\ \bibinfo {pages} {201401} (\bibinfo {year} {2023})},\ \Eprint {http://arxiv.org/abs/2306.17149} {arXiv:2306.17149 [astro-ph.CO]} \BibitemShut {NoStop}%
\bibitem [{\citenamefont {Dandoy}\ \emph {et~al.}(2023)\citenamefont {Dandoy}, \citenamefont {Domcke},\ and\ \citenamefont {Rompineve}}]{Dandoy:2023jot}%
  \BibitemOpen
  \bibfield  {author} {\bibinfo {author} {\bibfnamefont {V.}~\bibnamefont {Dandoy}}, \bibinfo {author} {\bibfnamefont {V.}~\bibnamefont {Domcke}}, \ and\ \bibinfo {author} {\bibfnamefont {F.}~\bibnamefont {Rompineve}},\ }\href {\doibase 10.21468/SciPostPhysCore.6.3.060} {\bibfield  {journal} {\bibinfo  {journal} {SciPost Phys. Core}\ }\textbf {\bibinfo {volume} {6}},\ \bibinfo {pages} {060} (\bibinfo {year} {2023})},\ \Eprint {http://arxiv.org/abs/2302.07901} {arXiv:2302.07901 [astro-ph.CO]} \BibitemShut {NoStop}%
\bibitem [{\citenamefont {Gorji}\ \emph {et~al.}(2023)\citenamefont {Gorji}, \citenamefont {Sasaki},\ and\ \citenamefont {Suyama}}]{Gorji:2023sil}%
  \BibitemOpen
  \bibfield  {author} {\bibinfo {author} {\bibfnamefont {M.~A.}\ \bibnamefont {Gorji}}, \bibinfo {author} {\bibfnamefont {M.}~\bibnamefont {Sasaki}}, \ and\ \bibinfo {author} {\bibfnamefont {T.}~\bibnamefont {Suyama}},\ }\href {\doibase 10.1016/j.physletb.2023.138214} {\bibfield  {journal} {\bibinfo  {journal} {Phys. Lett. B}\ }\textbf {\bibinfo {volume} {846}},\ \bibinfo {pages} {138214} (\bibinfo {year} {2023})},\ \Eprint {http://arxiv.org/abs/2307.13109} {arXiv:2307.13109 [astro-ph.CO]} \BibitemShut {NoStop}%
\bibitem [{\citenamefont {Balaji}\ \emph {et~al.}(2023)\citenamefont {Balaji}, \citenamefont {Dom\`enech},\ and\ \citenamefont {Franciolini}}]{Balaji:2023ehk}%
  \BibitemOpen
  \bibfield  {author} {\bibinfo {author} {\bibfnamefont {S.}~\bibnamefont {Balaji}}, \bibinfo {author} {\bibfnamefont {G.}~\bibnamefont {Dom\`enech}}, \ and\ \bibinfo {author} {\bibfnamefont {G.}~\bibnamefont {Franciolini}},\ }\href {\doibase 10.1088/1475-7516/2023/10/041} {\bibfield  {journal} {\bibinfo  {journal} {JCAP}\ }\textbf {\bibinfo {volume} {10}},\ \bibinfo {pages} {041} (\bibinfo {year} {2023})},\ \Eprint {http://arxiv.org/abs/2307.08552} {arXiv:2307.08552 [gr-qc]} \BibitemShut {NoStop}%
\bibitem [{\citenamefont {Maggiore}(2018)}]{Maggiore:2018sht}%
  \BibitemOpen
  \bibfield  {author} {\bibinfo {author} {\bibfnamefont {M.}~\bibnamefont {Maggiore}},\ }\href@noop {} {\emph {\bibinfo {title} {{Gravitational Waves. Vol. 2: Astrophysics and Cosmology}}}}\ (\bibinfo  {publisher} {Oxford University Press},\ \bibinfo {year} {2018})\BibitemShut {NoStop}%
\bibitem [{\citenamefont {Chamberlin}\ and\ \citenamefont {Siemens}(2012)}]{Chamberlin:2011ev}%
  \BibitemOpen
  \bibfield  {author} {\bibinfo {author} {\bibfnamefont {S.~J.}\ \bibnamefont {Chamberlin}}\ and\ \bibinfo {author} {\bibfnamefont {X.}~\bibnamefont {Siemens}},\ }\href {\doibase 10.1103/PhysRevD.85.082001} {\bibfield  {journal} {\bibinfo  {journal} {Phys. Rev. D}\ }\textbf {\bibinfo {volume} {85}},\ \bibinfo {pages} {082001} (\bibinfo {year} {2012})},\ \Eprint {http://arxiv.org/abs/1111.5661} {arXiv:1111.5661 [astro-ph.HE]} \BibitemShut {NoStop}%
\bibitem [{\citenamefont {Mingarelli}\ and\ \citenamefont {Mingarelli}(2018)}]{Mingarelli:2018kgp}%
  \BibitemOpen
  \bibfield  {author} {\bibinfo {author} {\bibfnamefont {C.~M.~F.}\ \bibnamefont {Mingarelli}}\ and\ \bibinfo {author} {\bibfnamefont {A.~B.}\ \bibnamefont {Mingarelli}},\ }\href {\doibase 10.1088/2399-6528/aae06d} {\bibfield  {journal} {\bibinfo  {journal} {J. Phys. Comm.}\ }\textbf {\bibinfo {volume} {2}},\ \bibinfo {pages} {105002} (\bibinfo {year} {2018})},\ \Eprint {http://arxiv.org/abs/1806.06979} {arXiv:1806.06979 [astro-ph.IM]} \BibitemShut {NoStop}%
\end{thebibliography}%
\end{document}